\pdfoutput=1
\documentclass[a4paper, 11pt, oneside]{article}

\usepackage{epsf,amsmath,amssymb,graphicx,dcolumn,slashed}
\usepackage{scalefnt,cite,hyperref,bbold}
\usepackage{cleveref,etoolbox,color,soul}
\usepackage{listings,booktabs}
\usepackage{epstopdf}
\usepackage{amsfonts}

\usepackage{indentfirst}
\usepackage{multirow}

\usepackage{float}

\allowdisplaybreaks

\newcommand{\half}{\tfrac{1}{2}}
\renewcommand {\Im} {\operatorname{Im}}
\renewcommand {\Re} {\operatorname{Re}}
\newcommand{\Lag}{\operatorname{\mathcal{L}}}

\newcommand{\beq}{\begin{equation}}  
\newcommand{\eeq}{\end{equation}}
\newcommand{\bea}{\begin{eqnarray}} 
\newcommand{\eea}{\end{eqnarray}}
\newcommand{\no}{\nonumber \\}
\DeclareMathOperator{\tr}{Tr}

\title{A GOOFy Three Higgs Doublet Model}

\author{
  \addtocounter{footnote}{2} 
  I.~de~Medeiros~Varzielas$^{(1)}$\thanks{\tt ivo.de@udo.edu},
  P.~M.~Ferreira$^{(2,3)}$\thanks{\tt pmmferreira@fc.ul.pt}\, and
  S.~Oliveira$^{(1,3)}$\thanks{\tt sara.m.r.oliveira@tecnico.ulisboa.pt}
    \\*[3mm]
   $^{(1)}\!$
  \small CFTP, Dep. F\'\i sica, Instituto Superior T\'ecnico, Lisboa, Portugal
  \\[2mm]
   $^{(2)}\!$
   \small Instituto~Superior~de~Engenharia~de~Lisboa~---~ISEL,
  1959-007~Lisboa, Portugal
  \\[2mm]
  $^{(3)}\!$
  \small Centro~de~F\'\i sica~Te\'orica~e~Computacional,
  Faculdade~de~Ci\^encias, Universidade~de~Lisboa, \\
  \small Av.~Prof.~Gama~Pinto~2, 1649-003~Lisboa, Portugal
  \\[2mm]
}

\date{\today}

\begin{document}

\maketitle

\begin{abstract}
Recent developments in two-Higgs doublet models identified regions of parameter space which 
are renormalization-group invariant to all orders of perturbation theory but cannot be described
by any of the known symmetries of those models. These were dubbed GOOFy symmetries, as they involve
scalar doublets and their complex conjugates transforming independently, and imaginary scaling of
spacetime coordinates is needed to ensure invariance of kinetic terms. We build a three-Higgs doublet model 
with a GOOFy symmetry including extensions to the fermion sector. We prove renormalization group invariance
through a direct calculation of the model's beta functions and analyse the corresponding phenomenology.
The model is shown to be compatible with current LHC observations, displaying interesting features 
which may be probed in the near future.
\end{abstract}

\section{Introduction}

The scalar sector of the Standard Model (SM) is responsible for electroweak symmetry breaking and the generation of particle masses. Despite its remarkable phenomenological success, the SM leaves several open questions, including the origin of neutrino masses and the stability of the electroweak scale. Extensions of the scalar sector provide a well-motivated framework for exploring physics beyond the SM. Among the simplest and most extensively studied possibilities are Multi-Higgs Doublet Models (NHDMs), which enlarge the scalar content of the theory by having $N$ hypercharge $+1$ $\mathrm{SU(2)}_L$ doublets.

The addition of scalar doublets drastically enlarges the parameter space of the theory: the two-Higgs Doublet Model (2HDM)~\cite{Lee:1973iz,Branco:2011iw} 
involves 14 independent parameters in its most general form, while the three-Higgs Doublet Model (3HDM) already contains 54.\footnote{After accounting for basis 
redefinitions in the scalar field space, these reduce to 11 and 46 physical parameters, respectively.} While this proliferation complicates theoretical and 
phenomenological analyses, the imposition of symmetries is primarily motivated by their physical implications rather than by mere parameter reduction. Moreover, 
symmetries encode structural principles of the theory, constrain the dynamics, and can enforce desirable phenomenological features. A well-known example is the 
imposition of a $\mathbb{Z}_2$ symmetry in the 2HDM, which forbids tree-level flavour-changing neutral currents (FCNCs)~\cite{Glashow:1976nt,Paschos:1976ay}. 
The resulting reduction of free parameters is therefore a consequence of symmetry, not its fundamental purpose.
For many years, the global symmetries of the $SU(2)\times U(1)$ 2HDM scalar potential were believed to be completely classified. These were 
grouped into two broad families: \textit{Higgs-family symmetries} and \textit{generalised CP symmetries}. The former correspond to unitary 
transformations in the scalar field space,
\begin{equation}
    \Phi_i' = U_{ij}\Phi_j, \qquad i,j=1,2, \qquad U \in U(2),
    \label{eq:hfs}
\end{equation}
while the latter involves generalised CP transformations of the form
\begin{equation}
    \Phi_i' = X_{ij}\Phi_j^{*}, \qquad i,j=1,2, \qquad X \in U(2).
    \label{eq:gcp}
\end{equation}
It was shown~\cite{Ivanov:2007de} that there were only {\em six} such possible symmetries -- three of the first kind and three of the second.
Recently, however, Ferreira, Grzadkowski, Ogreid, and Osland~\cite{Ferreira:2023dke} identified a previously unnoticed class of symmetries --- now 
referred to as GOOFy symmetries --- through an analysis of the renormalization group (RG) evolution of the scalar parameters. They observed that 
certain relations among the parameters of the scalar potential remain RG invariant to all orders in perturbation theory when both scalar and gauge
interactions are taken into account. The origin of these relations was traced to an imaginary rescaling of the Higgs doublet fields. While such a 
transformation by itself would spoil the invariance of the kinetic terms, this can be compensated by a corresponding imaginary rescaling of the 
spacetime coordinates, together with a consistent rescaling of the gauge fields, thereby restoring invariance of the full scalar and gauge Lagrangian. 
An extension of this construction to the Yukawa sector has also been achieved~\cite{Ferreira:2026ccg}.
There has been extensive work on this subject, with different interpretations of the cause of these RG invariant relations and their 
extension to several versions of the 2HDM and even 
3HDM~\cite{Doring:2024kdg,Haber:2025cbb,Trautner:2025yxz,Ferreira:2025ate,deBoer:2025jhc,Trautner:2025prm,Grzadkowski:2026gkx,deBoer:2026ktb,Kuncinas:2025uty,Pilaftsis:2026wyw,deMedeirosVarzielas:2026jsk}.

In this work, we extend the study of GOOFy symmetries to the three-Higgs Doublet Model, with the intention of obtaining a viable Yukawa sector. 
In Section~2 we review GOOFy symmetries and discuss their implementation within a 3HDM, including the fermionic sector. We also analyse electroweak
symmetry breaking in the model and present its renormalization group equations, demonstrating that the GOOFy parameter relations found are 
preserved under renormalization. Section~3 discusses the softly broken model we must introduce to have a realistic fermion sector. We further discuss
the several theoretical and experimental constraints we impose on the model and present results from a numerical scan of the model, including
predictions for Higgs phenomenology which may be measurable at the LHC. We conclude in Section 4.

\section{The model}

The three-Higgs doublet model (3HDM) is an extension of the Standard Model in which two additional $\mathrm{SU(2)}_L$ doublets with hypercharge $+1$
are added to the scalar sector. With this extra field content, the tree-level value for the $\rho$ parameter is found to be exactly $1$, so 
compliance with electroweak precision constraints is simplified. The 3HDM is more than simply an unnecessary complication of the scalar sector --
it brings with it new and interesting phenomenology which cannot be accounted for within other models, such as the 2HDM. For instance, a 3HDM
with a $\mathbb{Z}_2\times \mathbb{Z}_2$ symmetry was proposed by Weinberg~\cite{Weinberg:1976hu}, wherein CP violation occurs in the interactions
of the charged scalars. Coexisting charge and neutral breaking minima are possible in the 3HDM~\cite{Barroso:2006pa} but not in the 
2HDM~\cite{Ferreira:2004yd,Barroso:2005sm,Ivanov:2007de}, so the vacuum structure of the model acquires novel properties. A non-exhaustive
list of papers that deal with the 3HDM, including an exploration of its geometric structure involving the bilinear formalism; possible
Abelian symmetries; boundedness from below conditions; exploration of the phenomenology of $A_4$, $S_4$ and $\Delta(27)$ symmetries and associated CP properties; and Yukawa sector
phenomenology is shown  in~\cite{
Ivanov:2010ww,
deMedeirosVarzielas:2011zw, deMedeirosVarzielas:2012ylr, deMedeirosVarzielas:2012rxf, 
Bhattacharyya:2012pi,
deMedeirosVarzielas:2013dmf,
CarcamoHernandez:2013yiy,
Ivanov:2013bka,Ivanov:2014doa,Degee:2012sk,
deMedeirosVarzielas:2015ybd,
deMedeirosVarzielas:2016rii,
deMedeirosVarzielas:2017glw,
deMedeirosVarzielas:2017ote,
deMedeirosVarzielas:2019rrp,
Faro:2019vcd,Das:2021oik,
CarcamoHernandez:2021osw,
deMedeirosVarzielas:2021zqs,
Kalinowski:2021lvw,
Ferreira:2022zil,Carrolo:2022oyg,deMedeirosVarzielas:2022kbj,Ferreira:2022sno,
Barreto:2025jzx,
deMedeirosVarzielas:2025byb}.
These works, and many others, considered usual symmetry transformations  as the source
of symmetries with the 3HDM. Recently, however, a new type of ``GOOFy" symmetries were identified in the 2HDM. Let us first review those
new symmetries in order to apply them to the 3HDM. 

\subsection{GOOFy symmetries}

It is well known that parameter relations resulting from a given symmetry are invariant under renormalization to all orders. However, this statement
is not an equivalence, and no demonstration exists that the opposite is also true -- that renormalization-group (RG) invariant relations between parameters
in a given theory necessarily imply the existence of a symmetry. Such were the findings of Ref.~\cite{Ferreira:2023dke} where
certain relations for scalar potential parameters were found to be RG invariant to all orders of perturbation theory \textit{but} those relations
cannot be reproduced by any field transformations such as~\eqref{eq:hfs} or~\eqref{eq:gcp}. This raised the question of possible new symmetries having been found,
corresponding to different field transformations. The work of~\cite{Ferreira:2023dke} dealt with the 2HDM
potential, which in its most generic $\mathrm{SU(2)}_L \times U(1)_Y$ invariant and renormalizable form may be written as
\bea
V &=& m_{11}^2\Phi_1^\dagger\Phi_1+m_{22}^2\Phi_2^\dagger\Phi_2
-[m_{12}^2\Phi_1^\dagger\Phi_2+{\rm h.c.}]+\half\lambda_1(\Phi_1^\dagger\Phi_1)^2
+\half\lambda_2(\Phi_2^\dagger\Phi_2)^2
+\lambda_3(\Phi_1^\dagger\Phi_1)(\Phi_2^\dagger\Phi_2)\nonumber\\[8pt]
&&\quad
+\lambda_4(\Phi_1^\dagger\Phi_2)(\Phi_2^\dagger\Phi_1)
+\left\{\half\lambda_5(\Phi_1^\dagger\Phi_2)^2
+\big[\lambda_6(\Phi_1^\dagger\Phi_1)
+\lambda_7(\Phi_2^\dagger\Phi_2)\big]
\Phi_1^\dagger\Phi_2+{\rm h.c.}\right\}\,,
 \label{eq:pot}
\eea
with all parameters real with the exception of $m^2_{12}$ and $\lambda_{5,6,7}$. It was found that the parameter relations
\begin{equation}
    m_{11}^2 + m_{22}^2 = 0, 
    \qquad
    \lambda_1 = \lambda_2,
    \qquad
    \lambda_6 = -\lambda_7
    \label{eq:goofy2}
\end{equation}
remain invariant under RG evolution to all orders of perturbation theory when considering only the scalar and gauge sectors. A general proof of
RG invariance considering the Yukawa sector as well was not found in~\cite{Ferreira:2023dke}, although two (symmetry-imposed) Yukawa matrix textures were 
found, by direct calculation of the model's $\beta$-functions, to preserve the relations~\eqref{eq:goofy2} to two loops. 

The relations among quartic couplings in~\eqref{eq:goofy2} can be obtained by a ``standard" 2HDM symmetry, namely the CP2 symmetry wherein invariance of the
theory under the field transformations $\Phi_1 \to \Phi_2^*$, $\Phi_2 \to -\Phi_1^*$ is required. But the impact of that symmetry in the potential's
quadratic coefficients forces $m_{11}^2 = m_{22}^2$ and $m_{12}^2 = 0$, which are different from those in~\eqref{eq:goofy2}. Indeed, it can be shown
that no HF or GCP symmetry can yield the first of the conditions in~\eqref{eq:goofy2}. It was 
shown in~\cite{Ferreira:2023dke} that the relations~\eqref{eq:goofy2} could be obtained under the unorthodox field transformation
\beq
\begin{array}{rclcrcl}
\Phi_1 &\to& \Phi_2^* & , &
\Phi_1^\dagger &\to& -\Phi_2^T, \\
\Phi_2 &\to& -\Phi_1^* & , &
\Phi_2^\dagger &\to&  \Phi_1^T\,.
\end{array}
\label{eq:trandou}
\eeq
The transformations on the first column are identical to the CP2 case, but notice that the transformation of $\Phi_1^\dagger$ ($\Phi_2^\dagger$) 
is defined as \textit{minus} the hermitian conjugate of the transformation
of $\Phi_1$ ($\Phi_2$). In particular, these transformations imply $|\Phi_1|^2 + |\Phi_2|^2 \to -(|\Phi_1|^2 + |\Phi_2|^2)$, which 
yields the unusual relation between quadratic parameters in~\eqref{eq:goofy2}. It is also impossible to obtain such a transformation with
``normal" HF or GCP field transformations.

To explain the origin of such transformations, Ref.~\cite{Ferreira:2023dke} proposed that they arise from an \textit{imaginary scaling} of the real components of the scalar doublets. Writing the doublets in terms of real scalar fields $\phi_i$ as
\bea
\Phi_1=\frac{1}{\sqrt{2}}
\begin{pmatrix}
	\phi_1+i\phi_2 \\
	\phi_3+i\phi_4
\end{pmatrix}, \quad
\Phi_2=\frac{1}{\sqrt{2}}
\begin{pmatrix}
	\phi_5+i\phi_6 \\
	\phi_7+i\phi_8
\end{pmatrix}\,,
\eea
one may obtain the transformations of eq.~\eqref{eq:trandou} by imposing
\beq
\begin{array}{rcrcrcrcrcrcrcr}
 \phi_1 &\to& -i\phi_6 &,&  \phi_2 &\to& -i\phi_5 &,&  \phi_3 &\to& -i\phi_8 &,&  \phi_4 &\to& -i\phi_7, \\
\phi_5 &\to& i\phi_2 &,& \phi_6 &\to& i\phi_1 &,& \phi_7 &\to& i\phi_4 &,& \phi_8 &\to& i\phi_3. 
\end{array}
\label{eq:imss}
\eeq
In order to preserve the invariance of the kinetic Lagrangian under this transformation, a corresponding imaginary scaling of the spacetime coordinates is introduced, $x_\mu \rightarrow i x_\mu$, together with appropriate transformations of the gauge fields, to wit
\beq
B_\mu \to i B_\mu \quad , \quad  
W^1_\mu \to i W^1_\mu \quad ,\quad
W^2_\mu \to -i W^2_\mu \quad,\quad
W^3_\mu \to i W^3_\mu\,,
\label{eq:imsg}
\eeq
Analogous imaginary scalings are introduced for the fermionic fields to ensure invariance of the fermion's kinetic fields, their gauge
interactions and the Yukawa terms~\cite{Ferreira:2026ccg}. Considering only quarks, the 2HDM Yukawa lagrangian is given by 
\beq
-{\cal L_Y}\,=\, \overline{Q}_L\Gamma_1 \Phi_1 n_R\,+\,\overline{Q}_L\Gamma_2 \Phi_2 n_R\,+\,
\overline{Q}_L\Delta_1 \tilde{\Phi}_1 p_R\,+\,\overline{Q}_L\Delta_2 \tilde{\Phi}_2 p_R
+{\rm h.c.}\,,
\label{eq:yuk}
\eeq
where the $n_R$/ $p_R$ fields are the right-handed negative/positive quark fields which, when rotated to the mass basis,
give the right-handed down/up quarks. In the same manner the left handed $Q_L$ doublets are 3-vectors in flavour space as well and contain
the unrotated left quark fields. The $3\times 3$ complex matrices
$\Gamma_i$ and $\Delta_i$ are $3\times 3$ containing the down/up Yukawa couplings of the model. As usual $\tilde{\Phi}_i = i \sigma_2 \Phi_i^*$
is the charge conjugate of the scalar doublets. The extension to fermions of the imaginary scalings of eq.~\eqref{eq:imss} for the
scalars and~\eqref{eq:imsg} for the gauge fields are given by
\beq
\begin{array}{rclcrcl}
Q_L &\to &  X_\alpha\,\gamma^0 \,C\,Q_L^* & , &
\overline{Q}_L &\to & -\,i\,Q_L^T\,C\, X^\dagger_\alpha, \\
n_R &\to &  X_\beta\,\gamma^0\,C\,n_R^* & , &
\overline{n}_R &\to & -\,i\,n_R^T\,C\, X^\dagger_\beta, \\
p_R &\to &  X_\gamma\,\gamma^0\,C\,p_R^* & , &
\overline{p}_R &\to & -\,i\,p_R^T\,C\, X^\dagger_\gamma \,,
\end{array}
\label{eq:imsf}
\eeq
where, as in~\cite{Ferreira:2010bm}, the three unitary $3\times 3$ matrices $X_\theta$ act in flavour space and mix the different quarks, and
$C$ is the charge conjugation matrix (such that $C^2 = -1$, $C^{-1} = C^\dagger = -C$, and $C\gamma^\mu C^{-1} = -{\gamma^\mu}^T$). Notice how,
in analogy with the scalar transformations on eq.~\eqref{eq:trandou}, each field and its hermitian conjugate transform independently (in the former 
case a minus sign distinguishes those transformations, in the latter a factor of ``i"). These scalings imply constraints on the Yukawa matrices, 
which end up being  analogous to those obtained with other ``regular" symmetries. As shown in~\cite{Ferreira:2026ccg}, with the scalar transformations
of eq.~\eqref{eq:trandou} it is possible to choose the matrices $X_\theta$ such that the Yukawa matrices of the theory end up being exactly those of the CP2 
model~\cite{Ferreira:2010bm}.

The initials of the authors of~\cite{Ferreira:2023dke}, plus the bizarre nature of these field transformations, justified their being
dubbed ``GOOFy symmetries". Whether this GOOFy procedure is indeed being caused by a genuine symmetry or by something else 
(see~\cite{Doring:2024kdg,Haber:2025cbb,Trautner:2025yxz,Ferreira:2025ate,deBoer:2025jhc,Trautner:2025prm,Grzadkowski:2026gkx,deBoer:2026ktb,Kuncinas:2025uty,
Pilaftsis:2026wyw,deMedeirosVarzielas:2026jsk} for alternative 
explanations), the fact remains that it is an effective procedure to find hitherto unknown RG-invariant regions of parameter space within specific models, 
with interesting phenomenology. We will now apply these techniques to the case of the 3HDM. 

\subsection{Imposing a GOOFy symmetry in the Three Higgs Doublet Model }

The work presented here extends the ideas of the previous section to the three-Higgs Doublet Model. Let us recall that the most general  
$\mathrm{SU(2)}_L \times U(1)_Y$  renormalizable scalar potential of the 3HDM may be written as
\begin{align}
    V = 
    &
    - m_{11}^2 (\Phi_1^{\dagger} \Phi_1)
    - m_{22}^2 (\Phi_2^{\dagger} \Phi_2)
    - m_{33}^2 (\Phi_3^{\dagger} \Phi_3)-
        \big[ m_{12}^2 (\Phi_1^{\dagger} \Phi_2)
    +m_{13}^2 (\Phi_1^{\dagger} \Phi_3)
    +m_{23}^2 (\Phi_2^{\dagger} \Phi_3)\: +\:
       {\mathrm{h.c.}} \big] 
       \nonumber\\ &
    +\lambda_{1} (\Phi_1^{\dagger} \Phi_1)^2 
    + \lambda_{2} (\Phi_2^{\dagger} \Phi_2)^2 
    + \lambda_{3} (\Phi_3^{\dagger} \Phi_3)^2 +
         \lambda_{4} (\Phi_1^{\dagger} \Phi_1) (\Phi_2^{\dagger} \Phi_2) 
    + \lambda_{5} (\Phi_1^{\dagger} \Phi_1) (\Phi_3^{\dagger} \Phi_3) 
        \nonumber \\ &
    + \lambda_{6} (\Phi_2^{\dagger} \Phi_2) (\Phi_3^{\dagger} \Phi_3) 
    + \lambda_{7} (\Phi_1^{\dagger} \Phi_2) (\Phi_2^{\dagger} \Phi_1) 
    + \lambda_{8} (\Phi_1^{\dagger} \Phi_3) (\Phi_3^{\dagger} \Phi_1) 
    + \lambda_{9} (\Phi_2^{\dagger} \Phi_3) (\Phi_3^{\dagger} \Phi_2) 
         \nonumber \\ &
     + \big\{\, \lambda_{10} (\Phi_1^{\dagger} \Phi_2)^2 
    + \lambda_{11} (\Phi_1^{\dagger} \Phi_3)^2 
    + \lambda_{12}  (\Phi_2^{\dagger} \Phi_3)^2 
    + [\lambda_{13}  (\Phi_1^{\dagger} \Phi_1) 
    + \lambda_{14}  (\Phi_2^{\dagger} \Phi_2) 
    + \lambda_{15}  (\Phi_3^{\dagger} \Phi_3)] (\Phi_1^{\dagger} \Phi_2) 
         \nonumber \\ & 
    + [\lambda_{16}  (\Phi_1^{\dagger} \Phi_1) 
    + \lambda_{17}  (\Phi_2^{\dagger} \Phi_2) 
    + \lambda_{18}  (\Phi_3^{\dagger} \Phi_3)](\Phi_1^{\dagger} \Phi_3) 
    + [\lambda_{19}  (\Phi_1^{\dagger} \Phi_1) 
    + \lambda_{20}  (\Phi_2^{\dagger} \Phi_2) 
    + \lambda_{21}  (\Phi_3^{\dagger} \Phi_3)] (\Phi_2^{\dagger} \Phi_3) 
        \nonumber \\ & 
    + \lambda_{22}  (\Phi_1^{\dagger} \Phi_2) (\Phi_1^{\dagger} \Phi_3) 
    + \lambda_{23} (\Phi_2^{\dagger} \Phi_1) (\Phi_1^{\dagger} \Phi_3) 
    + \lambda_{24}    (\Phi_1^{\dagger} \Phi_3) (\Phi_2^{\dagger} \Phi_3) 
    + \lambda_{25}   (\Phi_1^{\dagger} \Phi_3) (\Phi_3^{\dagger} \Phi_2)  
    \nonumber \\ & 
    + \lambda_{26}  (\Phi_2^{\dagger} \Phi_1) (\Phi_2^{\dagger} \Phi_3) 
    + \lambda_{27} (\Phi_1^{\dagger} \Phi_2) (\Phi_2^{\dagger} \Phi_3) 
    +{\mathrm{h.c.}}\big\} \ \label{eq:3HDM_pot}
\end{align}
where the parameters $m^2_{11},\ m^2_{22},\ m^2_{33},\ \lambda_{1},\ \lambda_{2},\ \lambda_{3},\ \lambda_{4},\ \lambda_{5},\ \lambda_{6},\ \lambda_{7},\ 
\lambda_{8}\ \text{and}\ \lambda_{9}$ are real, while all the others are complex. In all, the potential seemingly has 54 real parameters, but 
freedom to choose a basis for the scalar doublets $\Phi_i$ reduces that number to 46 independent real ones. Still, that is a very large number 
of free parameters, which considerably limits the predictive power of the model. Imposing symmetries on the model is therefore an effective way
of reducing its parameter space, with the added benefit of providing us with interesting phenomenology. Examples of such symmetries include the 
aforementioned $\mathbb{Z}_2\times \mathbb{Z}_2$  
Weinberg model~\cite{Weinberg:1976hu}, the $A_4$ model studied in~\cite{Degee:2012sk,Carrolo:2022oyg} or a model~\cite{Das:2021oik,Ferreira:2022zil} 
with a symmetry inspired by the 2HDM BGL model~\cite{Branco:1996bq}. In all these cases, the global symmetries considered correspond to field transformations
analogous to those of eqs.~\eqref{eq:hfs} or~\eqref{eq:gcp}, which by construction preserve the theory's kinetic terms and gauge interactions.

Motivated by the structure of the transformations identified in the 2HDM case, we consider the following set of transformations among the scalar doublets:
\begin{align}
    \Phi_1 &\rightarrow \Phi_2^*, 
    &
    \Phi_1^\dagger &\rightarrow - \Phi_2^\top, \nonumber\\
    \Phi_2 &\rightarrow \Phi_3^*, 
    &
    \Phi_2^\dagger &\rightarrow - \Phi_3^\top, \nonumber\\
    \Phi_3 &\rightarrow \Phi_1^*, 
    &
    \Phi_3^\dagger &\rightarrow - \Phi_1^\top . 
    \label{eq: GOOFy_transfos}
\end{align}
If one were to consider only the transformations of the first column in the equations above, they would describe a 3HDM GCP symmetry,
wherein the standard CP transformation (complex conjugation of a field) is taken in tandem with a cyclic permutation between the two fields,
$\Phi_1 \to \Phi_2 \to \Phi_3 \to \Phi_1$. The GOOFy aspect is found in the second column above, with extra minus signs which differentiate
the transformation of each field's hermitian conjugate from the hermitian conjugate of its transformation. Notice, however, that this extra minus sign 
will have no bearing on the impact of the symmetry on the quartic couplings of the potential -- those multiply terms of the form 
$\Phi_i^\dagger\Phi_j\Phi_k^\dagger\Phi_l$, for which the extra minus signs in~\eqref{eq: GOOFy_transfos} will cancel out. Indeed, we 
see that the impact of the GOOFy symmetry on the quartic couplings will be identical to that of the GCP + cyclical permuation symmetry
mentioned above -- the differences arise in the quadratic sector. This has in fact been observed in all GOOFy models proposed thus far.

The field transformations of~\eqref{eq: GOOFy_transfos} can
also be understood as resulting from imaginary scalings of the doublets' real components (as in the 2HDM case, see eq.~\eqref{eq:imss}). 
Requiring that the scalar potential is invariant under
these transformations leads to a set of parameter relations that, as we will see further ahead, are stable under RG evolution. 
The resulting scalar potential is given by
\begin{align}
    V_{\text{GOOFy}} =\;& -M \big[ \Im(\phi_1^\dagger \phi_2)
    +\Im(\phi_2^\dagger \phi_3)
    +\Im(\phi_3^\dagger \phi_1) \big] +\Lambda_0 \left[ (\phi_1^\dagger \phi_1)
    +(\phi_2^\dagger \phi_2)
    +(\phi_3^\dagger \phi_3) \right]^2 \nonumber\\ 
    & +\Lambda_1 \Big[ (\phi_1^\dagger \phi_1)^2
    +(\phi_2^\dagger \phi_2)^2
    +(\phi_3^\dagger \phi_3)^2 -
    (\phi_1^\dagger \phi_1)(\phi_2^\dagger \phi_2) - (\phi_1^\dagger \phi_1)(\phi_3^\dagger \phi_3)
    - (\phi_2^\dagger \phi_2)(\phi_3^\dagger \phi_3)
    \Big] \nonumber\\
    &+ \Lambda_2
    \left[
    \Re(\phi_1^\dagger \phi_2)^2
    +\Re(\phi_2^\dagger \phi_3)^2
    +\Re(\phi_3^\dagger \phi_1)^2
    \right]+\Lambda_3
    \left[
    \Im(\phi_1^\dagger \phi_2)^2
    +\Im(\phi_2^\dagger \phi_3)^2
    +\Im(\phi_3^\dagger \phi_1)^2
    \right] \nonumber\\
    &+\Lambda_4
    \Big[
    \Re(\phi_1^\dagger \phi_2)\Re(\phi_2^\dagger \phi_3)+\Re(\phi_1^\dagger \phi_2)\Re(\phi_3^\dagger \phi_1)
    +\Re(\phi_2^\dagger \phi_3)\Re(\phi_3^\dagger \phi_1)
    \Big] \nonumber\\
    &+ \Lambda_5
    \Big[
    \Im(\phi_1^\dagger \phi_2)\Im(\phi_2^\dagger \phi_3)+\Im(\phi_1^\dagger \phi_2)\Im(\phi_3^\dagger \phi_1)
    +\Im(\phi_2^\dagger \phi_3)\Im(\phi_3^\dagger \phi_1)
    \Big]\nonumber\\
    &+\Lambda_6
    \left(\phi_1^\dagger \phi_1+\phi_2^\dagger \phi_2+\phi_3^\dagger \phi_3\right)
    \left[
    \Re(\phi_1^\dagger \phi_2)
    +\Re(\phi_2^\dagger \phi_3)
    +\Re(\phi_3^\dagger \phi_1)
    \right]\nonumber\\
    &+\Lambda_7
    \Big[
    \Re(\phi_1^\dagger \phi_2)
    (\phi_1^\dagger \phi_1+\phi_2^\dagger \phi_2-2\phi_3^\dagger \phi_3)
    +\Re(\phi_2^\dagger \phi_3)
    (-2\phi_1^\dagger \phi_1+\phi_2^\dagger \phi_2+\phi_3^\dagger \phi_3)
    \nonumber\\
    &+\Re(\phi_3^\dagger \phi_1)
    (\phi_1^\dagger \phi_1-2\phi_2^\dagger \phi_2+\phi_3^\dagger \phi_3)
    \Big]+ \nonumber\\
    &\Lambda_8
    \Big[
    \Re(\phi_1^\dagger \phi_2)
    (\phi_1^\dagger \phi_1-\phi_2^\dagger \phi_2) +\Re(\phi_3^\dagger \phi_1)
    (\phi_3^\dagger \phi_3-\phi_1^\dagger \phi_1)
    +\Re(\phi_2^\dagger \phi_3)
    (\phi_2^\dagger \phi_2-\phi_3^\dagger \phi_3)
    \Big] 
    \label{eq: 3HDM_GOOFy}
\end{align}
where all the parameters are real. These can be expressed in terms of those of the general 3HDM potential in Eq.~\eqref{eq:3HDM_pot} through the relations
\begin{equation}
    \begin{gathered}
        m^2_{12} = m^2_{23} = {m^2_{13}}^* = -\frac{iM}{2}, \qquad \lambda_1 = 2\Lambda_0, \qquad \lambda_4 = 2\Lambda_1, \qquad \lambda_7 = \Lambda_2+\Lambda_3, \qquad \lambda_{10} = \frac{\Lambda_2-\Lambda_3}{2}, \\
        \lambda_{13} = -\Lambda_6+\Lambda_7+\Lambda_8, \qquad \lambda_{14} = -\Lambda_6+\Lambda_7-\Lambda_8, \qquad \lambda_{15} = -\Lambda_6-2\Lambda_7, \\
        \lambda_{22} = \frac{\Lambda_4+\Lambda_5}{2}, \qquad \lambda_{23} = \frac{\Lambda_4-\Lambda_5}{2}\,,
    \end{gathered}
\label{eq:goofy3}
\end{equation}
with all other terms not explicitly written above equal to zero. This potential is in a different basis but corresponds to the GOOFy $S_3$ 
case of ref.~\cite{Kuncinas:2025uty}, wherein 
3HDM scalar potential symmetries were exhaustively classified, and where general necessary conditions for non-vanishing bi-linear terms were presented.

Let us now consider the invariance of the kinetic terms. 
Following the procedure outlined in~\cite{Ferreira:2023dke,Ferreira:2026ccg} consider the kinetic terms of the doublets, $\Lag_{\Phi K}$, given by 
\begin{equation}
    \Lag_{\Phi K} = (\partial_\mu\Phi_1)^\dagger(\partial^\mu\Phi_1)+(\partial_\mu\Phi_2)^\dagger(\partial^\mu\Phi_2)+(\partial_\mu\Phi_3)^\dagger(\partial^\mu\Phi_3) 
    \label{eq: lag_kin}
\end{equation}
Applying the transformations of Eq. \eqref{eq: GOOFy_transfos}, $\Lag_{\Phi K}$ would transform as $\Lag_{\Phi K} \rightarrow - \Lag_{\Phi K}$. This means
that, as in the 2HDM case, we need to perform an imaginary scaling of the spacetime coordinates: 
\begin{equation}
    x_\mu \rightarrow i x_\mu
\end{equation}
To account for the gauge boson interactions, the partial derivative $\partial_\mu$ must be replaced by the covariant derivative $D_\mu$, defined as follows,
\begin{equation}
    D_\mu = \partial_\mu + \frac{i}{2} g \sigma_a W_\mu^a - \frac{i}{2} g' Y B_\mu
    \label{eq:covder}
\end{equation}
where $\sigma_a$ are the Pauli matrices and $Y$ is the hypercharge. 
Taking the imaginary scaling of spacetime coordinates and the doublet transformations \eqref{eq: GOOFy_transfos} into account, invariance of 
the full gauge-kinetic lagrangian is achieved if once again the gauge fields are scaled by imaginary factors exactly as in~\eqref{eq:imsg}.
Indeed, it is found that each covariant derivative transforms as
\begin{align}
    D_\mu \Phi_1 \rightarrow i (D_\mu \Phi_2)^*, \qquad (D_\mu \Phi_1)^\dagger \rightarrow -i (D_\mu \Phi_2)^\top \nonumber\\
    D_\mu \Phi_2 \rightarrow i (D_\mu \Phi_3)^*, \qquad (D_\mu \Phi_2)^\dagger \rightarrow -i (D_\mu \Phi_3)^\top \nonumber\\
    D_\mu \Phi_3 \rightarrow i (D_\mu \Phi_1)^*, \qquad (D_\mu \Phi_3)^\dagger \rightarrow -i (D_\mu \Phi_1)^\top
\end{align}
which leaves the full gauge-kinetic terms invariant. And as in the 2HDM case, the imaginary scaling of gauge fields in~\eqref{eq:imsg}
leaves invariant the purely gauge part of the lagrangian,
\begin{equation}
    \Lag_{GK} = - \frac{1}{4} B_{\mu\nu} B^{\mu\nu} - \frac{1}{4} W^a_{\mu\nu} W^{a~\mu\nu}\,.
\end{equation}

\subsection{Electroweak symmetry breaking}

Before addressing the inclusion of fermions in the theory let us look at how electroweak symmetry occurs in this model. The 3HDM can have 
minima of different nature, in which vacuum expectation values (vevs) for the scalar fields break the $SU(2)_L\times U(1)_Y$ symmetry leaving the
usual electromagnetism symmetry group $U(1)_{em}$ intact, but with the CP symmetry unbroken, or not, by the vacuum; or charge-breaking vevs
occur and the photon acquires a mass; or discrete/continuous global symmetries of the potential are, or are not,
spontaneously broken. These different possible minima depend on the values of the potential's parameters. Since we are interested in realistic phenomenology,
 the most general neutral vev configuration preserving electromagnetic gauge invariance may be written as
\begin{equation}
    \langle\Phi_a\rangle =
    \frac{1}{\sqrt{2}}
    \begin{pmatrix}
        0 \\
        v_a e^{i\theta_a}
    \end{pmatrix},
    \label{eq:vevs}
\end{equation}
with
\begin{equation}
    v^2 = v_1^2 + v_2^2 + v_3^2 = (246~\mathrm{GeV})^2 
\end{equation}
and arbitrary real phases $\theta_a$, $a = 1,2,3$.
Using the global $\mathrm{U(1)}_Y$ hypercharge freedom, one overall phase can be removed allowing, without loss of generality, one of the vevs
(conventionally $v_1$) to be chosen real and positive.

In the case of the GOOFy potential 
of eq.~\eqref{eq: 3HDM_GOOFy}, the non-existence of the quadratic coefficients $m^2_{11}$, $m^2_{22}$ and $m^2_{33}$ makes it impossible for extrema with any $v_i = 0$ to occur -- all three doublets must have non-zero vevs. A direct
inspection of the minimization equations revealed that the only~\footnote{For the neutral vevs of eq.~\eqref{eq:vevs} there are five independent minimization 
conditions to solve. The analytical general solution of eq.~\eqref{eq:vevtr} was obtained via a direct calculation. Nevertheless, an exhaustive numerical 
search for other solutions was done, and no other solution was found, leading us to believe that this is indeed the only possible minimum of the potential.} 
possible vacuum of this model is of the form
\begin{equation}
    (\langle\Phi_1\rangle,\langle\Phi_2\rangle,\langle\Phi_3\rangle)\,=\,\frac{v}{\sqrt{6}}(1,\omega,\omega^2), \qquad \omega = e^{i\frac{2\pi}{3}}\,.
    \label{eq:vevtr}
\end{equation}
This vacuum clearly preserves the $S_3$ permutation symmetry, $\Phi_1\to\Phi_2\to\Phi_3\to\Phi_1$, and the value of 
$M$ is fixed by the minimization conditions,
\begin{equation}
    M = \frac{v^2}{6\sqrt{3}}(12\Lambda_0 + \Lambda_2 + 3 \Lambda_3 + \Lambda_4 + 3 \Lambda_5 - 6 \Lambda_6) \,.
\end{equation}
This yields the value of the potential at the minimum,
\begin{equation}
    V = -\frac{\sqrt{3}}{8}Mv^2\,.
\end{equation}
We can now compute the second derivatives of the potential, evaluate them at the solution of the minimization conditions~\eqref{eq:vevtr}
and compute the squared scalar masses of the model. The results reflect the restrictive symmetry both of the potential and of the vevs: we 
find, as expected, five different neutral scalars -- one which we call $h_{SM}$, and two pairs of degenerate states, $h_1$ and $h_2$ --
and two charged scalars, $H_1^\pm$ and $H_2^\pm$, with the masses of these particles given by
\begin{align}
    &m_{h_{SM}}^2 = \frac{v^2}{6}(12\Lambda_0 + \Lambda_2 + 3 \Lambda_3 + \Lambda_4 + 3 \Lambda_5 - 6 \Lambda_6)\nonumber\\
    & m_{h_{1,2}}^2 = \frac{v^2}{48} \Big(C \pm \sqrt{A^2 + 16(B^2 + 27 \Lambda_8^2)}\Big) \qquad \qquad \text{(twice degenerate)}\nonumber\\
    & m_{H_1^\pm}^2 = \frac{v^2}{6} (12\Lambda_0 + \Lambda_2 + \Lambda_4 - 6 \Lambda_6) \nonumber\\ 
    & m_{H_2^\pm}^2 = \frac{v^2}{12} (12\Lambda_0 - 2 \Lambda_2 - 2 \Lambda_4 + 3 \Lambda_6)
    \label{eq:massesex}
\end{align}
where
\begin{align}
    & A = 24 \Lambda_0 + 14 \Lambda_2 - 6 \Lambda_3 + 5 \Lambda_4 + 3 \Lambda_5 - 30 \Lambda_6 - 36 \Lambda_7 \nonumber \\ 
    & B = 6 \Lambda_1 - 2 \Lambda_2 + \Lambda_4 - 3 \Lambda_7  \nonumber \\
    & C = 72 \Lambda_0 + 24 \Lambda_1 + 10 \Lambda_2 + 6 \Lambda_3 - 5 \Lambda_4 - 3 \Lambda_5 - 18 \Lambda_6 - 12 \Lambda_7 \label{eq: ABC}\,.
\end{align}
We find the rotation matrix $R$ to the neutral mass eigenbasis, as presented in Appendix \ref{sec: rot_mass_basis}. With this we can 
determine how each of the neutral states couples to the gauge bosons in the usual manner: starting with the scalars' kinetic terms written 
in terms of covariant derivatives defined in~\eqref{eq:covder},
\beq
\Lag_{\Phi K} \, = \, \sum_{a=1}^3\,(D_\mu\Phi_a)^\dagger(D^\mu\Phi_a)\,,
\label{eq:LK}
\eeq
and rewriting each doublet as
\beq
\Phi_a\,=\,\begin{pmatrix}
    \phi_a^+ \\ \frac{1}{\sqrt{2}} (v_a e^{i\theta_a}  + \rho_a + i \eta_a)
\end{pmatrix}
\label{eq:doub}
\eeq
then, for each neutral scalar mass eigenstate $S$, the trilinear vertices between $S$ and a $Z$ boson are obtained 
expanding $\Lag_{\Phi K}$ and identifying the terms of the form $SZZ$. Due to the rotation matrix~\eqref{eq: physical_states_scalars}
and the vevs~\eqref{eq:vevtr}, we find
\begin{equation}
    \Lag_{SZZ} \,=\, \frac{g^2 + {g^\prime}^2}{8}
    \left(\eta_{2}-\eta_{3}+\frac{2 \rho_{1}}{\sqrt{3}}-\frac{\rho_{2}}{\sqrt{3}}-\frac{\rho_{3}}{\sqrt{3}}\right)v\,Z_\mu Z^\mu = 
    \frac{g^2 + {g^\prime}^2}{4} v\,h_{SM}\,Z_\mu Z^\mu \,.
\end{equation}
This means that the only neutral scalar which has trilinear vertices with $Z$ bosons is $h_{SM}$, and its couplings are identical to
those of the SM Higgs. The same thing is observed for interactions with the $W$ gauge bosons. Thus the states $h_1$ and $h_2$ are gaugephobic,
and \textit{a priori} might be good dark matter candidates. This is clearly a consequence of the unbroken $S_3$ symmetry of the vacuum
and would be quite an interesting feature of this model. The Yukawa sector, however, complicates matters.

\subsection{Including fermions}

The GOOFy symmetry now needs to be extended to the fermion sector. As in the original 2HDM GOOFy model, the 3HDM field transformations of 
eq.~\eqref{eq: GOOFy_transfos} are akin to a GCP symmetry, so GCP-like transformations for the fermions will also be considered. In fact, they are
identical in form to the ones defined for the 2HDM in eq.~\eqref{eq:imsf} for the quark fields (we do not treat leptons explicitly in what follows, 
but the reasoning is analogous). 

We proceed as done in~\cite{Ferreira:2026ccg} to find the conditions that the GOOFy transformation imposes on the Yukawa matrices. 
We consider only the case of the down quarks, the remaining fermions would be dealt with in the same manner. With the GOOFy transformations on 
the scalars, eq.~\eqref{eq: GOOFy_transfos}, and fermions, eq.~\eqref{eq:imsf}, the Yukawa lagrangian transforms as
\begin{align}
    -\Lag_{Y,d} &= \overline{Q}_L (\Gamma_1\Phi_1 + \Gamma_2\Phi_2 + \Gamma_3\Phi_3) n_R + \overline{n}_R (\Gamma^\dagger_1\Phi^\dagger_1 + \Gamma^\dagger_2\Phi^\dagger_2 + \Gamma^\dagger_3\Phi^\dagger_3) Q_L \rightarrow \label{eq:lYd} \\
    & \rightarrow i \overline{Q}_L X^\top_\alpha(\Gamma^*_1\Phi_2 + \Gamma^*_2\Phi_3 + \Gamma^*_3\Phi_1) X^*_\beta n_R + i \overline{n}_R X^\top_\beta (\Gamma^\top_1\Phi^\dagger_2 + \Gamma^\top_2\Phi^\dagger_3 + \Gamma^\top_3\Phi^\dagger_1) X^*_\alpha Q_L\,.
\end{align}
From this we obtain the following conditions to be satisfied by the $3\times 3$ complex Yukawa matrices $\Gamma$,
\bea
    \Gamma^*_1 &=&- i X_\alpha^\dagger \Gamma_3 X_\beta \nonumber \\
    \Gamma^*_2 &=& - i X_\alpha^\dagger \Gamma_1 X_\beta \nonumber \\
    \Gamma^*_3 &=& - i X_\alpha^\dagger \Gamma_2 X_\beta  \label{eq: Yukawa_couplings}\,.
\eea
Recall that the $X_\theta$ are unitary $3\times 3$ matrices which will in principle mix quark fields. We follow the procedure 
of~\cite{Ferreira:2010bm} to determine which choices of $X_\theta$ matrices may produce an acceptable fermionic sector.

\subsubsection{The Trivial Case: $X_\alpha = X_\beta = \mathbb{1}$} 

The simplest possibility is to take all $X_\theta$ matrices to be the identity, which implies no mixing between generations in the
GCP-like GOOFy transformations. With this the equations~\eqref{eq: Yukawa_couplings} yield
\begin{equation}
    \Gamma = \Gamma_1 = \Gamma_2 = \Gamma_3 = 
    e^{i\pi/4}
    \begin{pmatrix}
        a_{11} & a_{12} & a_{13}\\
        a_{21} & a_{22} & a_{23}\\
        a_{31} & a_{32} & a_{33}\\
    \end{pmatrix}, \qquad a_{ij} \in \mathbb{R},~j = 1,2,3 \,.
    \label{eq: massless_Yuk}
\end{equation}
The $e^{i\pi/4}$ phase can readily be absorbed by an overall rephasing of (for instance) the right-handed fermionic fields, which means thet $\Gamma$
matrix above is simply an arbitrary, real $3\times 3$ matrix. The down-type quark mass matrix thus becomes
\bea
    M_n &=& \frac{1}{\sqrt{2}}\,(\Gamma_1 v_1 + \Gamma_2 v_2 e^{i\theta_2}+ \Gamma_3 v_3 e^{i\theta_3})  \label{eq:Mn} \\
    &=& \frac{v \Gamma(1+\omega+\omega^2)}{\sqrt{6}}\,
\eea
and since $\omega = e^{2i \pi/3}$ we obtain $1+\omega+\omega^2 = 0$, which means that this model cannot generate fermion masses. 
Even though this is clearly not realistic it is still an interesting case, since it corresponds to a model of massless fermions that interact with scalars, 
defying the idea that everything that couples to the Higgs must be massive. Models such as these one are not new, in fact it has been shown in 
\cite{deMedeirosVarzielas:2011gly} that exact Yukawa alignment leads to at least two massless fermions for the most general set of vevs. Even though the 
mechanisms that lead to mass cancellation are different in both cases, they present the same feature of vanishing masses.

\subsubsection{Generic $X_\alpha$ and $X_\beta$}  \label{sec:yuk}

Following the methods of~\cite{Ferreira:2010bm} we use the work of~\cite{Ecker:1987qp} where it was shown that in GCP fermionic transformations it is 
always possible to choose a fermion basis such that, without loss of generality, 
\begin{equation}
    X_\theta = 
    \begin{pmatrix}
        \cos \theta & \sin \theta & 0\\
        -\sin \theta & \cos \theta & 0\\
        0 & 0 & 1\\
    \end{pmatrix}, \qquad \theta \in [0\,,\,\pi/2].
    \label{eq:X}
\end{equation}
This simplifies considerably the treatment of eqs.~\eqref{eq: Yukawa_couplings}~\footnote{The arguments of~\cite{Ecker:1987qp} rely only on
the freedom to choose fermionic basis, and will not be affected by the fact that our GOOFy GCP transformations, eq.~\eqref{eq:imsf}, are 
different from regular ones.}, which can be manipulated to yield
\bea
    X_\alpha \Gamma_1^* + i\Gamma_3 X_\beta &=& 0 \nonumber \\
    X_\alpha \Gamma_2^* + i\Gamma_1 X_\beta&=& 0 \no
    X_\alpha \Gamma_3^* + i\Gamma_2 X_\beta &=& 0\,.
\eea
Following~\cite{Ferreira:2026ccg},
we redefine the Yukawa matrices such that $\Gamma_a \to e^{-i {\pi\over4}} \Gamma_a, ~a = 1,2,3$. This eliminates the factor of $i$ in the above 
equations~\footnote{This could also be achieved by a basis transformation on the $n_R$ fields.} so we are left with
\bea
    X_\alpha \Gamma_1^* + \Gamma_3 X_\beta &=& 0 \no
    X_\alpha  \Gamma_2^* + \Gamma_1 X_\beta &=& 0 \no
    X_\alpha  \Gamma_3^* + \Gamma_2 X_\beta &=& 0\,.
    \label{eq:sysG}
\eea
This represents a set of 18 linear equations (since each matrix has 9 entries, which have real and imaginary parts).  The specific form of the $X$ matrices
in eq.~\eqref{eq:X} helps to simplify the problem: we start by considering the blocks $mn, 3m, n3$ and $33 , ~m,n = 1,2$ of said equations and for each of
these blocks, there are two sets of homogeneous equations, one for the real ($Re$) and another for the imaginary ($Im$) components of 
$\Gamma_1$, $\Gamma_2$  and $\Gamma_3$. These equations can easily be expressed in matrix form and we know that if the determinant of the matrices 
-- which will depend on the angles $\alpha$ and $\gamma$ for the down sector -- 
describing the system is non-vanishing, the corresponding parameters are 0. The results for each block are then summarised in table~\ref{tab:summ}. We
\begin{table}[htpb!]
    \centering
    \caption{Summary of the conditions for vanishing parameters for each block of the system of equations~\eqref{eq:sysG}.}
    \begin{tabular}{cccc}
        \toprule
        $\Gamma_a$& \multirow{2}{*}{Component} & \multirow{2}{*}{Determinant} & Condition for vanishing \\
        matrix element &&& determinant ($ k \in \mathbb{Z}$)\\
        \midrule
        \multirow{2}{*}{33} & Re & 2 & Never\\
        & Im & 0 & Always\\
        \multirow{2}{*}{13,23} & Re & $4 \cos^2(\frac{3 \alpha}{2})$ & $\alpha = \frac{\pi}{3} + \frac{2\pi k}{3}$\\
        & Im & $4 \sin^2(\frac{3 \alpha}{2})$ & $\alpha = \frac{2\pi k}{3}$\\
        \multirow{2}{*}{31,32} & Re & $4 \cos^2(\frac{3 \gamma}{2})$ & $\gamma = \frac{\pi}{3} + \frac{2\pi k}{3}$\\
        & Im & $4 \sin^2(\frac{3 \gamma}{2})$ & $\gamma = \frac{2\pi k}{3}$\\
        \multirow{2}{*}{11,12,21,22} & Re & $4 [\cos (3 \alpha )+\cos (3 \gamma )]^2$ & $\alpha = \frac{\pi}{3} \pm \gamma + \frac{2\pi k}{3}$\\
        & Im & $4 [\cos (3 \alpha )-\cos (3 \gamma )]^2$ & $\alpha = \pm \gamma + \frac{2\pi k}{3}$\\        
        \bottomrule
    \end{tabular}
    \label{tab:summ}
\end{table}
see, for instance, that the determinant of the $3\times 3$ matrix multiplying $\Re(\Gamma_1)_{33}$, $\Re(\Gamma_3)_{33}$ and $\Re(\Gamma_3)_{33}$ is non-zero
for any values of the angles $\alpha$ and $\gamma$, which immediately implies those quantities must be zero so that eqs.~\eqref{eq:sysG} are
satisfied. The same does not happen for the imaginary parts of those matrix entries, though, and one finds 
$\Im(\Gamma_1)_{33} = \Im(\Gamma_3)_{33} = \Im(\Gamma_3)_{33}$. Similar analyses must be conducted for all other entries of the matrix equations~\eqref{eq:sysG}.

There would appear to be infinite possibilities for the Yukawa textures. However, the above conclusions for the $(3,3)$ entries of the $\Gamma$ matrices lead to
\begin{equation}
    (M_n)_{33} = \frac{1}{\sqrt{12}}\,\left[\Im(\Gamma_1)_{33} + \omega \Im(\Gamma_2)_{33} + \omega^2 \Im(\Gamma_3)_{33}\right] = 
    \frac{\Im(\Gamma_1)_{33}}{\sqrt{12}}\,\left[1 + \omega  + \omega^2 \right]= 0. 
\end{equation}
Therefore, if either the third column or row of the down quark mass matrix $M_n$ are zero, i.e. for 
$\alpha \neq k \pi/3$ or $\gamma \neq k \pi/3,~k\in\mathbb{Z}$, at least one of the quark generations is massless. This puts severe restrictions in the allowed 
combinations of $\alpha$ and $\gamma$. Indeed, the simple requirement of 3 massive quarks imposes  $\alpha,\gamma = \frac{k \pi}{3}$ and either 
$\alpha = \frac{\pi}{3} \pm \gamma + \frac{2\pi k}{3}$ or $\alpha = \pm \gamma + \frac{2\pi k}{3},~k \in \mathbb{Z}$. In total, we have 24 different possible solutions
of eqs.~\eqref{eq:sysG} that can be grouped into three classes as shown in Table~\ref{tab: algle_class}.
\begin{table}[htbp!]
    \centering
    \caption{Number of massless fermion generations for each pair of angles $(\alpha,\gamma)=(i\pi/3,j\pi/3)$.}
    \label{tab: algle_class}
    \begin{tabular}{cl}
        \toprule
        Massless generations & $(i,~j)$ \\
        \midrule
        3 & $(0,0),\ (3,0),\ (0,3)$ \\[0.5em]
        2 & $(0,1),\ (0,5),\ (3,1),\ (3,4),\ (3,5),$ \\
          & $(1,0),\ (2,0),\ (4,0),\ (5,0),\ (5,3)$ \\[0.5em]
        0 & $(1,1),\ (1,5),\ (2,1),\ (2,4),\ (2,5),$ \\
          & $(4,1),\ (4,5),\ (5,1),\ (5,2),\ (5,5),\ (4,2)$ \\
        \bottomrule
    \end{tabular}
    \label{tab:text}
\end{table}
The third class is the only one that agrees with experimental evidence of three massive quarks, so all discussion from this point deals with
those specific solutions. The corresponding Yukawa matrices are presented in Appendix~\ref{ap:yukmat}. Here we will only present the one that we use in 
the analysis of the next chapter, which is the $(1,1)$ case above so that, for $\alpha=\gamma = \frac{\pi}{3}$, we obtain, for $\Gamma_1$,
\beq
\Gamma_1 = \begin{pmatrix}
        i a_{11} & i a_{12} & a_{13} \\
        i a_{21} & i a_{22} & a_{23} \\
        a_{31} & a_{32} & i a_{33}
    \end{pmatrix} \,.
\eeq
We also obtain expressions for $\Gamma_2$ and $\Gamma_3$, which include factors of ``$i$" exactly in the same positions where they
occur in $\Gamma_1$ (\textit{i.e.}, the $\{11,12,21,22\}$ block and the $(3,3)$ element). It is then a trivial matter to rephase the quark fields 
to get rid of these imaginary units~\footnote{The basis transformation in question acts on the different flavours of quarks as $n_{L,1}\to -i n_{L,1}$, 
$n_{L,2}\to -i n_{L,2}$, so that the first and second columns of the matrix above are multiplied by $-i$, and $n_{R,3}\to -i n_{R,3}$ which does the same thing
for its third line}. All Yukawa matrices then become
\begin{align}
    &\Gamma_1 = \begin{pmatrix}
        a_{11} &  a_{12} & a_{13} \\
        a_{21} &  a_{22} & a_{23} \\
        a_{31} & a_{32} &  a_{33}
    \end{pmatrix} \,,\nonumber\\[1em]
    &\Gamma_2 = \frac{1}{2}\begin{pmatrix}
         \frac{1}{2} \left(a_{11}-\sqrt{3} a_{12}- \sqrt{3} a_{21}+3 a_{22}\right) & \frac{1}{2} \left( \sqrt{3} a_{11}+a_{12}-3 a_{21}-\sqrt{3} a_{22}\right) &  -a_{13} + \sqrt{3} a_{23} \\
         \frac{1}{2} \left( \sqrt{3} a_{11}- 3 a_{12}+a_{21}- \sqrt{3} a_{22}\right) & \frac{1}{2} \left(3 a_{11}+ \sqrt{3} a_{12}+ \sqrt{3} a_{21}+a_{22}\right) & - \sqrt{3} a_{13}-a_{23} \\
        -a_{31} + \sqrt{3} a_{32} & - \sqrt{3} a_{31}-a_{32} & 2 a_{33} 
    \end{pmatrix}\,, \nonumber \\[1em]
    &\Gamma_3 = \frac{1}{2}\begin{pmatrix}
        \frac{1}{2} \left(a_{11}+\sqrt{3} a_{12}+ \sqrt{3} a_{21}+3 a_{22}\right) & \frac{1}{2} \left(- \sqrt{3} a_{11}+a_{12}-3 a_{21}+\sqrt{3} a_{22}\right) &  -a_{13}-\sqrt{3} a_{23} \\
         \frac{1}{2} \left( -\sqrt{3} a_{11}- 3 a_{12}+a_{21}+ \sqrt{3} a_{22}\right) & \frac{1}{2} \left(3 a_{11}- \sqrt{3} a_{12}- \sqrt{3} a_{21}+a_{22}\right) & \sqrt{3} a_{13}-a_{23} \\
        -a_{31} - \sqrt{3} a_{32} &  \sqrt{3} a_{31}-a_{32} & 2 a_{33} 
    \end{pmatrix} \,,
    \label{eq:Yukawa_texture} 
\end{align}
where all nine coefficients $a_{ij}$ are real. The same textures are found for the up-quark and lepton Yukawa matrices, and the fact that a fermionic 
rephasing can render those matrices reals holds for all possible choices of angles for the third class in table~\ref{tab:text}.

If one now uses the vevs from eq.~\eqref{eq:vevtr} to compute the down-quark mass matrix, 
eq.~\eqref{eq:Mn}, we obtain three massive quarks, and have enough freedom in the choice of parameters to correctly 
reproduce the experimental values for those masses. The same is true for the up-quark mass matrix $M_p = \sum_{a=1}^3 \Delta_a v_a e^{-i\theta_a}/\sqrt{2}$ : 
the $\Delta$ Yukawa matrices have textures 
analogous to those  of~\eqref{eq:Yukawa_texture} and now depends on 9 real $b_{ij}$ parameters, so that along with the vevs from~\eqref{eq:vevtr}
yield a mass matrix whose eigenvalues can easily be fitted to the know values of the up, charm and top masses.

One may now move on to the calculation of the CKM matrix: as usual, the quark mass matrices must be rotated to the mass basis via unitary transformations
on the left and right quark fields, 
\bea
M_d \,=\, \mbox{diag}(m_d\,,\,m_s\,,\,m_b) & = & {U_L^n}^\dagger \,M_n\, U_R^n \,, \no
M_u \,=\, \mbox{diag}(m_u\,,\,m_c\,,\,m_t) & = & {U_L^p}^\dagger \,M_p\, U_R^p \,,
\eea
with the CKM matrix depending only on the left rotation matrices. The high level of symmetry both on the Yukawa matrices and the vev leads to a
remarkable, if frustrating, result:  the left rotation matrices for the up and dwon sector  $U_L$ are \textit{constant}. Indeed,
a direct calculation shows that
\begin{equation}
    U_L^n =\begin{pmatrix}
            -\frac{i}{\sqrt{2}} & \frac{i}{\sqrt{2}} & 0 \\
            \frac{1}{\sqrt{2}} & \frac{1}{\sqrt{2}} & 0 \\
            0 & 0 & 1
          \end{pmatrix} \qquad \quad  U_L^p = \begin{pmatrix}
            \frac{i}{\sqrt{2}} & -\frac{i}{\sqrt{2}} & 0 \\
            \frac{1}{\sqrt{2}} & \frac{1}{\sqrt{2}} & 0 \\
            0 & 0 & 1
          \end{pmatrix}
\end{equation}
so that the CKM matric V is found to be
\beq
    V = ({U_L^p}^\dagger) U_L^n =\begin{pmatrix}
        0 & 1 & 0\\
        1 & 0 & 0\\
        0 & 0 & 1\\
    \end{pmatrix}
    \,,
\eeq
and with a trivial basis rotation on the fields, this means that $V$ is the identity matrix -- meaning, in this model there is no
mixing between generations due to the $W$ gauge boson~\footnote{There will be some mixing due to the neutral scalars of the model, as there are
tree-level flavour-changing neutral currents in this theory.}. The model is therefore in disagreement with a wealth of experimental data and must
be changed. This feature is common to the other possible textures with three massive quarks in table~\ref{tab:text}, which bring nothing new 
to this discussion (indeed, it is well possible that many of thoose textures are related via basis changes).

Since the CKM matrix has off-diagonal elements with magnitudes much smaller than its diagonal ones (by themselves with magnitudes close
to 1), the simplest way to generalize the model is to softly break the GOOFy symmetry via extra quadratic terms in the 
potential~\eqref{eq: 3HDM_GOOFy}. This will allow us to obtain a more generic form for the vevs, which we study 
in section~\ref{sec:soft}. Before, however, let us address the RG invariance of this GOOFy model.

\subsection{Renormalization group invariance of GOOFy parameter relations} 

In the original GOOFy paper~\cite{Ferreira:2023dke} it was possible to show, using arguments of dimensional analysis and
basis invariance, that the relations between parameters of eq.~\eqref{eq:goofy2} were renormalization group (RG) invariant to all orders of perturbation
theory when only scalar and gauge couplings were included. RG invariance up to two loops was proven for specific Yukawa textures as well (those
corresponding to a CP2 and CP3 symmetry) by direct computation of the model's $\beta$-functions, though an all-order demonstration could not be obtained.
The interest of GOOFy symmetries is precisely their implying new, hitherto unknown RG-invariant regions of parameter space, which cannot be reproduced
by common symmetries. There isn't yet a demonstration that this will hold to all orders of perturbation theory, particularly when the Yukawa sector is considered.
There are good arguments for it in refs.~\cite{Trautner:2025prm,Trautner:2025yxz}, where GOOFy symmetry relations emerge from independent transformations on fields 
and their hermitian conjugates, and it is argued that gauge kinetic terms which break those transformations are tantamount to soft breakings and therefore do not 
change the renormalizability of the model. We now wish to analyse whether the GOOFy parameter relations of the 3HDM we built -- to wit, the scalar sector relations
of eqs.~\eqref{eq: 3HDM_GOOFy} and~\eqref{eq:goofy3}, and the Yukawa textures of eqs.~\eqref{eq: massless_Yuk} and~\eqref{eq:Yukawa_texture} -- are preserved under renormalization.

To begin with we need not worry about the RG invariance of the dimensionless couplings. Indeed, if we think about the ``$S_3$ GCP" symmetry $\Phi_1\to \Phi_2^*$, 
$\Phi_2\to \Phi_3^*$ and $\Phi_3\to \Phi_1^*$,  it will yield exactly the quartic scalar couplings relations of eq.~\eqref{eq:goofy3}. Likewise, this 
symmetry requires that Yukawa matrices are like those of~\eqref{eq: massless_Yuk}~\footnote{Minus the $e^{i\pi/4}$ overall phase, which we explained 
earlier was spurious.} or those of~\eqref{eq:Yukawa_texture}. Thus, since the GOOFy relations on dimensionless couplings are identical to those 
obtained by a regular symmetry, their RG invariance is automatically guaranteed. This had already been anticipated and explained in the discussion 
following eq.~\eqref{eq: GOOFy_transfos}. Our GOOFy symmetry differs from  the regular $S_3$ GCP one in the quadratic couplings, and we will now show 
the RG invariance up to two loops in the GOOFy quadratic parameter relations.

We computed the one-loop 3HDM $\beta$-functions using the results from~\cite{Cheng:1973nv,Machacek:1984zw,Luo:2002ti,Schienbein:2018fsw,Sartore:2020pkk}
which give, for the $m^2_{ij}$ parameters, 
\bea
\beta_{m_{11}^2} &=&
12 \lambda_1 m_{11}^2
+ 2( 2\lambda_4  + \lambda_7 ) \,m_{22}^2
+ 2( 2\lambda_5 +\lambda_8 )m_{33}^2 \nonumber\\
& &+ \, \left[ 6 \lambda_{13} m_{12}^{2\,*} + 6 \lambda_{16} m_{13}^{2\,*} + 2 (2\lambda_{19} + \lambda_{23}) m_{23}^{2\,*}
\,+\, \mbox{h.c.} \right] \,-\,  \frac{3}{2} (g_1^{2} +3 g_2^{2} )\,m_{11}^2\nonumber \\
& & \, 
+ \,2 \left[ 3 \tr\left(\Delta_1^{\dagger} \Delta_1 \right) + 3 \tr\left(\Gamma_1^{\dagger} \Gamma_1 \right) + \tr\left(\Pi_1^{\dagger} \Pi_1 \right)\right] m_{11}^2
 \nonumber \\
& &
\,+\,\left\{ \left[ 3  \tr\left(\Delta_1^{\dagger} \Delta_2 \right)
+ 3  \tr\left(\Gamma_2^{\dagger} \Gamma_1 \right)
+  \tr\left(\Pi_2^{\dagger} \Pi_1 \right) \right]  m_{12}^2 \right. \nonumber \\
& &
+ \left. \left[ 3 \tr\left(\Delta_1^{\dagger} \Delta_3 \right)
+ 3  \tr\left(\Gamma_3^{\dagger} \Gamma_1 \right)
+ \tr\left(\Pi_3^{\dagger} \Pi_1 \right)\right]  m_{13}^2 \,+\, \mbox{h.c.} \right\}\\
\nonumber\\
\beta_{m_{22}^2} &=& 2(2\lambda_4 + \lambda_7) m_{11}^2 + 12\lambda_2 m_{22}^2 + 2(2\lambda_6 + \lambda_9) m_{33}^2 \nonumber\\
&&+ \left[6 \lambda_{14}^{*} m_{12}^2 + (4\lambda_{17}^{*} + 2\lambda_{27}^{*}) m_{13}^2 + 6 \lambda_{20}^{*} m_{23}^2 + \mbox{h.c.}\right] - \frac{3}{2} (g_1^{2} + 3g_2^{2}) m_{22}^2 \nonumber\\
&& + 2 \left[3\tr\left(\Gamma_2^{\dagger} \Gamma_2 \right) + 3\tr\left(\Delta_2^{\dagger} \Delta_2 \right) + \tr\left(\Pi_2^{\dagger} \Pi_2 \right)
\right] m_{22}^2 \nonumber \\
&&+ \Big\{\left[3\tr\left(\Gamma_2^{\dagger} \Gamma_1 \right)
+ 3 \tr\left(\Delta_1^{\dagger} \Delta_2 \right)
+ \tr\left(\Pi_2^{\dagger} \Pi_1 \right)\right]m_{12}^2\nonumber\\
&&+ \left[3\tr\left(\Gamma_3^{\dagger} \Gamma_2 \right)
+ 3 \tr\left(\Delta_2^{\dagger} \Delta_3 \right)
+ \tr\left(\Pi_3^{\dagger} \Pi_2 \right)\right] m_{23}^2 + h.c. \Big\} \\
\nonumber\\
\beta_{m_{33}^2} &=& 2(2\lambda_5 + \lambda_8) m_{11}^2 + 2(2\lambda_6 + \lambda_9) m_{22}^2 + 12 \lambda_3 m_{33}^2 \nonumber\\
&&+ [(4\lambda_{15}^{*} + 2\lambda_{25}^{*}) m_{12}^2 + 6 m_{13}^2 \lambda_{18}^{*} + 6 \lambda_{21}^{*} m_{23}^2 + \mbox{h.c.}] - \frac{3}{2} (g_1^{2} + 3g_2^{2}) m_{33}^2  \nonumber\\
&& + 2\left[3\tr\left(\Gamma_3^{\dagger} \Gamma_3 \right) + 3\tr\left(\Delta_3^{\dagger} \Delta_3 \right) + \tr\left(\Pi_3^{\dagger} \Pi_3 \right)\right]  m_{33}^2 \nonumber \\
&&+ \Big\{  \left[3\tr\left(\Gamma_3^{\dagger} \Gamma_1 \right)
+ 3 \tr\left(\Delta_1^{\dagger} \Delta_3 \right)
+ \tr\left(\Pi_3^{\dagger} \Pi_1 \right)\right] m_{13}^2\nonumber\\
&&+  \left[3\tr\left(\Gamma_3^{\dagger} \Gamma_2 \right)
+ 3 \tr\left(\Delta_2^{\dagger} \Delta_3 \right)
+ \tr\left(\Pi_3^{\dagger} \Pi_2 \right)\right] m_{23}^2 + h.c.\Big\} \\
\nonumber\\
\beta_{m_{12}^2} &=& 6 \lambda_{13} m_{11}^2 + 6 \lambda_{14} m_{22}^2 + 2(2\lambda_{15} + \lambda_{25}) m_{33}^2 + 2(\lambda_4 + 2\lambda_7) m_{12}^2 + 12 \lambda_{10} m_{12}^{2\,*}+ 2 \lambda_{19}^{*} m_{13}^2\nonumber\\
&&+ 6 \lambda_{22} m_{13}^{2\,*} + 4 \lambda_{23}^{*} m_{13}^2+ 2 \lambda_{17} m_{23}^{2\,*} + 6 \lambda_{26}^{*} m_{23}^2 + 4 \lambda_{27} m_{23}^{2\,*} - \frac{3}{2} (g_1^{2} + 3g_2^{2}) m_{12}^2 \nonumber\\
&&+ \left[3\tr\left(\Gamma_1^{\dagger} \Gamma_2 \right)+ 3 \tr\left(\Delta_2^{\dagger} \Delta_1 \right) + \tr\left(\Pi_1^{\dagger} \Pi_2 \right)\right] \left(m_{11}^2 + m_{22}^2\right)\nonumber\\
&&+ \Big[3\tr\left(\Gamma_1^{\dagger} \Gamma_1 \right) + 3\tr\left(\Delta_1^{\dagger} \Delta_1\right) + \tr\left(\Pi_1^{\dagger} \Pi_1 \right)  \nonumber\\
&&+ 3\tr\left(\Gamma_2^{\dagger} \Gamma_2 \right) + 3\tr\left(\Delta_2^{\dagger} \Delta_2 \right) + \tr\left(\Pi_2^{\dagger} \Pi_2 \right)\Big] m_{12}^2\nonumber\\
&&+ \Big[3\tr\left(\Gamma_3^{\dagger} \Gamma_2 \right)
+ 3 \tr\left(\Delta_2^{\dagger} \Delta_3 \right)
+ \tr\left(\Pi_3^{\dagger} \Pi_2 \right)\Big] m_{13}^2 \nonumber\\
&&+ \Big[3\tr\left(\Gamma_1^{\dagger} \Gamma_3 \right)
+ 3 \tr\left(\Delta_3^{\dagger} \Delta_1 \right)
+ \tr\left(\Pi_1^{\dagger} \Pi_3 \right)\Big] m_{23}^{2\,*}\\
\nonumber\\
\beta_{m_{13}^2} &=& 6 \lambda_{16} m_{11}^2 + 2(2\lambda_{17} + \lambda_{27}) m_{22}^2 + 6 \lambda_{18} m_{33}^2 + 2 \lambda_{19} m_{12}^2 + 6 \lambda_{22} m_{12}^{2\,*} + 4 \lambda_{23} m_{12}^2 + 2 \lambda_5 m_{13}^2 \nonumber\\
&&+ 4 \lambda_8 m_{13}^2 + 12 \lambda_{11} m_{13}^{2\,*} + 2 \lambda_{15} m_{23}^2 + 6 \lambda_{24} m_{23}^{2\,*} + 4 \lambda_{25} m_{23}^2 - \frac{3}{2} (g_1^{2} + 3g_2^{2}) m_{13}^2 \nonumber\\
&&+ \Big[3\tr\left(\Gamma_1^{\dagger} \Gamma_3 \right)+ 3 \tr\left(\Delta_3^{\dagger} \Delta_1 \right) + \tr\left(\Pi_1^{\dagger} \Pi_3 \right)\Big] \left(m_{11}^2 + m_{33}^2\right)\nonumber\\
&&+ \left[3\tr\left(\Gamma_2^{\dagger} \Gamma_3 \right) + 3\tr\left(\Delta_3^{\dagger} \Delta_2\right) + \tr\left(\Pi_2^{\dagger} \Pi_3 \right)\right] m_{12}^2 \nonumber\\
&&+ \Big [3\tr\left(\Gamma_1^{\dagger} \Gamma_1 \right)
+ 3 \tr\left(\Delta_1^{\dagger} \Delta_1 \right)
+ \tr\left(\Pi_1^{\dagger} \Pi_1 \right) \nonumber\\
&&+ 3\tr\left(\Gamma_3^{\dagger} \Gamma_3 \right)
+ 3 \tr\left(\Delta_3^{\dagger} \Delta_3 \right)
+ \tr\left(\Pi_3^{\dagger} \Pi_3 \right)\Big] m_{13}^2\nonumber\\
&&+ \left[3\tr\left(\Gamma_1^{\dagger} \Gamma_2 \right)
+ 3 \tr\left(\Delta_2^{\dagger} \Delta_1 \right)
+ \tr\left(\Pi_1^{\dagger} \Pi_2 \right)\right] m_{23}^2 \\
\nonumber\\
\beta_{m_{23}^2} &=~& 2(2\lambda_{19} + \lambda_{23}) m_{11}^2 + 6 \lambda_{20} m_{22}^2 + 6 \lambda_{21} m_{33}^2 + 2 \lambda_{17} m_{12}^{2\,*} + 6 \lambda_{26} m_{12}^2 + 4 \lambda_{27} m_{12}^{2\,*} + 2 \lambda_{15}^{*} m_{13}^2\nonumber\\
&& + 6 \lambda_{24} m_{13}^{2\,*} + 4 \lambda_{25}^{*} m_{13}^2 + 2 \lambda_6 m_{23}^2 + 4 \lambda_9 m_{23}^2 + 12 \lambda_{12} m_{23}^{2\,*} - \frac{3}{2} (g_1^{2} + 3g_2^{2}) m_{23}^2 \nonumber \\
&& + \left[3\tr\left(\Gamma_2^{\dagger} \Gamma_3 \right)+ 3 \tr\left(\Delta_3^{\dagger} \Delta_2 \right) + \tr\left(\Pi_2^{\dagger} \Pi_3 \right)\right] \left(m_{22}^2 + m_{33}^2\right) \nonumber\\
&&+ \left[3\tr\left(\Gamma_1^{\dagger} \Gamma_3 \right) + 3\tr\left(\Delta_3^{\dagger} \Delta_1\right) + \tr\left(\Pi_1^{\dagger} \Pi_3 \right)\right] m_{12}^{2\,*}\nonumber\\
&&+ \left[3\tr\left(\Gamma_2^{\dagger} \Gamma_1 \right)
+ 3 \tr\left(\Delta_1^{\dagger} \Delta_2 \right)
+ \tr\left(\Pi_2^{\dagger} \Pi_1 \right)\right] m_{13}^2 \nonumber\\
&&+ \Big[3\tr\left(\Gamma_2^{\dagger} \Gamma_2 \right)
+ 3 \tr\left(\Delta_2^{\dagger} \Delta_2 \right)
+ \tr\left(\Pi_2^{\dagger} \Pi_2 \right)  \nonumber\\
&&+ 3\tr\left(\Gamma_3^{\dagger} \Gamma_3 \right)
+ 3 \tr\left(\Delta_3^{\dagger} \Delta_3 \right)
+ \tr\left(\Pi_3^{\dagger} \Pi_3 \right)\Big] m_{23}^2~.
\eea
Let us now write these expressions with the scalar quartic couplings of the GOOFy model, eq.~\eqref{eq:goofy3}, but keeping generic quadratic parameters 
and (for the moment) the general form for the Yukawa contributions. This gives us
\bea
\beta_{m_{11}^2} &=& 
24 ~ (\Lambda_0 - \Lambda_1)~m_{11}^2 
+ 2 ~ (8\Lambda_0 + 4\Lambda_1 + \Lambda_2 + \Lambda_3) ~(m_{22}^2 +  m_{33}^2) \nonumber\\ 
&&+ \Big[
- 6 (\Lambda_6 - \Lambda_7 - \Lambda_8) ~ m_{12}^2
- 6(\Lambda_6 - \Lambda_7 + \Lambda_8) ~ m_{13}^2 \nonumber\\ 
&& + (\Lambda_4 - \Lambda_5 - 4\Lambda_6 - 8\Lambda_7) ~ m_{23}^2 + \text{h.c.}\Big]
- \frac{3}{2} (g_1^{2} +3 g_2^{2} )\,m_{11}^2\nonumber \\
&&+ \,2 \left[ 3 \tr\left(\Gamma_1^{\dagger} \Gamma_1 \right) + 3 \tr\left(\Delta_1^{\dagger} \Delta_1 \right) + \tr\left(\Pi_1^{\dagger} \Pi_1 \right)\right] m_{11}^2
 \nonumber \\
&&
\,+\,\left\{ \left[ 
3  \tr\left(\Gamma_2^{\dagger} \Gamma_1 \right)
+ 3  \tr\left(\Delta_1^{\dagger} \Delta_2 \right)
+\tr\left(\Pi_2^{\dagger} \Pi_1 \right) \right]  m_{12}^2 \right. \nonumber \\
& &
+ \left. \left[
3  \tr\left(\Gamma_3^{\dagger} \Gamma_1 \right)
+ 3 \tr\left(\Delta_1^{\dagger} \Delta_3 \right)
+\tr\left(\Pi_3^{\dagger} \Pi_1 \right)\right]  m_{13}^2 \,+\, \mbox{h.c.} \right\}\\
\nonumber\\
\beta_{m_{22}^2} &=& 
24 ~ (\Lambda_0 - \Lambda_1) ~ m_{22}^2 
+ 2 ~ (8\Lambda_0 + 4\Lambda_1 + \Lambda_2 + \Lambda_3) ~ (m_{11}^2 +  m_{33}^2) \nonumber\\ 
&&+ \Big[
- 6 (\Lambda_6 - \Lambda_7 + \Lambda_8) m_{12}^2
+ (\Lambda_4 - \Lambda_5 - 4\Lambda_6 - 8\Lambda_7) m_{13}^2 \nonumber\\ 
&& - 6 (\Lambda_6 - 7 \Lambda_7 - \Lambda_8) m_{23}^2 + \text{h.c.}\Big]
- \frac{3}{2} (g_1^{2} + 3g_2^{2}) m_{22}^2 \nonumber\\
&& + 2 \left[3\tr\left(\Gamma_2^{\dagger} \Gamma_2 \right) + 3\tr\left(\Delta_2^{\dagger} \Delta_2 \right) + \tr\left(\Pi_2^{\dagger} \Pi_2 \right)\right] m_{22}^2 \nonumber \\
&&+ \Big\{\left[3\tr\left(\Gamma_2^{\dagger} \Gamma_1 \right)
+ 3 \tr\left(\Delta_1^{\dagger} \Delta_2 \right)
+ \tr\left(\Pi_2^{\dagger} \Pi_1 \right)\right] m_{12}^2\nonumber\\
&&+ \left[3\tr\left(\Gamma_3^{\dagger} \Gamma_2 \right)
+ 3 \tr\left(\Delta_2^{\dagger} \Delta_3 \right)
+ \tr\left(\Pi_3^{\dagger} \Pi_2 \right)\right] m_{23}^2 + h.c. \Big\} \\
\nonumber\\
\beta_{m_{33}^2} &=& 
24~ (\Lambda_0 - \Lambda_1)~ m_{33}^2 
+ 2~ (8\Lambda_0 + 4\Lambda_1 + \Lambda_2 + \Lambda_3)~ (m_{11}^2 +  m_{22}^2) \nonumber\\ 
&&+ \Big[
(\Lambda_4 - \Lambda_5 - 4\Lambda_6 - 8\Lambda_7) m_{12}^2
- 6 (\Lambda_6 - \Lambda_7 - \Lambda_8) m_{13}^2 \nonumber\\ 
&& - 6 (\Lambda_6 - 7 \Lambda_7 + \Lambda_8) m_{23}^2 + \text{h.c.}\Big]
- \frac{3}{2} (g_1^{2} + 3g_2^{2}) m_{33}^2 \nonumber\\
&& + 2 \left[3\tr\left(\Gamma_3^{\dagger} \Gamma_3 \right) + 3\tr\left(\Delta_3^{\dagger} \Delta_3 \right) + \tr\left(\Pi_3^{\dagger} \Pi_3 \right)\right] m_{33}^2 \nonumber \\
&&+ \Big\{\left[3\tr\left(\Gamma_3^{\dagger} \Gamma_1 \right)
+ 3 \tr\left(\Delta_1^{\dagger} \Delta_3 \right)
+ \tr\left(\Pi_3^{\dagger} \Pi_1 \right)\right] m_{13}^2\nonumber\\
&&+ \left[3\tr\left(\Gamma_3^{\dagger} \Gamma_2 \right)
+ 3 \tr\left(\Delta_2^{\dagger} \Delta_3 \right)
+ \tr\left(\Pi_3^{\dagger} \Pi_2 \right)\right] m_{23}^2 + h.c.\Big\} \\
\nonumber\\
\beta_{m_{12}^2} &=& 
-6 ~ (\Lambda_6 - \Lambda_7) ~(m_{11}^2 + m_{22}^2)
+6 \Lambda_8~(m_{11}^2 - m_{22}^2) + (\Lambda_4 - \Lambda_5 -4\Lambda_6 - 8\Lambda_7)~m_{33}^2 \nonumber\\
&& +\, 4 ~ (2 \Lambda_0 + \Lambda_1 + \Lambda_2 + \Lambda_3)~m_{12}^2 + 6 ~ (\Lambda_2 - \Lambda_3)~m_{12}^{2~*}\nonumber\\
&&+\,2 ~ (\Lambda_4 - \Lambda_5 - \Lambda_6 - 2 \Lambda_7)~m_{13}^2 + 3 ~ (\Lambda_4 + \Lambda_5)~m_{13}^{2~*} \nonumber\\
&&+ \,3 ~ (\Lambda_4 + \Lambda_5) ~ m_{23}^2
+ 2 ~(\Lambda_4 - \Lambda_5 - \Lambda_6 -2\Lambda_7 ) ~ m_{23}^{2~*}
- \frac{3}{2} (g_1^{2} + 3g_2^{2}) m_{12}^2 \nonumber\\
&&+ \left[3\tr\left(\Gamma_1^{\dagger} \Gamma_2 \right)+ 3 \tr\left(\Delta_2^{\dagger} \Delta_1 \right) + \tr\left(\Pi_1^{\dagger} \Pi_2 \right)\right] \left(m_{11}^2 + m_{22}^2\right) \nonumber\\
&&+ \Big[3\tr\left(\Gamma_1^{\dagger} \Gamma_1 \right) + 3\tr\left(\Delta_1^{\dagger} \Delta_1\right) + \tr\left(\Pi_1^{\dagger} \Pi_1 \right)  \nonumber\\
&&+ 3\tr\left(\Gamma_2^{\dagger} \Gamma_2 \right) + 3\tr\left(\Delta_2^{\dagger} \Delta_2 \right) + \tr\left(\Pi_2^{\dagger} \Pi_2 \right)\Big] m_{12}^2\nonumber\\
&&+ \Big[3\tr\left(\Gamma_3^{\dagger} \Gamma_2 \right)
+ 3 \tr\left(\Delta_2^{\dagger} \Delta_3 \right)
+ \tr\left(\Pi_3^{\dagger} \Pi_2 \right)\Big] m_{13}^2 \nonumber\\
&&+ \Big[3\tr\left(\Gamma_1^{\dagger} \Gamma_3 \right)
+ 3 \tr\left(\Delta_3^{\dagger} \Delta_1 \right)
+ \tr\left(\Pi_1^{\dagger} \Pi_3 \right)\Big] m_{23}^{2\,*}\label{eq:betam12}\\
\nonumber\\
\beta_{m_{13}^2} &=& 
-6 ~ (\Lambda_6 - \Lambda_7) ~(m_{33}^2 + m_{11}^2)
+6 \Lambda_8~(m_{33}^2 - m_{11}^2) + (\Lambda_4 - \Lambda_5 -4\Lambda_6 - 8\Lambda_7)~m_{22}^2 \nonumber\\
&&
+\, 2~(\Lambda_4 - \Lambda_5 - \Lambda_6 - 2\Lambda_7 ) ~ m_{12}^2 +3~(\Lambda_4 + \Lambda_5) ~ m_{12}^{2\,*}\nonumber\\
&&+\, 4~(2 \Lambda_0 + \Lambda_1 + \Lambda_2 + \Lambda_3)~m_{13}^2 + 6~(\Lambda_2 - \Lambda_3)~m_{13}^{2~*} 
 \nonumber\\
&& + \,2~(\Lambda_4 - \Lambda_5 - \Lambda_6 - 2 \Lambda_7)~m_{23}^2 +3~(\Lambda_4 + \Lambda_5)~m_{23}^{2~*} 
- \frac{3}{2} (g_1^{2} + 3g_2^{2}) m_{13}^2 \nonumber\\
&&+ \Big[3\tr\left(\Gamma_1^{\dagger} \Gamma_3 \right)+ 3 \tr\left(\Delta_3^{\dagger} \Delta_1 \right) + \tr\left(\Pi_1^{\dagger} \Pi_3 \right)\Big] \left(m_{11}^2 + m_{33}^2\right) \nonumber\\
&&+ \left[3\tr\left(\Gamma_2^{\dagger} \Gamma_3 \right) + 3\tr\left(\Delta_3^{\dagger} \Delta_2\right) + \tr\left(\Pi_2^{\dagger} \Pi_3 \right)\right] m_{12}^2\nonumber\\
&&+ \Big [3\tr\left(\Gamma_1^{\dagger} \Gamma_1 \right)
+ 3 \tr\left(\Delta_1^{\dagger} \Delta_1 \right)
+ \tr\left(\Pi_1^{\dagger} \Pi_1 \right) \nonumber\\
&&+ 3\tr\left(\Gamma_3^{\dagger} \Gamma_3 \right)
+ 3 \tr\left(\Delta_3^{\dagger} \Delta_3 \right)
+ \tr\left(\Pi_3^{\dagger} \Pi_3 \right)\Big]  m_{13}^2\nonumber\\
&&+ \left[3\tr\left(\Gamma_1^{\dagger} \Gamma_2 \right)
+ 3 \tr\left(\Delta_2^{\dagger} \Delta_1 \right)
+ \tr\left(\Pi_1^{\dagger} \Pi_2 \right)\right] m_{23}^2\\
\nonumber\\
\beta_{m_{23}^2} &=~& 
-6 ~ (\Lambda_6 - \Lambda_7) ~(m_{22}^2 + m_{33}^2)
+ 6 \Lambda_8~(m_{22}^2 - m_{33}^2) + (\Lambda_4 - \Lambda_5 -4\Lambda_6 - 8\Lambda_7)~m_{11}^2 \nonumber\\
&&+\, 3~(\Lambda_4 + \Lambda_5)~m_{12}^2 + 2~(\Lambda_4 - \Lambda_5 - \Lambda_6 - 2 \Lambda_7)~m_{12}^{2~*} \nonumber\\
&& +\, 2~(\Lambda_4 - \Lambda_5 - \Lambda_6 - 2\Lambda_7 ) ~ m_{13}^2 + 3~(\Lambda_4 + \Lambda_5) ~ m_{13}^{2~*} \nonumber\\
&& + \,4~(2 \Lambda_0 + \Lambda_1 + \Lambda_2 + \Lambda_3)~m_{23}^2  + 6~(\Lambda_2 - \Lambda_3)~m_{23}^{2~*}
- \frac{3}{2} (g_1^{2} + 3g_2^{2}) m_{23}^2 \nonumber \\
&& + \left[3\tr\left(\Gamma_2^{\dagger} \Gamma_3 \right)+ 3 \tr\left(\Delta_3^{\dagger} \Delta_2 \right) + \tr\left(\Pi_2^{\dagger} \Pi_3 \right)\right] \left(m_{22}^2 + m_{33}^2\right) \nonumber\\
&&+ \left[3\tr\left(\Gamma_1^{\dagger} \Gamma_3 \right) + 3\tr\left(\Delta_3^{\dagger} \Delta_1\right) + \tr\left(\Pi_1^{\dagger} \Pi_3 \right)\right] m_{12}^{2\,*} \nonumber\\
&&+ \left[3\tr\left(\Gamma_2^{\dagger} \Gamma_1 \right)
+ 3 \tr\left(\Delta_1^{\dagger} \Delta_2 \right)
+ \tr\left(\Pi_2^{\dagger} \Pi_1 \right)\right] m_{13}^2\nonumber\\
&&+ \Big[3\tr\left(\Gamma_2^{\dagger} \Gamma_2 \right)
+ 3 \tr\left(\Delta_2^{\dagger} \Delta_2 \right)
+ \tr\left(\Pi_2^{\dagger} \Pi_2 \right)  \nonumber\\
&&+ 3\tr\left(\Gamma_3^{\dagger} \Gamma_3 \right)
+ 3 \tr\left(\Delta_3^{\dagger} \Delta_3 \right)
+ \tr\left(\Pi_3^{\dagger} \Pi_3 \right)\Big] m_{23}^2~.
\eea
Let us look at $\beta_{m_{11}^2}$ -- the scalar and gauge contributions become automatically zero if $m^2_{11} = m^2_{22} = m^2_{33} = 0$
and $m^2_{12}$, $m^2_{13}$ and $m^2_{23}$ are pure imaginary numbers. As for the Yukawa terms, the contribution proportional to $m^2_{11}$ is trivially 
zero since that quadratic coupling is itself zero; and the term proportional to $m^2_{12}$ is (taking only the down Yukawa matrices)
\beq
3  \left[\tr\left(\Gamma_2^{\dagger} \Gamma_1 \right)  m_{12}^2\,+\,
\tr\left(\Gamma_1^{\dagger} \Gamma_2 \right)  {m_{12}^2}^*\right]\,.
\label{eq:m12}
\eeq
We now observe that for both Yukawa textures considered -- those of eq.~\eqref{eq: massless_Yuk}, which yield massless quarks, and eq.~\eqref{eq:Yukawa_texture},
which yield massive quarks but a CKM matrix equal to the identity -- the $\Gamma$ matrices obey the following conditions:
\begin{align}
    &\tr\left(\Gamma_1^{\dagger} \Gamma_1 \right) = \tr\left(\Gamma_2^{\dagger} \Gamma_2 \right) = \tr\left(\Gamma_3^{\dagger} \Gamma_3 \right) \label{eq: Yukawa_identities1}\\
    & \tr\left(\Gamma_1^{\dagger} \Gamma_2 \right) = \tr\left(\Gamma_2^{\dagger} \Gamma_3 \right) = \tr\left(\Gamma_1^{\dagger} \Gamma_3 \right) \label{eq: Yukawa_identities2}\\
    & \tr\left(\Gamma_i^{\dagger} \Gamma_j \right) = \tr\left(\Gamma_j^{\dagger} \Gamma_i \right),~i,j=1,2,3 \label{eq: Yukawa_identities3}\,.
\end{align}
The last of these equations is a trivial consequence of the fact that, as explained earlier, all GOOFy Yukawa matrices are real. 
Therefore, for the GOOFy-symmetric Yukawa sector of our 3HDM we may rewrite~\eqref{eq:m12} as
\beq
3  \tr\left(\Gamma_2^{\dagger} \Gamma_1 \right)  \,( m_{12}^2\,+\,{m_{12}^2}^*)\,=\,0\,,
\label{eq:m12s}
\eeq
where the final equality is due to the fact that the GOOFy symmetry implies Re$( m_{12}^2) = 0$. Since the three sectors are considered to have Yukawa 
matrices with the same texture, the identities of eqs.~(\ref{eq: Yukawa_identities1}-\ref{eq: Yukawa_identities3}) are valid for the up quark and lepton 
sectors and therefore the result of eq.~\eqref{eq:m12s} holds when one considers all contributions from the Yukawa sector. And it is trivial to show that,
for the same reasons, the $m_{13}^2$ contribution to $\beta_{m_{11}^2}$ is also zero. Following the same reasoning, one finds the same conclusion for
$\beta_{m_{22}^2}$ and $\beta_{m_{33}^2}$. Thus we conclude that
\beq
m^2_{11} = m^2_{22} = m^2_{33} = \mbox{Re}(m^2_{12}) = \mbox{Re}(m^2_{13}) = \text{Re}(m^2_{23}) = 0\;\Longrightarrow
\beta_{m_{11}^2} = \beta_{m_{22}^2} = \beta_{m_{33}^2}\,=\,0\,.
\label{eq:betasii}
\eeq
Notice that this conclusion holds even without using the first of the conditions of~\eqref{eq:goofy3},  $\text{Im}(m^2_{12}) = \mbox{Im}(m^2_{23}) = 
-\mbox{Im}(m^2_{13}) = - M/2$. This will be important shortly, when we consider the softly broken model.

Let us now consider $\beta_{m_{12}^2}$ -- we start with the (partial) GOOFy conditions listed above in eq.~\eqref{eq:betasii}. Immediately, the 
real part of the right-hand-side of eq.~\eqref{eq:betam12} is found to be zero~\footnote{Recall that, thanks to eq.~\eqref{eq: Yukawa_identities3},
all contributions from the Yukawa matrices are real.}. We then obtain, keeping only the terms involving the $\Gamma$ matrices for simplicity,
\bea
\beta_{\text{Im}(m_{12}^2)} &=& 
2  (4 \Lambda_0 + 2 \Lambda_1 - \Lambda_2 + 5 \Lambda_3) \text{Im}(m_{12}^2)  + ( \Lambda_4 + 5 \Lambda_5 +  2  \Lambda_6 + 4 \Lambda_7) 
\left[\mbox{Im}(m_{23}^2) - \mbox{Im}(m_{13}^2) \right]\nonumber\\
&& \,+ \left[ 6\tr\left(\Gamma_1^{\dagger} \Gamma_1 \right) - \frac{3}{2} (g_1^{2} + 3g_2^{2}) \right] \mbox{Im}(m_{12}^2)  
+ 3\tr\left(\Gamma_1^{\dagger} \Gamma_2 \right)\,\left[ \mbox{Im}(m_{13}^2) - \mbox{Im}(m_{23}^2)\right]\,,
\eea
where we have already used eqs.~\eqref{eq: Yukawa_identities1}--\eqref{eq: Yukawa_identities3} to simplify the result. Likewise, we find
\bea
\beta_{\text{Im}(m_{13}^2)} &=& 
2  (4 \Lambda_0 + 2 \Lambda_1 - \Lambda_2 + 5 \Lambda_3) \text{Im}(m_{13}^2) - ( \Lambda_4 + 5 \Lambda_5 +  2  \Lambda_6 + 4 \Lambda_7) 
\left[\mbox{Im}(m_{12}^2) + \mbox{Im}(m_{23}^2) \right]\nonumber\\
&& \,- \left[ 6\tr\left(\Gamma_1^{\dagger} \Gamma_1 \right) - \frac{3}{2} (g_1^{2} + 3g_2^{2}) \right] \mbox{Im}(m_{13}^2)  
+ 3\tr\left(\Gamma_1^{\dagger} \Gamma_2 \right)\,\left[\mbox{Im}(m_{12}^2) + \mbox{Im}(m_{23}^2) \right]\,,\no
\beta_{\text{Im}(m_{23}^2)} &=& 
2  (4 \Lambda_0 + 2 \Lambda_1 - \Lambda_2 + 5 \Lambda_3) \text{Im}(m_{23}^2)  + ( \Lambda_4 + 5 \Lambda_5 +  2  \Lambda_6 + 4 \Lambda_7) 
\left[\mbox{Im}(m_{12}^2) - \mbox{Im}(m_{13}^2) \right]\nonumber\\
&& \,+ \left[ 6\tr\left(\Gamma_1^{\dagger} \Gamma_1 \right) - \frac{3}{2} (g_1^{2} + 3g_2^{2}) \right] \mbox{Im}(m_{23}^2)  
+ 3\tr\left(\Gamma_1^{\dagger} \Gamma_2 \right)\,\left[\mbox{Im}(m_{13}^2) - \mbox{Im}(m_{12}^2) \right]\,,
\eea
It then becomes simple to see that if $\text{Im}(m_{12}^2) = - \text{Im}(m_{13}^2) = \text{Im}(m_{23}^2)$ one obtains 
$\beta_{\text{Im}(m_{12}^2)} = - \beta_{\text{Im}(m_{13}^2)}= \beta_{\text{Im}(m_{23}^2)}$, thereby proving the 1-loop RG invariance of the first
of the GOOFy parameter relations in~\eqref{eq:goofy3}. It is equally simple to prove that if $\text{Im}(m_{12}^2) = - \text{Im}(m_{13}^2) = M_1$
and $\text{Im}(m_{23}^2) = M_2$, one would still obtain  $\beta_{\text{Im}(m_{12}^2)} = - \beta_{\text{Im}(m_{13}^2)}$ -- and these are exactly 
the soft breaking parameters we will consider in the next section~\footnote{There is nothing special about this choice -- given the cyclic nature of the GOOFy
symmetry, we can choose any two of $\{\text{Im}(m_{12}^2) , -\text{Im}(m_{13}^2) , 
\text{Im}(m_{23}^2)\}$ to be equal as a soft breaking, RG invariance would still hold.}. 

We further point out that all of the above results were were confirmed at two-loops using the symbolic calculation package PyR@TE~\cite{Sartore:2020gou}. 
We do not show the explicit expressions for the $\beta$ functions at two loops because they are too long to write here, but the RG-invariance of the 
quadratic terms in eq.~\eqref{eq:goofy3} is easily checked with that code. Likewise, the soft symmetry breaking conditions discussed in the 
previous paragraph were also found to be RG-invariant up to at least two loops.

\section{The softly broken model}
\label{sec:soft}

As we saw in the previous section, the 3HDM with the GOOFy symmetry of eq.~\eqref{eq: GOOFy_transfos} yields a model where spontaneous electroweak
symmetry breaking can be achieved with a vacuum with vevs~\eqref{eq:vevtr}, which preserve a residual $S_3$ symmetry. The scalar spectrum 
of the model includes a SM-like state with electroweak gauge interactions identical to those of the SM, and two pairs of degenerate and gauge-phobic 
neutral scalars, as well as two pairs of charged scalars. The problems arise when the GOOFy symmetry is extended to the fermion sector -- the Yukawa matrices
of the model must obey eq.~\eqref{eq: Yukawa_couplings}, and we found  several textures satisfying these conditions: (A) a set of three identical real  
matrices for each fermion of the same charge, eq.~\eqref{eq: massless_Yuk}; and (B) a set of solutions (the third class of textures in Table~\ref{tab:text}),
an example of which is shown in eq.~\eqref{eq:Yukawa_texture}. However, all such textures clash with experimental data: the class (A) produces massless fermions,
and all textures in class (B), though producing three generations of massive fermions, yield a CKM matrix equal to the identity. The only way to improve
this situation is to move away from the vacuum of eq.~\eqref{eq:vevtr}, which implies adding soft breaking terms to the GOOFy scalar potential. 

\subsection{Soft breaking terms and vacuum}
\label{sec:sssb}

We consider
the minimal modification of the potential that is needed to produce a valid fermion sector and maintaining a GOOFy-like structure. 
To wit, the quadratic terms in the scalar potential, $V_2$, instead of their form in eq.~\eqref{eq: 3HDM_GOOFy}, now become 
\beq
V_2\, \to \, -M_1 \big[ \Im(\phi_1^\dagger \phi_2)
    +\Im(\phi_3^\dagger \phi_1) \big] - M_2 \Im(\phi_2^\dagger \phi_3)\,.
\eeq
We have verified that this choice of soft breaking terms is RG-invariant at least up to two loops in perturbation theory. It is easy to see, for instance, 
that to all orders of perturbation theory one will certainly have $m^2_{11} = m^2_{22} = m^2_{33} = 0$ -- indeed, all terms in the lagrangian with 
squared mass dimension are imaginary (the imaginary components of $m^2_{12}$, $m^2_{13}$ and $m^2_{23}$) and all dimensionless couplings (scalar, gauge
and Yukawa) are real, which means that contributions to $\beta_{m^2_{11}}$, $\beta_{m^2_{22}}$ and $\beta_{m^2_{33}}$ are guaranteed to be zero to all orders. 
With this choice for soft breaking the GOOFy symmetry the vacuum's vevs become more elaborate: instead of those of eq.~\eqref{eq:vevtr}, we will now have
\begin{equation}
    (\langle\Phi_1\rangle,\langle\Phi_2\rangle,\langle\Phi_3\rangle)\,=\,\frac{1}{\sqrt{2}}(v_1,v_2e^{i\alpha_1},v_3 e^{i\alpha_2})\,=\,
    \frac{v}{\sqrt{2}}\,(c_{\beta_2}s_{\beta_1},s_{\beta_2} s_{\beta_1} e^{\alpha_1},c_{\beta_1} e^{\alpha_2})\,,
    \label{eq:vevgen}
\end{equation}
where we defined $\tan\beta_1 = \sqrt{v_1^2 + v_2^2}/v_3$ and $\tan\beta_2 = v_2/v_1$. We also adopted the notation $s_x \equiv \sin x$ and $c_x\equiv \cos x$.
Clearly the vevs no longer preserve an $S_3$ symmetry, since for instance the two phases
$\alpha_1$ and $\alpha_2$ are independent. 

These more generic vevs allow us to obtain a realistic fermionic sector, to wit with Yukawa textures~\footnote{Of course, the introduction of soft breaking 
terms in the potential does not alter the extension of symmetries to the Yukawa sector, or the textures discussed in the previous section.} 
of class (B), eq.~\eqref{eq:Yukawa_texture}. As we
shall soon see, those textures and the vevs above can correctly reproduce the fermion mass spectrum (both quarks and charged leptons) as well as
the CKM matrix, complex phases included. The real Yukawa matrices of class (A), eq.~\eqref{eq: massless_Yuk}, do not yield an acceptable fermion
sector, even with the general vevs of eq.~\eqref{eq:vevgen}: in order to see this consider the down quark sector, for which the
Yukawa matrices are real and $\Gamma_1 = \Gamma_2 = \Gamma_3$; the down quark mass matrix is therefore of the form
\beq
M_n \,=\,\frac{1}{\sqrt{2}}\,\left(\Gamma_1 v_1  + \Gamma_2 v_2 e^{i\alpha_1}  + \Gamma_3 v_3 e^{i\alpha_2}\right)\,=\,\frac{v}{\sqrt{2}}\,\bar{\Gamma}\,
e^{i\theta_d}\,,
\eeq
where $\bar{\Gamma}$ is a $3\times 3$ real matrix and $\theta_d$ is an overall spurious complex phase which can be absorbed in the right-handed down quark fields.
$M_n$ is therefore a real matrix, and the same argument applies to the up-quark mass matrix, $M_p$, which is found to be a real
matrix multiplied by an overall phase $\theta_u$ which can be absorbed by a basis change -- and if both of those matrices are real, then their
diagonalization will necessarily produce a real CKM matrix as well. 

\subsection{Scalar spectrum and interactions}

With the soft breaking terms considered, for which the $m^2_{ij}$ (with $i\neq j$) parameters are pure imaginary numbers, the scalar sector (and therefore
the whole lagrangian) has soft explicit CP violation. There are 5 independent minimization equations, whose analytical solution is possible but produces
very unwieldy expressions. We chose to supply the model with the vevs of eq.~\eqref{eq:vevgen} and obtain five of the parameters of the model,
$M_1$, $M_2$, $\Lambda_4$, $\Lambda_5$ and $\Lambda_6$, from the minimization equations by taking the remaining $\Lambda$ parameters as numerical inputs 
-- this choice was convenient to handle
the bounded from below conditions, explained below, in a more efficient manner. Computing the squared mass matrices for the scalars in the usual manner, we obtain
the correct number of Goldstone bosons; the charged masses and neutral ones have unwieldy expressions that we will not reproduce here, but are easily 
obtained numerically. We can however write relatively straightforward expressions for the \textit{traces} of the neutral and charged mass matrices, 
obtaining 
\bea
    \mathrm{Tr}(M^2_\mathrm{neutral}) &=& (8 \Lambda_0 + 2 \Lambda_1 + \Lambda_2 + \Lambda_3)~v^2 \nonumber\\
    & &+ \frac{1}{2} \big(\Lambda_4 - \Lambda_5 + 10 \Lambda_6 + 
    4 \Lambda_7\big) \left[c_{\alpha_1 - \alpha_2}\, v_2 v_3 + c_{\alpha_1} v_1 v_2 + c_{\alpha_2} v_1 v_3\right] \,, \no
    \mathrm{Tr}(M^2_\mathrm{charged}) &=& 6 \Lambda_0 v^2 + 3\Lambda_6 \left[c_{\alpha_1 - \alpha_2} v_2 v_3 + c_{\alpha_1} v_1 v_2 + c_{\alpha_2} v_1 v_3\right]\,
\eea
where we observe that all the squared scalar masses of the model are therefore of the order $\sim \Lambda v^2$, with $\Lambda$ a linear combination of the 
quartic couplings. This occurs because the model only has two quadratic mass parameters -- $M_1$ and $M_2$ -- which, due to the minimization conditions, 
are themselves constrained to be of the order $\sim \Lambda v^2$. Now, since the model must obey unitarity bounds (see section~\ref{sec:bounds}), the 
quartic couplings of the model are constrained to be at most of order 10, which gives a rough upper bound of 800 GeV on the masses of the scalars. As we 
will see in our numerical analysis, they end up being even lighter than that. We thus conclude that this (softly broken) GOOFy 3HDM has no decoupling 
limit -- it is not possible, within the parameter space of the model to make the masses of the scalars arbitrarily large. 

As we have already stated, the model is explicitly CP breaking -- one way of seeing this is realizing that the complex number $i$ affecting the off-diagonal
quadratic terms cannot be rephased away without reappearing in the quartic couplings. This impacts the interactions of the scalars with gauge bosons. 
Indeed, the neutral scalars 
are the eigenvalues of a  $6\times 6$ mass matrix which has most of its entries different from zero, signalling mixing between CP-even and CP-odd component 
fields. This matrix is diagonalized via a real orthogonal $6 \times 6$ matrix $O$ such that the physical eigenstates -- a state with mass equal to 125 GeV 
that will be our SM-like scalar, $h_{SM}$, 4 heavier neutral states $h_{1,2,3,4}$ and
the neutral Goldstone $G$ -- are related to the neutral doublet components from eq.~\eqref{eq:doub} by the rotation
\begin{equation}
 \begin{pmatrix}
        \rho_1\\ \rho_2\\ \rho_3\\ \eta_1\\ \eta_2\\ \eta_3
    \end{pmatrix}
    \,=\, O\,
    \begin{pmatrix}
        h_{SM}\\ h_1\\ h_2\\ h_3\\ h_4 \\  G\\
    \end{pmatrix}
        \,.    
        \label{eq:O}
\end{equation}
Thus, analysing the triple vertices between a neutral scalar and a pair of $Z$ bosons emerging from the kinetic terms of eq.~\eqref{eq:LK}, we find that
\begin{align}
g_{h_{SM}ZZ} &= \left[ \frac{}{} c_{\beta_2} s_{\beta_1} O_{1,1} \,+\, s_{\beta_2} s_{\beta_1} ( c_{\alpha_2} O_{2,1} + s_{\alpha_2} O_{5,1} )
\,+\, c_{\beta_1} (c_{\alpha_3} O_{3,1} + s_{\alpha_3} O_{6,1} ) \right] \,g^{SM}_{hZZ} \, \no
g_{h_i ZZ} &= \left[ \frac{}{} c_{\beta_2}s_{\beta_1} O_{i+1,1} \,+\,s_{\beta_2} s_{\beta_1} (c_{\alpha_2} O_{2,i+1} + s_{\alpha_2} O_{5,i+1} )
\,+\,c_{\beta_1} (c_{\alpha_3} O_{3,i+1}   + \, s_{\alpha_3} O_{6,i+1}) \right]\,g^{SM}_{hZZ} \,
\label{eq:hVV}
\end{align}
where $g^{SM}_{hZZ}$ is the SM Higgs-ZZ coupling. We see that \textit{all} the neutral scalars couple to $Z$ pairs, which is a hallmark of CP 
violation~\footnote{If CP was conserved we would expect three CP-even states, which would couple to $Z$'s, and two pseudoscalar ones, which would not.}. 
The expressions among square brackets in the previous equation are the \textit{coupling modifiers} of the vertices neutral scalar-Z-Z in this model. 
For the SM-like state, we define $\kappa_V = g_{h_{SM}ZZ}/g^{SM}_{hZZ}$. Notice that within a 3HDM we will also have $\kappa_V = g_{h_{SM}WW}/g^{SM}_{WW}$.
The current LHC results push us towards the so-called alignment limit,  wherein we must have $\kappa_V \simeq 1$. 

CP violation also affects the Yukawa interactions of the neutral scalars to fermions. Using the down quark sector as an example, the interactions
between neutral scalars and down quarks may be obtained from the respective Yukawa lagrangian, eq.~\eqref{eq:lYd}. Expressing the
doublets as in eq.~\eqref{eq:doub} and extracting the neutral components we obtain
\beq
 -\Lag_{Y,d} = \overline{Q}_L  \Gamma_a \Phi_a  n_R + \overline{n}_R \Gamma^\dagger_a \Phi^\dagger_a Q_L \to 
 \frac{1}{\sqrt{2}} \left[ (\overline{n}_L \Gamma_a n_R + \overline{n}_R \Gamma_a^\dagger n_L) \rho_a + i (\overline{n}_L \Gamma_a n_R - \overline{n}_R \Gamma_a^\dagger n_L) \eta_a \right]\,,
\eeq
where a sum in the $a$ index is assumed. The $n_{L,R}$ are the unrotated negative quark fields, related to the mass eigenstates by
\beq
n_L\,=\, U^d_L \,d_L\;\;\; , \;\;\; n_R\,=\, U^d_R\,,
\eeq
where $d_{L,R}\equiv (d,s,b)_{L,R}$ is a 3-vector in flavour space. We define the matrices $\Gamma'_a = {U^d_L}^\dagger \Gamma_a U^d_R$ and
use the rotation between the component fields $\rho_a$, $\eta_a$ to obtain
\bea
 -\Lag_{h^0 \bar{d}d} &=& \; \frac{1}{\sqrt{2}} \bar{d}\,\left\{ 
 \frac{1}{2}\,\left[ \left(\Gamma'_a + {\Gamma'}^\dagger_a\right) O_{a,1} +
 i \left(\Gamma'_a - {\Gamma'}^\dagger_a\right) O_{a+3,1}\right]  \right. \no
 & & +\, 
 \left. \frac{1}{2}\,\left[ \left(\Gamma'_a - {\Gamma'}^\dagger_a\right) O_{a,1} +
 i \left(\Gamma'_a + {\Gamma'}^\dagger_a\right) O_{a+3,1}\right]\,\gamma_5 \right\}\, d\,h_{SM}\no
 & & +\, \frac{1}{\sqrt{2}} \bar{d}\,\left\{ \frac{1}{2}\,\left[ \left(\Gamma'_a + {\Gamma'}^\dagger_a\right) O_{a,j+1} +
 i \left(\Gamma'_a - {\Gamma'}^\dagger_a\right) O_{a+3,j+1}\right] \right. \no 
 & & +\, 
 \left. \frac{1}{2}\,\left[ \left(\Gamma'_a - {\Gamma'}^\dagger_a\right) O_{a,j+1} +
 i \left(\Gamma'_a + {\Gamma'}^\dagger_a\right) O_{a+3,j+1}\right]\,\gamma_5 \right\}\, d\,h_j\,.
 \eea
An analogous expression is found for the $h^0\bar{u}u$ interaction and also for the interactions with charged leptons. 
In this manner, we see that all Yukawa interactions between fermions of the same flavour have vertices of the form $h^0 \bar{f} (a + i b \gamma_5) f$,
with real numbers $a$ and $b$ -- this is also a clear sign of CP violation and indicates that each neutral scalar $h^0$ is a CP admixture between
CP-even and CP-odd states, since a purely CP-even scalar would have vertices with $b = 0$, and a purely CP-odd pseudoscalar would imply vertices with $a = 0$.
We thus define, for each fermion $f$ of mass $m_f$, the fermionic coupling modifiers of the Higgs-like particle as
\beq
-\Lag_{h_{SM}\bar{f}f} \,=\, \frac{m_f}{v} \,\bar{f}\,(\kappa_f^e\,+\, i \kappa_f^o\gamma_5)\,f\,h_{SM}\,,
\label{eq:kf}
\eeq
with real CP-even couplings $\kappa^e$ and CP-odd ones $\kappa^o$.

We also observe that the matrices $\Gamma^\prime$ are not diagonal, and therefore there are flavour changing neutral currents (FCNC) mediated by the neutral scalars.
These induce the possibility of non-SM decays for the different scalars, for instance of the form $h_{SM}\to \bar{b}s$. It also implies that
the neutral scalars will have tree-level contributions to CP violating observables such as the mass difference between neutral $B$ and $K$ mesons, 
$\Delta M_{B_d}$, $\Delta M_{B_s}$ and $\Delta M_{K}$, or the $\epsilon_K$ parameter in the neutral Kaon system. 

\subsection{Numerical analysis and phenomenology}
\label{sec:pheno}

Our purpose with this chapter is to investigate whether this model is phenomenologically viable -- given that we imposed a very restrictive
GOOFy symmetry, it is not guaranteed that an acceptable scalar spectrum, with a Higgs-like particle, is possible to obtain. We also wish to
verify whether the model makes any distinctive predictions that might be tested experimentally. We have an immediate problem, in that
the complexity of the softly broken model means that analytical results are no longer possible and we must resort to numerical 
calculations to probe the phenomenology of the sector. In fact we will use a semi-analytical approach: we obtain analytical expressions
for the minimization conditions and the scalar mass matrices, in terms of the model's parameters and the vevs of eq.~\eqref{eq:vevgen}, 
but in order to obtain the scalar mass eigenstates we must diagonalize those matrices numerically. Likewise, though we possess analytical 
expressions for the GOOFy-symmetric Yukawa matrices, their diagonalization is made numerically. 

\subsubsection{Scanning procedure}

The total number of parameters of the model is sufficiently large (despite the very constraining symmetries imposed) to in principle 
be able to successfully fit all current experimental constraints -- indeed, we have a total of 10 parameters in the scalar potential (down 
from the 46 of the most general 3HDM); 9 parameters for the down quark Yukawa matrices; another 9 for the up quark Yukawa matrices; and 
another 9 for the charged lepton sector. 
Our scan of this parameter space  was carried out as follows: starting with the minimisation equations, it was possible to obtain conditions 
that express $M_1,~M_2,~\Lambda_4,~\Lambda_5,~\Lambda_6$ in terms of the remaining parameters of the potential and the vevs of eq.~\eqref{eq:vevgen},
which were tsaken as input parameters. This choice was not arbitrary:
$\Lambda_0,~\Lambda_1,~\Lambda_2,~\Lambda_3$ were specifically chosen as input parameters to make the requirement of boundedness from below 
(see below) more easily satisfied. A numerical fit procedure was applied to ensure some basic requirements:
\begin{itemize}
\item The lightest scalar has a mass of about 125 GeV~\footnote{There is of course also the possibility of scalars lighter than 125 GeV which haven't yet
been detected, but we chose to disregard that scenario in this analysis.}. This will be $h_{SM}$, the candidate to play the role of the Higgs boson 
observed at the LHC. 
\item We demand that $h_{SM}$ has interactions with gauge bosons and fermions similar to those of the SM. To this end, we require that:  
(a) $0.9 \leq |\kappa_V| \leq 1$, with  $\kappa_V$ defined below eq.~\eqref{eq:hVV};  (b) for the bottom and top quarks, we require 
$0.8 \leq |\kappa_f| \leq 1.2$ $(f = b, t)$, where 
\beq
\kappa_f = \sqrt{{\kappa_f^e}^2 + {\kappa_f^o}^2}\,
\eeq
with the CP-even and odd coupling modifiers defined in eq.~\eqref{eq:kf}.
\item Since the Higgs has been observed at the LHC as behaving very much like a CP-even particle, we must ensure that the CP-odd
components of the Yukawa interactions are not too big. Thus we also required, for each fermion, $|\kappa_f^o/\kappa_f^e|\in [0,0.35]$.
This is in rough agreement with current LHC results as in \cite{ATLAS:2020ior,CMS:2021sdq,ATLAS:2022akr}.
\item In paralell, our numerical fit also ensures that the six quarks have their know values and that quark mixing is also in agreement
with experimental results -- we compared the magnitudes of all elements of the CKM matrix with their measured values and ensured that the
resulting value for the Jarlskog invariant $J_{CKM}$ was within its current uncertainty interval. All values for quark masses, CKM matrix elements,
$J_{CKM}$ and corresponding experimental uncertainties were taken from the Particle Data Group listings~\cite{ParticleDataGroup:2024cfk}.
\end{itemize}
In table \ref{tab:ranges} we show  the ranges considered for the input parameters, which have been optimised for greater computational speed.
Indeed, the numerical fit is quite involved, as we are attempting to reproduce simultaneously quantities from the scalar and Yukawa sector. 
A three-stage minimisation
 was required, with different cost functions at each stage. In the first stage, the scalar sector is fitted using a \textit{L-BFGS-B} method in Python,
 to make sure the $h_{SM}$ mass is correct, and that the remaining scalars had masses above 130 GeV (and also that all eigenvalues of the squared mass
 matrices were positive, with the exception of zero-mass Goldstone bosons, to ensure the existence of a minimum). 
 This is particularly well-suited for this stage because it is fast while keeping the cost function low. 
Then, using the points that pass stage one, the observables that concern the Yukawa sector are fitted individually, using a different minimiser and a 
\textit{trf} method. Here, the 
change in minimisers is required by the increased number of residuals, making it very difficult for to obtain a low cost. Besides changing the minimiser, it 
was necessary to employ a staged minimisation within this phase. This means that each restriction is turned on progressively in order to obtain the lowest
cost possible. Finally, using the results from both stages as seeds, a joint fit of the two sectors is performed to further refine the result. This was done again 
with the \textit{trf} method. A normalisation of the variables was made before the minimisation took place. This was necessary because the different 
observables (such as the quark masses) span very different orders of magnitude, which meant that, without it, the minimiser would prioritise the 
mass of the top much more than that of the down, for example.
\begin{table}[htbp]
    \centering
    \caption{Scan ranges of the input parameters.}
    \label{tab:ranges}

    \begingroup
    \renewcommand{\arraystretch}{1.2}
    \setlength{\tabcolsep}{12pt}
    \begin{tabular}[t]{@{}lc@{}}
        \toprule
        Parameter & Range \\
        \midrule
        $\Lambda_0,\Lambda_1,\Lambda_2,$
        & $[-1,\,1]$ \\
 %           & $[\num{1e-5},\,1]$ \\
        $\Lambda_3,\Lambda_7,\Lambda_8$
            & $[-1,\,1]$ \\
        $\alpha_1,\alpha_2$
            & $[-\pi,\,\pi]$ \\
        $\beta_1,\beta_2$
            & $[-\pi,\,\pi]$ \\
        \bottomrule
    \end{tabular}
    \hspace{0.6cm}
    \begin{tabular}[t]{@{}lc@{}}
        \toprule
        Parameter & Range \\
        \midrule
        $\Gamma_{1j},~\Gamma_{2j}$
            & $[-0.02,\,0.02]$ \\
        $\Gamma_{3j}$
            & $[-0.10,\,0.10]$ \\
        $\Delta_{1j},~\Delta_{2j}$
            & $[-0.10,\,0.10]$ \\
        $\Delta_{3j}$
            & $[-1.50,\,1.50]$ \\
        \bottomrule
    \end{tabular}
    \endgroup
\end{table}

Note that Table \ref{tab:ranges} does not comprise values for the Yukawa matrices of the leptonic sector. The scalar-lepton Yukawa
lagrangian would be analogous to the down quark one, eq.~\eqref{eq:lYd}, with matrices $\Pi_a$ ($a = 1,2,3$) replacing the $\Gamma_a$. In our analysis,
we will take the muon and electron as massless for simplicity, and consider only interactions between scalars and the $\tau$ leptons. This corresponds 
to the $\Pi_a$ matrices -- which are constrained by the GOOFy symmetry to have  the same textures as the $\Gamma$ matrices of
eq.~\eqref{eq:Yukawa_texture} -- having only a non-zero entry, in the $(3,3)$ position. The charged lepton mass matrix therefore becomes
\begin{equation}
    M_\ell = \frac{\Pi_1 v_1 + \Pi_2 e^{i\alpha_1} v_2 + \Pi_3 e^{i\alpha_2} v_3}{\sqrt{2}} ~ \Rightarrow ~ m_\tau = 
    \frac{\pi_{33}}{\sqrt{2}} (v_1 + e^{i\alpha_1} v_2 + e^{i\alpha_2} v_3)\,.
\end{equation}
The Yukawa couplings for scalars and the $\tau$ leptons thus depends on the real parameter $\pi_{33}$, which can easily be chosen 
to give SM-like Yukawa interactions for $h_{SM}$ -- as for the quarks, there will be CP-even $\kappa^e_\tau$ and CP-odd $\kappa^o_\tau$ 
coupling modifiers, upon which we imposed the same constraints of the quark coupling modifers. 

\subsubsection{Experimental and theoretical constraints}
\label{sec:bounds}

Even though the minimisation process already included some experimental constraints, valid parameter points must also satisfy the bounds imposed by many more LHC measurements, as well as guarantee that the theoretical constraints are met. 

\begin{itemize}
\item Perturbative unitarity
\end{itemize}

The quartic couplings of the scalar potential cannot be too large, otherwise unitarity in scalar-scalar scattering processes
would be violated. This is a restriction already applied to the SM, and was used, prior to the Higgs' discovery at the LHC, to 
establish upper bounds on its mass. For multiscalar models, the procedure to obtain the perturbative unitarity conditions is known and 
summarised in \cite{Bento:2017eti,Bento:2022vsb}. Following the procedure outlined in those works one may find the scattering matrices both for 
the most general 3HDM as well as the ones for the present model. The unitarity conditions are then obtained by computing the eigenvalues, $A_i$, of these matrices and requiring
\begin{equation}
    |A_i| < 16 \pi~.
\end{equation}
These conditions were implemented numerically as the calculations of the eigenvalues of such large matrices are far too complicated. In practice, these conditions
limit the magnitude of the $\Lambda$ couplings in the potential of eq.~\eqref{eq: 3HDM_GOOFy} and motivate the ranges for these quantities considered in 
Table~\ref{tab:ranges}.

\begin{itemize}
\item Boundedness from Below
\end{itemize}

Since the scalar potential must have a minimum somewhere, it must be \textit{bounded from below} (BFB), \textit{i.e.} it cannot tend to $-\infty$ along
any possible direction in field space, which implies the potential is overall a convex function. In the SM this implies that the single scalar quartic 
coupling, $\lambda$, must be positive. In the 2HDM, the variety of possible field directions along which the potential could tend to $-\infty$ 
implied that for a long time only necessary or sufficient BFB conditions were known, until necessary-and-sufficient ones were established~\cite{Ivanov:2007de}
(see \cite{Ferreira:2015rha} for a simple implementation of these conditions). These conditions, which require obtaining the eigenvalues of $4\times 4$ 
matrices, already require numerical studies.

However, in the 3HDM the situation is different. To this day, there is no set of necessary and sufficient BFB conditions for the most general 3HDM
scalar potential. Specific models
with simplifying symmetries may have such BFB conditions -- for instance, Ivanov and Faro developed a technique that guarantees BFB for the $U(1) \times U(1)$ 
model \cite{Faro:2019vcd}. The method consists on finding the copositivity conditions~\cite{Kannike:2012pe} for matrices built from the model parameters. 
For a class of more complicated models, Boto \textit{et al} built on Faro and Ivanov's method, in which BFB is guaranteed by introducing a potential that is a
lower bound on the one under study, for which the copositivity conditions can be extended \cite{Boto:2022uwv}.

For the GOOFy model under study here, however, these methods become too cumbersome. We can obtain of course some simple necessary conditions for
some of the quartic couplings of eq.~\eqref{eq: 3HDM_GOOFy}. Following the same procedure used in the 2HDM to find necessary conditions for 
the couplings $\lambda_1$, $\lambda_2$ and $\lambda_3$ of that model, we find that in the GOOFy model $\Lambda_0$ and $\Lambda_1$ must obey
\beq
\Lambda_0 + \Lambda_1 \geq 0 \quad\quad , \quad \quad  4\Lambda_0 + \Lambda_1 \geq 0\,.
\eeq
This however leaves out all the remaning $\Lambda$ couplings, so in our analysis BFB is ensured by a successive global minimisation of 
the quartic potential, which starts from different values of the scalar fields chosen at random. %To ensure that this condition is more easily satisfied, we 
%only consider points for which $\Lambda_0,~\Lambda_1,~\Lambda_2$ and $\Lambda_3$ are positive. This comes directly from the (sufficient) copositivity conditions for the 
%$U(1) \times U(1)$ model, which must also be satisfied in our case, as all the terms of the $U(1) \times U(1)$ model are also present here. 
This numerical procedure is not enough to ensure that BFB is satisfied, but the minimisation procedure followed is quite thorough (it is indeed 
one of the main sources of delay in the overall numerical scan of the parameter space) and the likelyhood that unbounded from below potentials survive it is
expected to be quite low. 

\begin{itemize}
\item Oblique parameters
\end{itemize}

In any SM extension with extra doublets there may be substantial contributions to $W$ and $Z$ self energies. Electroweak precision measurements
of $W$ and $Z$ properties imply severe constraints on the $S$, $T$ and $U$ Peskin–Takeuchi parameters~\cite{Peskin:1990zt,Peskin:1991sw},
also called \textit{oblique parameters}. To compute these quantities we followed\cite{Grimus:2008nb} wherein expressions for $S$, $T$ and $U$ 
for generic scalar potentials, which depend on the masses of the scalars and the rotation matrices which diagonalize the scalars' mass matrices.
Their impact on the scalar spectum is typically to reduce the size of the possible mass splittings between scalars. 

\begin{figure}[htb!]
    \centering
    \includegraphics[width=0.9\linewidth]{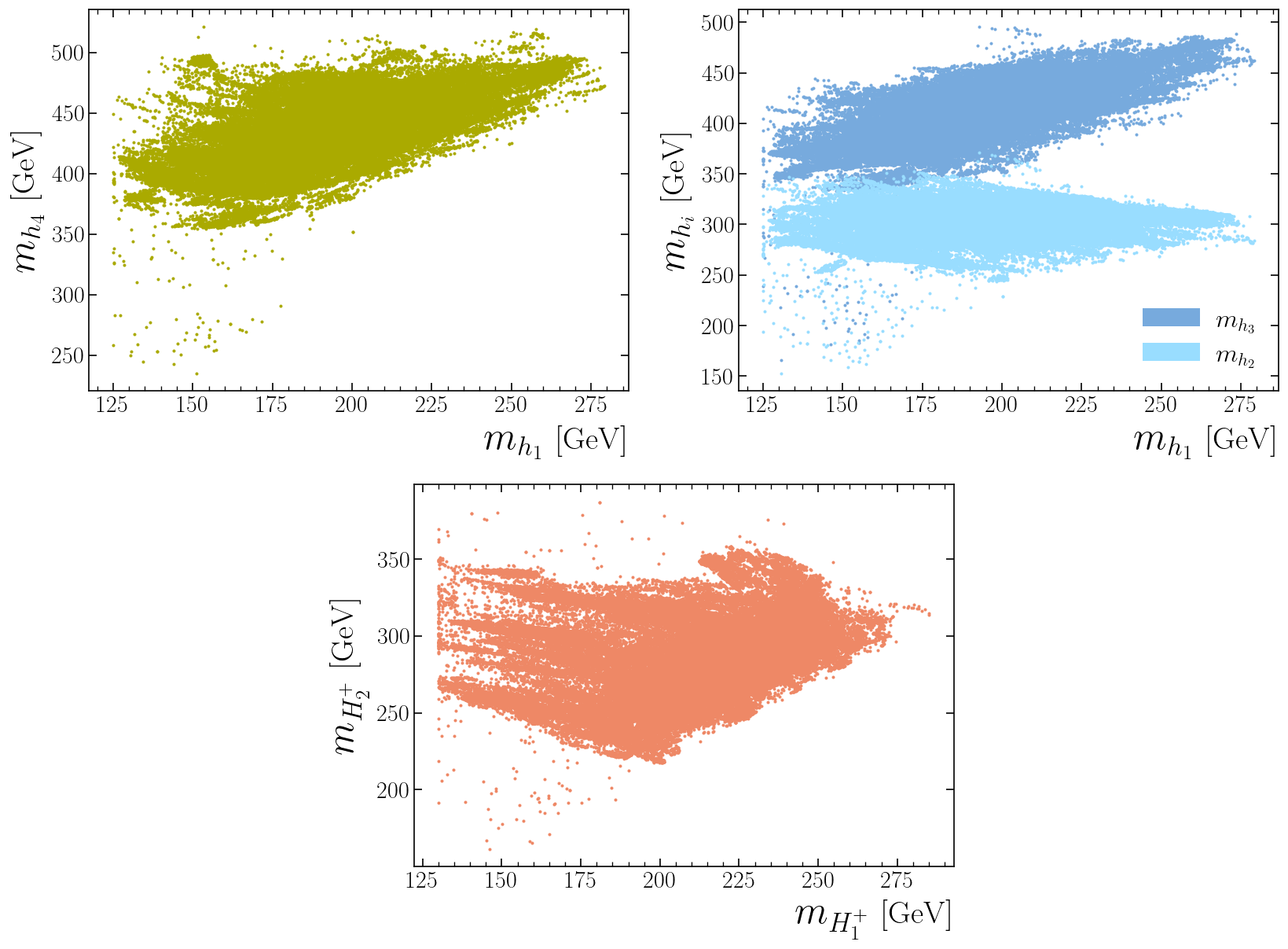}
    \caption{Mass distributions of the generated parameter points. }
    \label{fig:masses}
\end{figure}

With these basic constraints on the scalar sector, the masses we obtain are show in fig.~\ref{fig:masses} for an extensive scalar
of the parameter space which obeyed all constraints listed thus far. 
One may easily verify that the masses of the different particles are not entirely independent. In fact, the masses of the states 
$h_2,~h_3$ and $h_4$ tend to increase with the $h_1$ mass (recall that $h_1$ is the first scalar heavier than $h_{SM}$). We also observe that 
there is a rough upper bound of $\sim$ 500 GeV on all extra scalar masses -- this a consequence of the non-existence of a \textit{decoupling limit} 
in this model. Indeed, there are only two massive parameters, $M_1$ and $M_2$, in the  GOOFy-symmetric scalar potential of~\eqref{eq: 3HDM_GOOFy}, 
and since the vacuum of the model, eq.~\eqref{eq:vevgen}, implies that one must solve five minimization equations, those parameters are
limited in size to be of the order $\sim \Lambda v^2$, with $\Lambda$ of the order of the quartic couplings. Since the quartic couplings are
themselves limited in magnitude by the unitarity conditions described above, this explains that all masses of the model are predicted to be
below 500 GeV. We also see that the mass splittings between different scalars are limited in size, a consequence of the electroweak precision constraints. 
Finally, the $h_1$ mass is found to be at most $\sim$ 280 GeV. Our GOOFy 3HDM thus predicts a fairly tight scalar mass spectrum, with the first
heavier neutral scalar with mass below 280 GeV. 

\subsubsection{Direct searches at LHC}
\label{sec:dirs}

The LHC collaborations have searched for extra scalars beyond the Higgs boson. With comparably low masses for the extra scalar particles in our
model, it is necessary to verify that their existence has not been ruled out by current experimental bounds. To do so we looked at $h_1$, the first scalar 
heavier than the Higgs, and computed its gluon fusion production cross section at the LHC, as well as the branching ratios for $h_1$'s possible decays. Since these 
scalars are CP-mixed states, we used the formulae shown in the appendix of ref.~\cite{Barroso:2012wz}. Cross section numbers were obtained via 
{\tt SusHi}~\cite{Harlander:2012pb,Harlander:2016hcx}. QCD corrections to scalar decays to quarks were implemented as in~\cite{Djouadi:2005gi,Djouadi:2005gj}. 
\begin{figure}[htb!]
    \centering
    \includegraphics[width=0.9\linewidth]{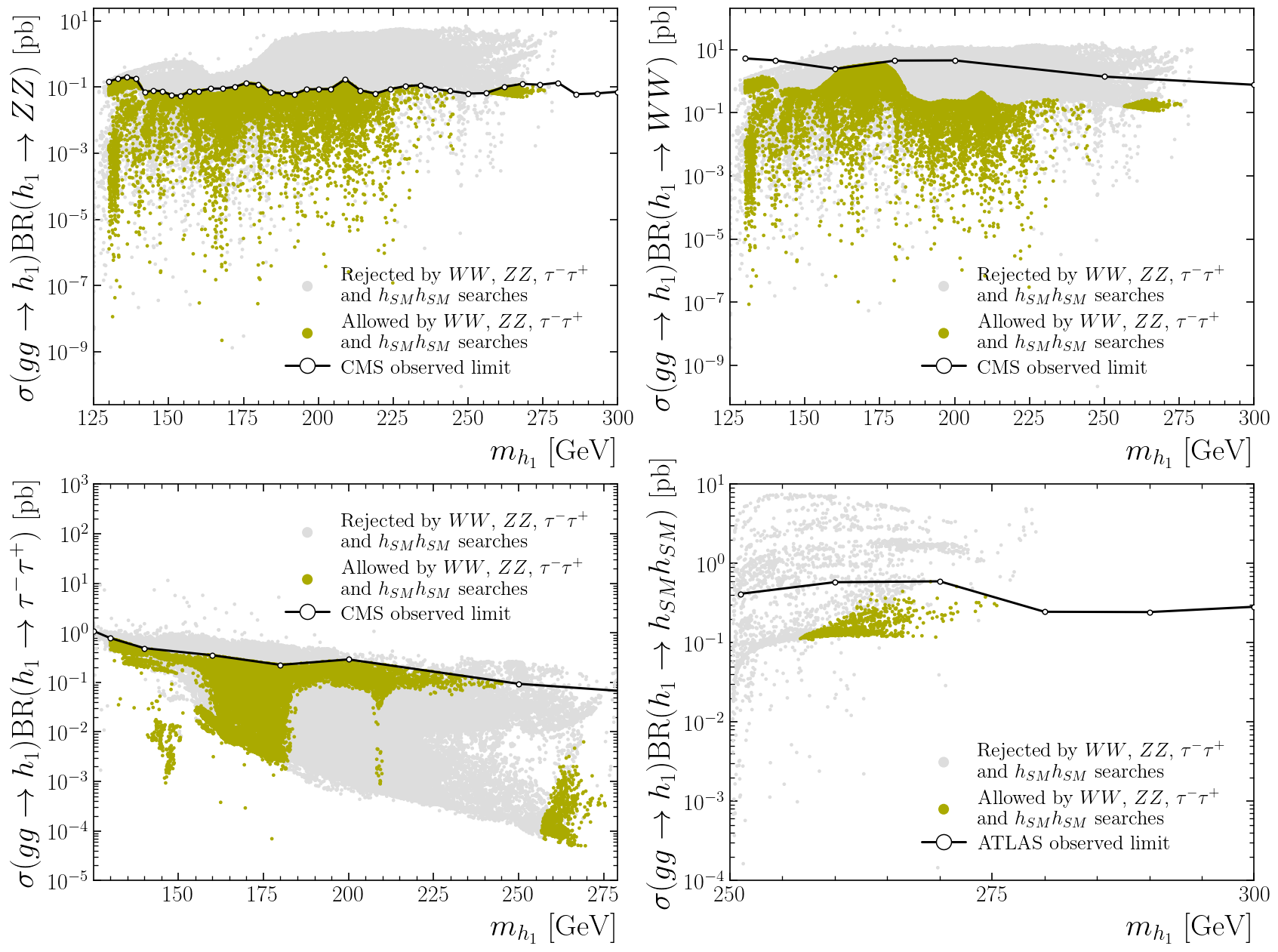}
    \caption{Gluon-fusion production cross-section times branching ratio of the lightest non-SM-like neutral scalar into $WW, \ ZZ, \ \tau\overline{\tau} 
    \text{ and } h_{SM}h_{SM}$. The coloured points simultaneously satisfy the LHC limits \cite{CMS:2026xbb, CMS:2022bcb, CMS:2022rbd, ATLAS:2023vdy} 
    from the four channels, while the grey points represent the complete generated sample.}
    \label{fig:production}
\end{figure}
In fig.~\ref{fig:production} we show the result of the calculations with the parameters of our scan which survived all the constraints outlined 
before (both for scalars and fermions) for the quantity
\beq
\sigma(gg\to h_1) BR(h_1\to XX)\,
\eeq
for the final states $WW$, $ZZ$, $\tau^+\tau^-$ and $h_{SM}h_{SM}$ (the latter is only relevant if $m_{h_1} > 2 m_{h_{SM}}$), as a function
of $m_{h_1}$. In all cases we show an exclusion line from the current observed limits on these quantities at the LHC. 
The conclusion is clear -- there is an ample region of parameter space which allows for the existence of the $h_1$ particle, although current LHC 
bounds on the several channels can already probe the properties of this state. Analogous conclusions would be found for the remaining
neutral scalars. We do not show their signal rates here, since we are not (yet) interested in a thorough analysis of the exclusion of parameter space
of the GOOFy model. However, it is clear that they will not be excluded by LHC searches themselves, since there is no enhancement of their couplings to 
quarks that might raise their expected production cross section or enhance decay branching ratios -- and their decays into electroweak gauge bosons will be 
suppressed compared to those of $h_{SM}$ or $h_1$, due to the sum rule on those couplings.

\subsubsection{$b\to s\gamma$ constraints}
\label{sec:btosg}

\begin{figure}[htb!]
    \centering
    \includegraphics[width=0.9\linewidth]{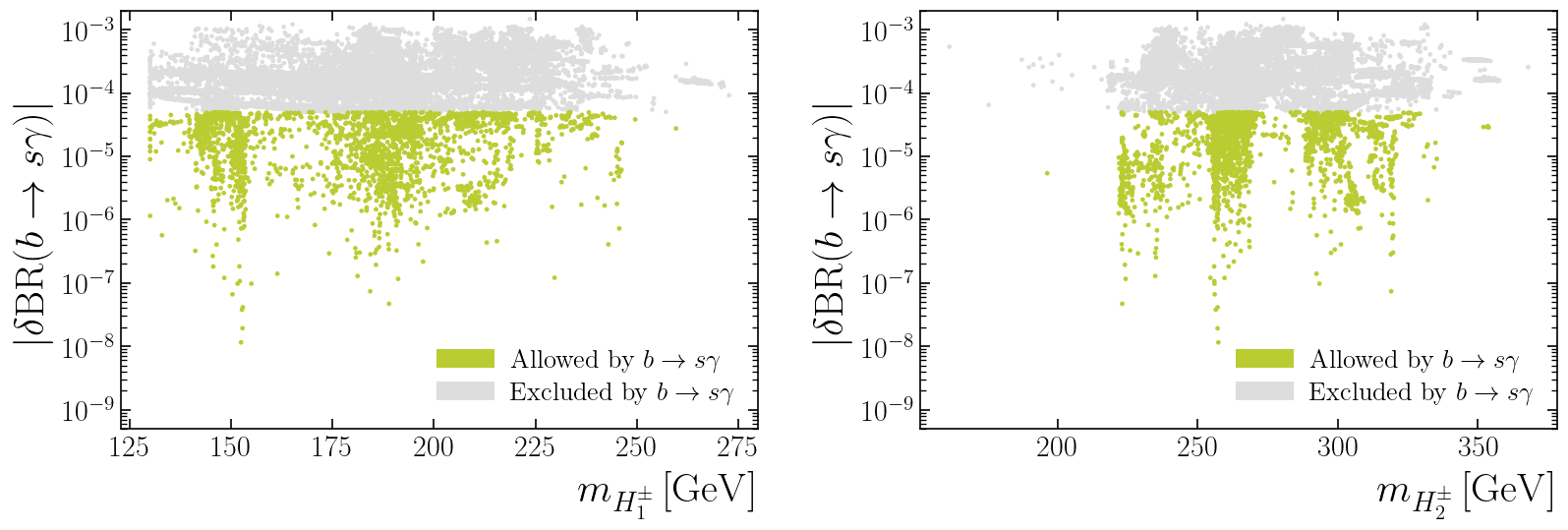}
    \caption{Parameter points that are accepted and rejected by the bound on the new physics contribution to the branching ratio of $b \to s \gamma$, 
    $|\delta BR(b \to s \gamma)| < 5\times10^{-5}$.}
    \label{fig:btosgamma}
\end{figure}

\begin{figure}[htb!]
    \centering
    \includegraphics[width=0.9\linewidth]{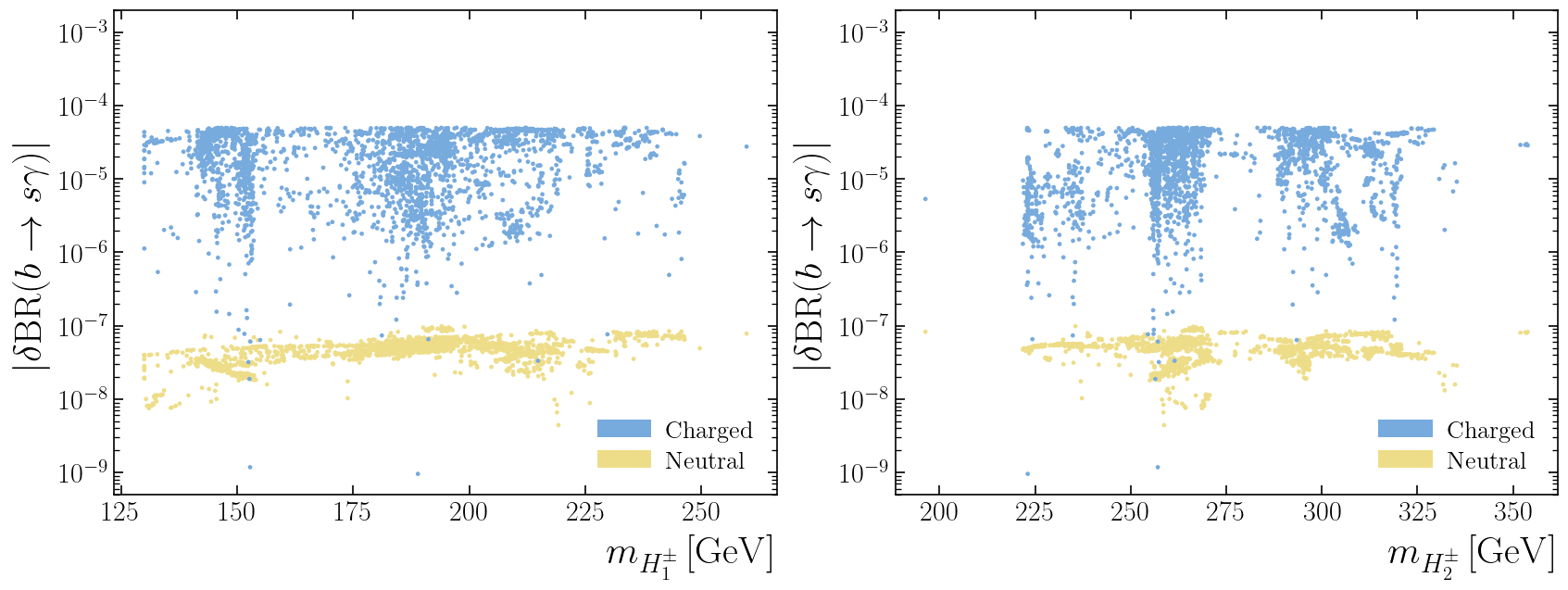}
    \caption{Neutral and charged contribution to the branching ratio of $b \to s \gamma$. }
    \label{fig:btosgamma_neutralvscharged}
\end{figure}

In theories with extra charged scalars there may be substantial extra contributions to the $b\to s\gamma$ branching ratio, other than
those present in the SM which arise due to the $W$ boson. In the 2HDM this process is one of those that most seriously curtails the model's
parameter space, originating large exclusion areas in the $m_{H^\pm}-\tan\beta$ plane. Given that we have two charged scalars in our GOOFy model,
and they have relatively small masses, it is crucial to verify whether the model is not excluded on these grounds. Further, in the model being
studied here the neutral scalars have FCNC interactions, which also contribute to the $b\to s\gamma$ constraints. The general formulae for
the $B\to X_s\gamma$ branching ratio in a general multi-higgs model were obtained in~\cite{Lavoura:2003xp}, and we follow the procedure outlined 
in~\cite{Ferreira:2019aps} to incorporate QCD corrections and evaluate the corresponding operators at the appropriate scales. We chose a 
bound of at most $5\times10^{-5}$ in the new physics contributions to $b \to s \gamma$ -- $|\delta BR(b \to s \gamma)| < 5\times10^{-5}$ -- 
which is fairly generous and based on current experimental values~\cite{Connell:2023jqq}~\footnote{Such an upper bound, if applied to a 2HDM of Type I or
II, correctly reproduces the well-known $b \to s \gamma$  constraints on those two models.}.
The results for a generic scan are shown in fig.~\ref{fig:btosgamma}, where the non-SM contrubutions to the branching ratio of the process 
$b\to s\gamma$ are shown as a function of the masses of both charged scalars. We see that there is an ample region of parameter space where the $b \to s \gamma$ bound is obeyed in our model, but we also witness the importance of this constraint, as there is a substantial exclusion of parameters as well. Further, we notice that there are no
specific values of charged masses preferred after the $b\to s \gamma$ bound is imposed, both for $H_1^\pm$ and for $H_2^\pm$. For completeness,
we also investigated the relative magnitude of the contributions from charged and neutral scalars to $b\to s \gamma$, which we show in 
fig.~\ref{fig:btosgamma_neutralvscharged}. As is usually the case in these models, the charged contributions are substantially larger than the
neutral ones. This also shows that in this model the FCNC interactions should be under control, though of course their contribution to 
several sensitive observables (such as $B$ meson mass differences) ought to be analysed.

\subsubsection{Precision Higgs data}

\begin{figure}[htb!]
    \centering
    \includegraphics[width=0.9\linewidth]{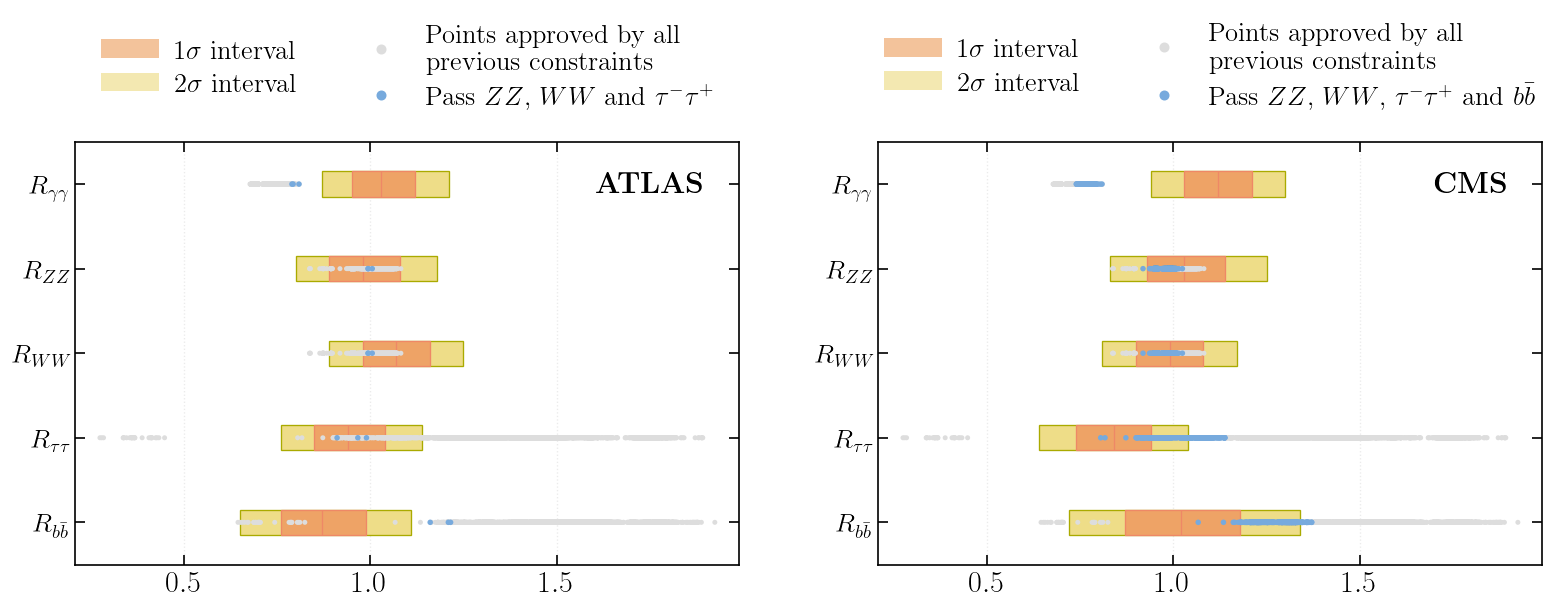}
    \caption{Higgs precision results for the $\gamma\gamma,~ZZ,~WW,~\tau^-\tau^+ \text{ and } b\bar{b}$ decay channels. The grey points pass all restrictions mentioned from Section~\ref{sec:bounds} to \ref{sec:btosg}, while the coloured points lie within the $2\sigma$ region for the $ZZ,~WW \text{ and }\tau^-\tau^+$ channels for the plot on the left and $ZZ,~WW~\tau^-\tau^+ \text{ and } b\bar{b}$ for the plot on the right \cite{ATLAS:2026pdi, CMS:2026zxu}.}
    \label{fig:ratios2s}
\end{figure}

Finally, we must analyse what the model implies for the properties of $h_{SM}$, the Higgs candidate within this model. The numerical scan we 
used already included constraints designed to force the interactions of $h_{SM}$ to be SM-like, in the form of limits imposed on the coupling modifiers
of this particle to the $W$ and $Z$ gauge bosons and to fermions as well. To verify that $h_{SM}$ is indeed behaving as the LHC is observing, we computed 
its gluon fusion production cross section and branching ratios, and compared them with SM expectations in the form of the signal ratios
\beq
R_X \,=\,\frac{\sigma(gg\to h_{SM}) BR(h_{SM} \to XX)}{\left[\sigma(gg\to h) BR(h \to XX)\right]_{SM}}\,,
\eeq
for four different decay channels: $h_{SM} \to \gamma\gamma$; $h_{SM} \to ZZ$; $h_{SM} \to WW$; and $h_{SM} \to \tau^+\tau^-$. 
The results are shown in fig.~\ref{fig:ratios2s}, where we plot the values for the four $R_X$ considering the current 1 and 2-$\sigma$ bounds from the
ATLAS and CMS collaborations. $R_X = 1$ would of course mean exact SM behaviour. We observe some small deviations in preferences for central values
in both collaborations, which is why we plot our results for both of their sets of results. We must emphasise that the points going into this plot
obey \textit{all} of the constraints outlined thus far: a correct fit to the fermion sector, satisfying the unitarity, BFB and electroweak
precision bounds, complying with non-observation of $h_1$ in the four channels considered in~\ref{sec:dirs} and with the $b\to s \gamma$ contraints
of~\ref{sec:btosg}. 

In fig.~\ref{fig:ratios2s} we observe that the GOOFy 3HDM we are studying has a preference for lower values of $R_{\gamma\gamma}$, and higher values of
$R_{bb}$. These seem to be features of the model and not simply an artifact of the scan, wherein some bias might be present in the choice of 
parameters. We cannot exclude the latter possibility, though we tried to be as exhaustive as possible in our covering of the (vast) parameter space. 
We found an interesting feature of the model which helps to explain the preference for lower values of $R_{\gamma\gamma}$ -- after all constraints were
taken into account, we found that the coupling modifier to $W$ bosons, $\kappa_V$ (see eq.~\eqref{eq:hVV}) and the CP-even top coupling modifier
$\kappa^e_t$ (see eq.~\eqref{eq:kf}) always have the same relative signal than the SM case; and the relative sign of the couplings of $h_{SM}$ to
$H_1^\pm$ and $H_2^\pm$ is such that the charged scalars' contribution to the $h_{SM}\to \gamma\gamma$ decay width is \textit{always} a destructive 
interference. We have further verified that the relative sign of the $h_{SM}WW$ and $h_{SM}H_i^+H_i^-$ vertices is not fixed \textit{a priori} by the 
model: these vertices can have any signs before theoretical bounds are imposed, but their relative sign is fixed by a combination of the unitarity and
BFB bounds. An even more thorough scan of the parameter space would be necessary to confirm that this is indeed a feature of the model. 

\begin{figure}[htb!]
    \centering
    \includegraphics[width=0.9\linewidth]{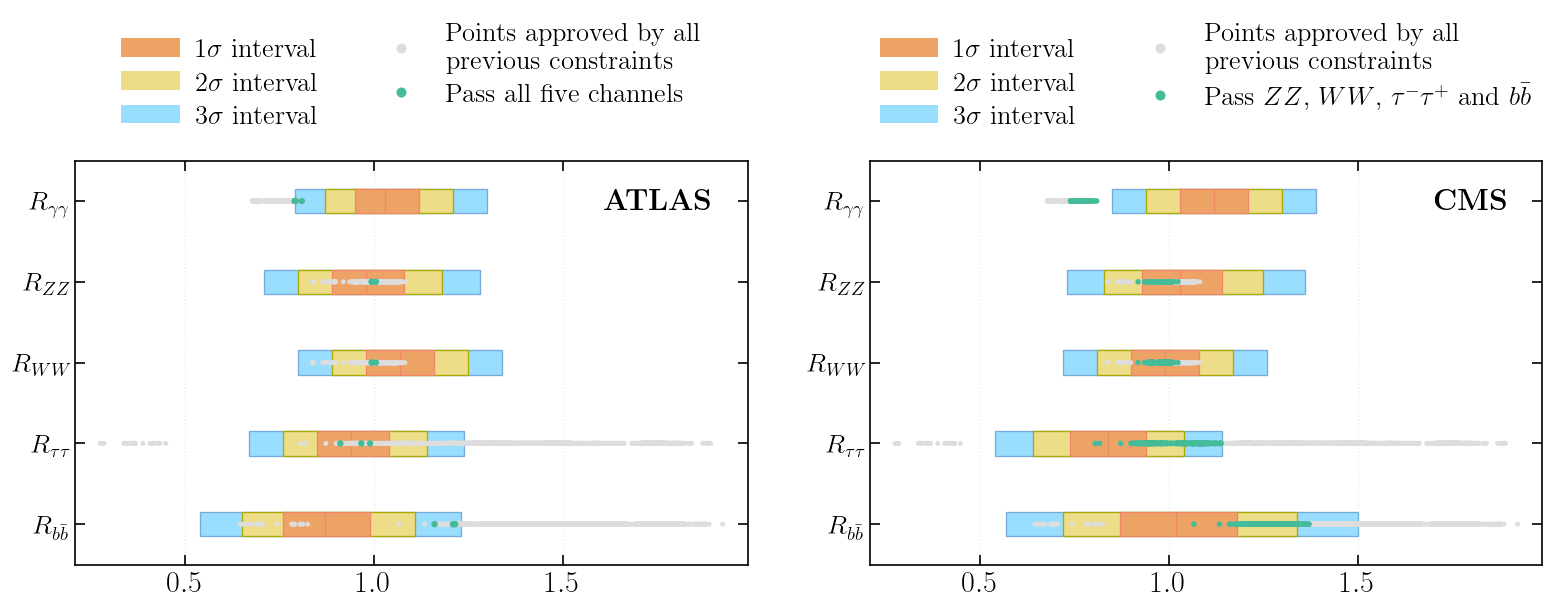}
    \caption{Higgs precision results for the $\gamma\gamma,~ZZ,~WW,~\tau^-\tau^+ \text{ and } b\bar{b}$ decay channels. The grey points pass all restrictions mentioned from Section~\ref{sec:bounds} to \ref{sec:btosg}, while the coloured points lie within the $3\sigma$ region for all channels for the plot on the left and $ZZ,~WW~\tau^-\tau^+ \text{ and } b\bar{b}$ for the plot on the right \cite{ATLAS:2026pdi, CMS:2026zxu}.}
    \label{fig:ratios3s}
\end{figure}

The green points in fig.~\ref{fig:ratios2s} are those that comply with three of the signal rates in the ATLAS case, and four in the CMS one. The model
has a clash with the $\gamma\gamma$ and $bb$ signal rates for the ATLAS results; and only with the $\gamma\gamma$ result for the CMS results. However, if one takes the 3-$\sigma$ limits of both collaborations, we find agreement with ATLAS and disagreement with CMS only in the $\gamma\gamma$
signal, as may be observed in fig.~\eqref{fig:ratios3s}. 

\section{Conclusions}

We constructed a 3HDM with a so-called GOOFy symmetry, which transforms the doublets and respective conjugates differently, with a permutation of the 
three up to negative signs and conjugations. The respective scalar potential is stable under RG evolution and features a pair of mass degenerate 
neutral scalars and a very symmetrical VEV, with equal magnitude for the 3 doublets (in the symmetry basis). The Yukawa sector is not realistic, which motivates the soft-breaking of the symmetry. The soft-breaking lifts the degeneracy of the two physical neutral scalars, 
and allows for a realistic Yukawa sector including a physical CP phase for the quark sector. We studied the RG stability of this complete model, and 
verified that the GOOFy parameter relations held to at least two loops via a direct inspection of the model's $\beta$-functions. An argument was even presented
for an all-order RG invariance of a more generic soft breaking pattern. We also performed a detailed phenomenological analysis. We find that the model 
is within the bounds of collider searches, despite having light scalars -- this indeed is a feature of the GOOFy symmetry, given that it implies the non
existence of a decoupling limit, due to the low number of dimensionful parameters it allows in the scalar potential. The model permits a good fit to quark
masses, their mixing and CP violation through the CKM matrix. There is a large allowed
parameter space even after stringent constraints, such as unitarity, boundedness from below, electroweak precisions constraints, extra 
neutral scalars' searches, $b\to s \gamma$ limits, are imposed. The model boasts a good Higgs candidate with SM-like interactions, though the results of our 
numerical fit show that it displays a small deviation from the SM expectation for the SM in the diphoton (always below the SM prediction) 
and $bb$ (always above the SM prediction) channels , and the model has good fits to the quark masses, mixing and CP.

Further features of the model include soft explicit CP breaking due to the imaginary soft breaking parameters introduced, 
yielding scalars which are CP admixtures, neither scalar nor pseudoscalar; and FCNC 
interactions mediated by the neutral scalars. These two properties will have contributions to observables such as the electric dipole moment
of the electron and neutron; and to $B$ and $K$ meson mixing. That analysis was not performed in this work. Nor did we exhaust the parameter space
we scanned, or imposed constraints from the non-observation of the extra neutral scalars $h_{2,3,4}$. A complete phenomenological analysis of the model
is therefore recommended. We have however shown that the GOOFy symmetries can yield interesting 3HDMs with testable phenomenology at current and
future colliders. The GOOFy symmetry topic is far from being a curiosity limited to the 2HDM, and quite clearly can be tested experimentally. 
The model herein proposed deserves to be further probed and confronted with all available data, and the techniques we employed in this paper 
may be used to construct other GOOFy 3HDMs -- our proposal is merely one among many different possibilities.

\subsection*{Acknowledgments:} 

IdMV thanks the University of Basel for hospitality.
IdMV acknowledges funding from Fundação para a Ciência e a Tecnologia (FCT) through the FCT Mobility program, and through
the project CFTP-FCT Unit 
UID/00777/2025 (\url{https://doi.org/10.54499/UID/00777/2025}).
PMF is supported by \textit{Funda\c c\~ao para a Ci\^encia e a Tecnologia} under contract
UIDP/00618/2025, UIDB/00618/2025  and through the PRR (Recovery and Resilience
Plan), within the scope of the investment “RE-C06-i06 - Science Plus Capacity Building”, measure 
``RE-C06-i06.m02 - Reinforcement of financing for International Partnerships in Science,
Technology and Innovation of the PRR”, under the project with the reference 2024.03328.CERN.

{\footnotesize
\bibliographystyle{utphys}
\bibliography{refs}

@article{Ivanov:2010ww,
    author = "Ivanov, I. P. and Nishi, C. C.",
    title = "{Properties of the general NHDM. I. The Orbit space}",
    eprint = "1004.1799",
    archivePrefix = "arXiv",
    primaryClass = "hep-th",
    doi = "10.1103/PhysRevD.82.015014",
    journal = "Phys. Rev. D",
    volume = "82",
    pages = "015014",
    year = "2010"
}

@article{Sartore:2020pkk,
    author = "Sartore, Lohan",
    title = "{General renormalization group equations for dimensionful couplings in the $\overline{\mathrm{MS}}$ scheme}",
    eprint = "2006.12307",
    archivePrefix = "arXiv",
    primaryClass = "hep-ph",
    doi = "10.1103/PhysRevD.102.076002",
    journal = "Phys. Rev. D",
    volume = "102",
    number = "7",
    pages = "076002",
    year = "2020"
}

@article{deMedeirosVarzielas:2011zw,
    author = "de Medeiros Varzielas, Ivo and Emmanuel-Costa, D.",
    title = "{Geometrical CP Violation}",
    eprint = "1106.5477",
    archivePrefix = "arXiv",
    primaryClass = "hep-ph",
    reportNumber = "CFTP-10-007, DO-TH-11-19",
    doi = "10.1103/PhysRevD.84.117901",
    journal = "Phys. Rev. D",
    volume = "84",
    pages = "117901",
    year = "2011"
}

@article{deMedeirosVarzielas:2012ylr,
    author = "de Medeiros Varzielas, Ivo and Emmanuel-Costa, David and Leser, Philipp",
    title = "{Geometrical CP Violation from Non-Renormalisable Scalar Potentials}",
    eprint = "1204.3633",
    archivePrefix = "arXiv",
    primaryClass = "hep-ph",
    reportNumber = "CFTP-12-005, DO-TH-12-12",
    doi = "10.1016/j.physletb.2012.08.008",
    journal = "Phys. Lett. B",
    volume = "716",
    pages = "193--196",
    year = "2012"
}

@article{deMedeirosVarzielas:2012rxf,
    author = "de Medeiros Varzielas, Ivo",
    title = "{Geometrical CP violation in multi-Higgs models}",
    eprint = "1205.3780",
    archivePrefix = "arXiv",
    primaryClass = "hep-ph",
    reportNumber = "DO-TH-12-18",
    doi = "10.1007/JHEP08(2012)055",
    journal = "JHEP",
    volume = "08",
    pages = "055",
    year = "2012"
}

@article{Bhattacharyya:2012pi,
    author = "Bhattacharyya, Gautam and de Medeiros Varzielas, Ivo and Leser, Philipp",
    title = "{A common origin of fermion mixing and geometrical CP violation, and its test through Higgs physics at the LHC}",
    eprint = "1210.0545",
    archivePrefix = "arXiv",
    primaryClass = "hep-ph",
    reportNumber = "SINP-TNP-2012-12, DO-TH-12-28",
    doi = "10.1103/PhysRevLett.109.241603",
    journal = "Phys. Rev. Lett.",
    volume = "109",
    pages = "241603",
    year = "2012"
}

@article{deMedeirosVarzielas:2013dmf,
    author = "de Medeiros Varzielas, Ivo and Pidt, Daniel",
    title = "{Geometrical CP violation with a complete fermion sector}",
    eprint = "1307.6545",
    archivePrefix = "arXiv",
    primaryClass = "hep-ph",
    reportNumber = "DO-13-20",
    doi = "10.1007/JHEP11(2013)206",
    journal = "JHEP",
    volume = "11",
    pages = "206",
    year = "2013"
}

@article{CarcamoHernandez:2013yiy,
    author = {Carcamo Hernandez, Antonio Enrique and de Medeiros Varzielas, Ivo and Kovalenko, S. G. and P{\"a}s, H. and Schmidt, Ivan},
    title = "{Lepton masses and mixings in an $A_4$ multi-Higgs model with a radiative seesaw mechanism}",
    eprint = "1307.6499",
    archivePrefix = "arXiv",
    primaryClass = "hep-ph",
    doi = "10.1103/PhysRevD.88.076014",
    journal = "Phys. Rev. D",
    volume = "88",
    number = "7",
    pages = "076014",
    year = "2013"
}

@article{deMedeirosVarzielas:2015ybd,
    author = "de Medeiros Varzielas, Ivo and Fischer, Oliver and Maurer, Vinzenz",
    title = "{$ {\mathbb{A}}_4 $ symmetry at colliders and in the universe}",
    eprint = "1504.03955",
    archivePrefix = "arXiv",
    primaryClass = "hep-ph",
    doi = "10.1007/JHEP08(2015)080",
    journal = "JHEP",
    volume = "08",
    pages = "080",
    year = "2015"
}

@article{deMedeirosVarzielas:2016rii,
    author = "de Medeiros Varzielas, Ivo and King, Stephen F. and Luhn, Christoph and Neder, Thomas",
    title = "{CP-odd invariants for multi-Higgs models: applications with discrete symmetry}",
    eprint = "1603.06942",
    archivePrefix = "arXiv",
    primaryClass = "hep-ph",
    doi = "10.1103/PhysRevD.94.056007",
    journal = "Phys. Rev. D",
    volume = "94",
    number = "5",
    pages = "056007",
    year = "2016"
}

@article{deMedeirosVarzielas:2017glw,
    author = "de Medeiros Varzielas, Ivo and King, Stephen F. and Luhn, Christoph and Neder, Thomas",
    title = "{Minima of multi-Higgs potentials with triplets of $\Delta(3n^2)$ and $\Delta(6n^2)$}",
    eprint = "1704.06322",
    archivePrefix = "arXiv",
    primaryClass = "hep-ph",
    reportNumber = "QFET-2017-05, SI-HEP-2017-06",
    doi = "10.1016/j.physletb.2017.11.005",
    journal = "Phys. Lett. B",
    volume = "775",
    pages = "303--310",
    year = "2017"
}

@article{deMedeirosVarzielas:2017ote,
    author = "de Medeiros Varzielas, Ivo and King, Stephen F. and Luhn, Christoph and Neder, Thomas",
    title = "{Spontaneous CP violation in multi-Higgs potentials with triplets of $\Delta(3n^2)$ and $\Delta(6n^2)$}",
    eprint = "1706.07606",
    archivePrefix = "arXiv",
    primaryClass = "hep-ph",
    doi = "10.1007/JHEP11(2017)136",
    journal = "JHEP",
    volume = "11",
    pages = "136",
    year = "2017"
}

@article{deMedeirosVarzielas:2019rrp,
    author = "de Medeiros Varzielas, I. and Ivanov, Igor P.",
    title = "{Recognizing symmetries in a 3HDM in a basis-independent way}",
    eprint = "1903.11110",
    archivePrefix = "arXiv",
    primaryClass = "hep-ph",
    reportNumber = "CFTP/19-005",
    doi = "10.1103/PhysRevD.100.015008",
    journal = "Phys. Rev. D",
    volume = "100",
    number = "1",
    pages = "015008",
    year = "2019"
}

@article{CarcamoHernandez:2021osw,
    author = "C{\'a}rcamo Hern{\'a}ndez, A. E. and de Medeiros Varzielas, Ivo and L{\'o}pez-Ib{\'a}{\~n}ez, M. L. and Melis, Aurora",
    title = "{Controlled fermion mixing and FCNCs in a $\Delta(27)$ 3+1 Higgs Doublet Model}",
    eprint = "2102.05658",
    archivePrefix = "arXiv",
    primaryClass = "hep-ph",
    doi = "10.1007/JHEP05(2021)215",
    journal = "JHEP",
    volume = "05",
    pages = "215",
    year = "2021"
}

@article{deMedeirosVarzielas:2021zqs,
    author = "de Medeiros Varzielas, Ivo and Ivanov, Igor P. and Levy, Miguel",
    title = "{Exploring multi-Higgs models with softly broken large discrete symmetry groups}",
    eprint = "2107.08227",
    archivePrefix = "arXiv",
    primaryClass = "hep-ph",
    doi = "10.1140/epjc/s10052-021-09681-w",
    journal = "Eur. Phys. J. C",
    volume = "81",
    number = "10",
    pages = "918",
    year = "2021"
}

@article{Kalinowski:2021lvw,
    author = "Kalinowski, J. and Kotlarski, W. and Rebelo, M. N. and de Medeiros Varzielas, I.",
    title = "{3HDM with {\ensuremath{\Delta}}(27) symmetry and its phenomenological consequences}",
    eprint = "2112.12699",
    archivePrefix = "arXiv",
    primaryClass = "hep-ph",
    reportNumber = "CERN-TH-2021-182",
    doi = "10.1007/JHEP02(2023)231",
    journal = "JHEP",
    volume = "02",
    pages = "231",
    year = "2023"
}

@article{Barreto:2025jzx,
    author = "Barreto, Gon{\c{c}}alo and de Medeiros Varzielas, Ivo",
    title = "{3HDM with softly broken $\Delta (54)$ and $\Sigma (36)$}",
    eprint = "2509.04369",
    archivePrefix = "arXiv",
    primaryClass = "hep-ph",
    doi = "10.1140/epjc/s10052-025-15140-7",
    journal = "Eur. Phys. J. C",
    volume = "85",
    number = "12",
    pages = "1416",
    year = "2025"
}

@article{deMedeirosVarzielas:2025byb,
    author = "de Medeiros Varzielas, Ivo and Liu, Ming-Shau and Sengupta, Amartya and Talbert, Jim",
    title = "{Residual symmetries and scalar multiplet vacuum alignment in non-Abelian flavour models}",
    eprint = "2512.19789",
    archivePrefix = "arXiv",
    primaryClass = "hep-ph",
    reportNumber = "LA-UR-25-31854",
    doi = "10.1007/JHEP07(2026)022",
    journal = "JHEP",
    volume = "07",
    pages = "022",
    year = "2026"
}

@article{Haber:2025cbb,
    author = "Haber, Howard E. and Ferreira, P. M.",
    title = "{RG-stable parameter relations of a scalar field theory in absence of a symmetry}",
    eprint = "2502.11011",
    archivePrefix = "arXiv",
    primaryClass = "hep-ph",
    doi = "10.1140/epjc/s10052-025-14570-7",
    journal = "Eur. Phys. J. C",
    volume = "85",
    number = "5",
    pages = "541",
    year = "2025",
    note = "[Erratum: Eur.Phys.J.C 85, 867 (2025), Erratum: Eur.Phys.J.C 86, 245 (2026)]"
}

@article{Ferreira:2026ccg,
    author = "Ferreira, P. M.",
    title = "{GOOFy fermions}",
    eprint = "2604.07458",
    archivePrefix = "arXiv",
    primaryClass = "hep-ph",
    doi = "10.1007/JHEP09(2026)007",
    journal = "JHEP",
    volume = "09",
    pages = "007",
    year = "2026"
}

@article{Bento:2017eti,
    author = "Bento, Miguel P. and Haber, Howard E. and Rom{\~a}o, J. C. and Silva, Jo{\~a}o P.",
    title = "{Multi-Higgs doublet models: physical parametrization, sum rules and unitarity bounds}",
    eprint = "1708.09408",
    archivePrefix = "arXiv",
    primaryClass = "hep-ph",
    reportNumber = "CFTP-17-005, SCIPP-17-10",
    doi = "10.1007/JHEP11(2017)095",
    journal = "JHEP",
    volume = "11",
    pages = "095",
    year = "2017"
}

@article{Bento:2022vsb,
    author = "Bento, Miguel P. and Rom{\~a}o, Jorge C. and Silva, Jo{\~a}o P.",
    title = "{Unitarity bounds for all symmetry-constrained 3HDMs}",
    eprint = "2204.13130",
    archivePrefix = "arXiv",
    primaryClass = "hep-ph",
    doi = "10.1007/JHEP08(2022)273",
    journal = "JHEP",
    volume = "08",
    pages = "273",
    year = "2022"
}

@article{Grimus:2008nb,
    author = "Grimus, W. and Lavoura, L. and Ogreid, O. M. and Osland, P.",
    title = "{The Oblique parameters in multi-Higgs-doublet models}",
    eprint = "0802.4353",
    archivePrefix = "arXiv",
    primaryClass = "hep-ph",
    reportNumber = "UWTHPH-2008-4",
    doi = "10.1016/j.nuclphysb.2008.04.019",
    journal = "Nucl. Phys. B",
    volume = "801",
    pages = "81--96",
    year = "2008"
}

@article{Peskin:1990zt,
    author = "Peskin, Michael E. and Takeuchi, Tatsu",
    title = "{A New constraint on a strongly interacting Higgs sector}",
    reportNumber = "SLAC-PUB-5272",
    doi = "10.1103/PhysRevLett.65.964",
    journal = "Phys. Rev. Lett.",
    volume = "65",
    pages = "964--967",
    year = "1990"
}

@article{Peskin:1991sw,
    author = "Peskin, Michael E. and Takeuchi, Tatsu",
    title = "{Estimation of oblique electroweak corrections}",
    reportNumber = "SLAC-PUB-5618",
    doi = "10.1103/PhysRevD.46.381",
    journal = "Phys. Rev. D",
    volume = "46",
    pages = "381--409",
    year = "1992"
}

@article{Boto:2022uwv,
    author = "Boto, Rafael and Rom{\~a}o, Jorge C. and Silva, Jo{\~a}o P.",
    title = "{Bounded from below conditions on a class of symmetry constrained 3HDM}",
    eprint = "2208.01068",
    archivePrefix = "arXiv",
    primaryClass = "hep-ph",
    reportNumber = "CFTP/22-004",
    doi = "10.1103/PhysRevD.106.115010",
    journal = "Phys. Rev. D",
    volume = "106",
    number = "11",
    pages = "115010",
    year = "2022"
}

@article{Djouadi:2005gi,
    author = "Djouadi, Abdelhak",
    title = "{The Anatomy of electro-weak symmetry breaking. I: The Higgs boson in the standard model}",
    eprint = "hep-ph/0503172",
    archivePrefix = "arXiv",
    reportNumber = "LPT-ORSAY-05-17",
    doi = "10.1016/j.physrep.2007.10.004",
    journal = "Phys. Rept.",
    volume = "457",
    pages = "1--216",
    year = "2008"
}

@article{Djouadi:2005gj,
    author = "Djouadi, Abdelhak",
    title = "{The Anatomy of electro-weak symmetry breaking. II. The Higgs bosons in the minimal supersymmetric model}",
    eprint = "hep-ph/0503173",
    archivePrefix = "arXiv",
    reportNumber = "LPT-ORSAY-05-18",
    doi = "10.1016/j.physrep.2007.10.005",
    journal = "Phys. Rept.",
    volume = "459",
    pages = "1--241",
    year = "2008"
}

@article{Kannike:2012pe,
    author = "Kannike, Kristjan",
    title = "{Vacuum Stability Conditions From Copositivity Criteria}",
    eprint = "1205.3781",
    archivePrefix = "arXiv",
    primaryClass = "hep-ph",
    doi = "10.1140/epjc/s10052-012-2093-z",
    journal = "Eur. Phys. J. C",
    volume = "72",
    pages = "2093",
    year = "2012"
}

@article{Lee:1973iz,
      author         = "Lee, T. D.",
      title          = "{A Theory of Spontaneous T Violation}",
      journal        = "Phys. Rev.",
      volume         = "D8",
      year           = "1973",
      pages          = "1226-1239",
      doi            = "10.1103/PhysRevD.8.1226",
      SLACcitation   = "%%CITATION = PHRVA,D8,1226;%%"
}

@article{Lavoura:2003xp,
    author = "Lavoura, L.",
    title = "{General formulae for f(1) ---{\ensuremath{>}} f(2) gamma}",
    eprint = "hep-ph/0302221",
    archivePrefix = "arXiv",
    doi = "10.1140/epjc/s2003-01212-7",
    journal = "Eur. Phys. J. C",
    volume = "29",
    pages = "191--195",
    year = "2003"
}

@article{Branco:2011iw,
      author         = "Branco, G. C. and Ferreira, P. M. and Lavoura, L. and
                        Rebelo, M. N. and Sher, Marc and Silva, Jo{\~a}o P.",
      title          = "{Theory and phenomenology of two-Higgs-doublet models}",
      journal        = "Phys. Rept.",
      volume         = "516",
      year           = "2012",
      pages          = "1-102",
      doi            = "10.1016/j.physrep.2012.02.002",
      eprint         = "1106.0034",
      archivePrefix  = "arXiv",
      primaryClass   = "hep-ph",
      SLACcitation   = "%%CITATION = ARXIV:1106.0034;%%"
}

@article{Glashow:1976nt,
      author         = "Glashow, Sheldon L. and Weinberg, Steven",
      title          = "{Natural Conservation Laws for Neutral Currents}",
      journal        = "Phys. Rev.",
      volume         = "D15",
      year           = "1977",
      pages          = "1958",
      doi            = "10.1103/PhysRevD.15.1958",
      reportNumber   = "HUTP-76-A158",
      SLACcitation   = "%%CITATION = PHRVA,D15,1958;%%"
}

@article{Paschos:1976ay,
      author         = "Paschos, E. A.",
      title          = "{Diagonal Neutral Currents}",
      journal        = "Phys. Rev.",
      volume         = "D15",
      year           = "1977",
      pages          = "1966",
      doi            = "10.1103/PhysRevD.15.1966",
      reportNumber   = "BNL-21870",
      SLACcitation   = "%%CITATION = PHRVA,D15,1966;%%"
}

@article{deMedeirosVarzielas:2011gly,
    author = "de Medeiros Varzielas, Ivo",
    title = "{Family symmetries and alignment in multi-Higgs doublet models}",
    eprint = "1104.2601",
    archivePrefix = "arXiv",
    primaryClass = "hep-ph",
    reportNumber = "DO-TH-11-09",
    doi = "10.1016/j.physletb.2011.06.042",
    journal = "Phys. Lett. B",
    volume = "701",
    pages = "597--600",
    year = "2011"
}

@article{Ferreira:2010bm,
    author = "Ferreira, P.M. and Silva, Jo{\~a}o P.",
    title = "{A Two-Higgs Doublet Model With Remarkable CP Properties}",
    eprint = "1001.0574",
    archivePrefix = "arXiv",
    primaryClass = "hep-ph",
    doi = "10.1140/epjc/s10052-010-1384-5",
    journal = "Eur. Phys. J. C",
    volume = "69",
    pages = "45--52",
    year = "2010"
}

@article{Cheng:1973nv,
    author = "Cheng, T. P. and Eichten, E. and Li, Ling-Fong",
    title = "{Higgs Phenomena in Asymptotically Free Gauge Theories}",
    reportNumber = "SLAC-PUB-1340",
    doi = "10.1103/PhysRevD.9.2259",
    journal = "Phys. Rev. D",
    volume = "9",
    pages = "2259",
    year = "1974"
}

@article{Machacek:1984zw,
    author = "Machacek, Marie E. and Vaughn, Michael T.",
    title = "{Two Loop Renormalization Group Equations in a General Quantum Field Theory. 3. Scalar Quartic Couplings}",
    reportNumber = "NUB-2653-REV, NUB-2653",
    doi = "10.1016/0550-3213(85)90040-9",
    journal = "Nucl. Phys. B",
    volume = "249",
    pages = "70--92",
    year = "1985"
}

@article{Luo:2002ti,
    author = "Luo, Ming-xing and Wang, Hua-wen and Xiao, Yong",
    title = "{Two loop renormalization group equations in general gauge field theories}",
    eprint = "hep-ph/0211440",
    archivePrefix = "arXiv",
    doi = "10.1103/PhysRevD.67.065019",
    journal = "Phys. Rev. D",
    volume = "67",
    pages = "065019",
    year = "2003"
}

@article{ParticleDataGroup:2024cfk,
    author = "Navas, S. and others",
    collaboration = "Particle Data Group",
    title = "{Review of particle physics}",
    doi = "10.1103/PhysRevD.110.030001",
    journal = "Phys. Rev. D",
    volume = "110",
    number = "3",
    pages = "030001",
    year = "2024"
}

@article{Ferreira:2015rha,
    author = "Ferreira, Pedro and Haber, Howard E. and Santos, Edward",
    title = "{Preserving the validity of the Two-Higgs Doublet Model up to the Planck scale}",
    eprint = "1505.04001",
    archivePrefix = "arXiv",
    primaryClass = "hep-ph",
    reportNumber = "SCIPP-15-07, SCIPP-15/07",
    doi = "10.1103/PhysRevD.92.033003",
    journal = "Phys. Rev. D",
    volume = "92",
    pages = "033003",
    year = "2015",
    note = "[Erratum: Phys.Rev.D 94, 059903 (2016)]"
}

@article{Sartore:2020gou,
    author = "Sartore, Lohan and Schienbein, Ingo",
    title = "{PyR@TE 3}",
    eprint = "2007.12700",
    archivePrefix = "arXiv",
    primaryClass = "hep-ph",
    doi = "10.1016/j.cpc.2020.107819",
    journal = "Comput. Phys. Commun.",
    volume = "261",
    pages = "107819",
    year = "2021"
}

@article{Doring:2024kdg,
    author = {D{\"o}ring, Christian and Trautner, Andreas},
    title = "{Symmetries from outer automorphisms and unorthodox group extensions}",
    eprint = "2410.11052",
    archivePrefix = "arXiv",
    primaryClass = "hep-ph",
    reportNumber = "ULB-TH/24-11",
    doi = "10.1088/1751-8121/ae17fa",
    journal = "J. Phys. A",
    volume = "58",
    number = "47",
    pages = "475401",
    year = "2025"
}

@article{Ivanov:2007de,
      author         = "Ivanov, Igor P.",
      title          = "{Minkowski space structure of the Higgs potential in
                        2HDM. II. Minima, symmetries, and topology}",
      journal        = "Phys. Rev.",
      volume         = "D77",
      year           = "2008",
      pages          = "015017",
      doi            = "10.1103/PhysRevD.77.015017",
      eprint         = "0710.3490",
      archivePrefix  = "arXiv",
      primaryClass   = "hep-ph",
      SLACcitation   = "%%CITATION = ARXIV:0710.3490;%%"
}

@article{Ferreira:2023dke,
    author = "Ferreira, P. M. and Grzadkowski, B. and Ogreid, O. M. and Osland, P.",
    title = "{New symmetries of the two-Higgs-doublet model}",
    eprint = "2306.02410",
    archivePrefix = "arXiv",
    primaryClass = "hep-ph",
    doi = "10.1140/epjc/s10052-024-12561-8",
    journal = "Eur. Phys. J. C",
    volume = "84",
    number = "3",
    pages = "234",
    year = "2024"
}

@article{Ferreira:2004yd,
      author         = "Ferreira, P. M. and Santos, R. and Barroso, A.",
      title          = "{Stability of the tree-level vacuum in two Higgs doublet
                        models against charge or CP spontaneous violation}",
      journal        = "Phys. Lett.",
      volume         = "B603",
      year           = "2004",
      pages          = "219-229",
      doi            = "10.1016/j.physletb.2004.10.022,
                        10.1016/j.physletb.2005.09.074",
      note           = "[Erratum: Phys. Lett.B629,114(2005)]",
      eprint         = "hep-ph/0406231",
      archivePrefix  = "arXiv",
      primaryClass   = "hep-ph",
      SLACcitation   = "%%CITATION = HEP-PH/0406231;%%"
}

@article{Barroso:2005sm,
      author         = "Barroso, A. and Ferreira, P. M. and Santos, R.",
      title          = "{Charge and CP symmetry breaking in two Higgs doublet
                        models}",
      journal        = "Phys. Lett.",
      volume         = "B632",
      year           = "2006",
      pages          = "684-687",
      doi            = "10.1016/j.physletb.2005.11.031",
      eprint         = "hep-ph/0507224",
      archivePrefix  = "arXiv",
      primaryClass   = "hep-ph",
      SLACcitation   = "%%CITATION = HEP-PH/0507224;%%"
}

@article{Ivanov:2013bka,
    author = "Ivanov, I. P. and Nishi, C. C.",
    title = "{Abelian symmetries of the N-Higgs-doublet model with Yukawa interactions}",
    eprint = "1309.3682",
    archivePrefix = "arXiv",
    primaryClass = "hep-ph",
    doi = "10.1007/JHEP11(2013)069",
    journal = "JHEP",
    volume = "11",
    pages = "069",
    year = "2013"
}

@article{Ivanov:2014doa,
    author = "Ivanov, I. P. and Nishi, C. C.",
    title = "{Symmetry breaking patterns in 3HDM}",
    eprint = "1410.6139",
    archivePrefix = "arXiv",
    primaryClass = "hep-ph",
    reportNumber = "UMD-PP-014-017",
    doi = "10.1007/JHEP01(2015)021",
    journal = "JHEP",
    volume = "01",
    pages = "021",
    year = "2015"
}

@article{Degee:2012sk,
    author = "Degee, A. and Ivanov, I. P. and Keus, V.",
    title = "{Geometric minimization of highly symmetric potentials}",
    eprint = "1211.4989",
    archivePrefix = "arXiv",
    primaryClass = "hep-ph",
    doi = "10.1007/JHEP02(2013)125",
    journal = "JHEP",
    volume = "02",
    pages = "125",
    year = "2013"
}

@article{Faro:2019vcd,
    author = "Faro, Francisco S. and Ivanov, Igor P.",
    title = "{Boundedness from below in the $U(1)\times U(1)$ three-Higgs-doublet model}",
    eprint = "1907.01963",
    archivePrefix = "arXiv",
    primaryClass = "hep-ph",
    reportNumber = "CFTP-19-021, CFTP/19-022",
    doi = "10.1103/PhysRevD.100.035038",
    journal = "Phys. Rev. D",
    volume = "100",
    number = "3",
    pages = "035038",
    year = "2019"
}

@article{Das:2021oik,
    author = "Das, Dipankar and Ferreira, P. M. and Morais, Ant{\'o}nio P. and Padilla-Gay, Ian and Pasechnik, Roman and Rodrigues, J. Pedro",
    title = "{A three Higgs doublet model with symmetry-suppressed flavour changing neutral currents}",
    eprint = "2106.06425",
    archivePrefix = "arXiv",
    primaryClass = "hep-ph",
    reportNumber = "LU TP 19-12",
    doi = "10.1007/JHEP11(2021)079",
    journal = "JHEP",
    volume = "11",
    pages = "079",
    year = "2021"
}

@article{Ferreira:2022zil,
    author = "Ferreira, Pedro M. and Freitas, Felipe F. and Gon{\c{c}}alves, Jo{\~a}o and Morais, Ant{\'o}nio P. and Pasechnik, Roman and Vatellis, Vasileios",
    title = "{Phenomenology of a flavored multiscalar Branco-Grimus-Lavoura-like model with three generations of massive neutrinos}",
    eprint = "2202.13153",
    archivePrefix = "arXiv",
    primaryClass = "hep-ph",
    doi = "10.1103/PhysRevD.106.075017",
    journal = "Phys. Rev. D",
    volume = "106",
    number = "7",
    pages = "075017",
    year = "2022"
}

@article{Carrolo:2022oyg,
    author = "Carrolo, Sergio and Romao, Jorge C. and Silva, Joao P.",
    title = "{Conditions for global minimum in the $A_4$ symmetric 3HDM}",
    eprint = "2207.02928",
    archivePrefix = "arXiv",
    primaryClass = "hep-ph",
    doi = "10.1140/epjc/s10052-022-10712-3",
    journal = "Eur. Phys. J. C",
    volume = "82",
    number = "8",
    pages = "749",
    year = "2022"
}

@article{deMedeirosVarzielas:2022kbj,
    author = "de Medeiros Varzielas, Ivo and Ivo, Diogo",
    title = "{Softly-broken $A_4$ or $S_4$ 3HDMs with stable states}",
    eprint = "2202.00681",
    archivePrefix = "arXiv",
    primaryClass = "hep-ph",
    doi = "10.1140/epjc/s10052-022-10331-y",
    journal = "Eur. Phys. J. C",
    volume = "82",
    number = "5",
    pages = "415",
    year = "2022"
}

@article{ATLAS:2020ior,
    author = "Aad, Georges and others",
    collaboration = "ATLAS",
    title = "{$CP$ Properties of Higgs Boson Interactions with Top Quarks in the $t\bar{t}H$ and $tH$ Processes Using $H \rightarrow \gamma\gamma$ with the ATLAS Detector}",
    eprint = "2004.04545",
    archivePrefix = "arXiv",
    primaryClass = "hep-ex",
    reportNumber = "CERN-EP-2020-046",
    doi = "10.1103/PhysRevLett.125.061802",
    journal = "Phys. Rev. Lett.",
    volume = "125",
    number = "6",
    pages = "061802",
    year = "2020"
}

@article{Connell:2023jqq,
    author = "Connell, Joseph M. and Ferreira, Pedro and Haber, Howard E.",
    title = "{Accommodating hints of new heavy scalars in the framework of the flavor-aligned two-Higgs-doublet model}",
    eprint = "2302.13697",
    archivePrefix = "arXiv",
    primaryClass = "hep-ph",
    reportNumber = "SCIPP-22/02",
    doi = "10.1103/PhysRevD.108.055031",
    journal = "Phys. Rev. D",
    volume = "108",
    number = "5",
    pages = "055031",
    year = "2023"
}

@article{Ferreira:2019aps,
    author = "Ferreira, P. M. and Lavoura, L.",
    title = "{No strong $CP$ violation up to the one-loop level in a two-Higgs-doublet model}",
    eprint = "1904.08438",
    archivePrefix = "arXiv",
    primaryClass = "hep-ph",
    reportNumber = "CFTP/19-013",
    doi = "10.1140/epjc/s10052-019-7053-4",
    journal = "Eur. Phys. J. C",
    volume = "79",
    number = "7",
    pages = "552",
    year = "2019"
}

@article{Ferreira:2022sno,
    author = "Ferreira, Pedro M. and Gon{\c{c}}alves, Jo{\~a}o and Morais, Ant{\'o}nio P. and Onofre, Ant{\'o}nio and Pasechnik, Roman and Vatellis, Vasileios",
    title = "{Collider phenomenology of new neutral scalars in a flavored multi-Higgs model}",
    eprint = "2211.10109",
    archivePrefix = "arXiv",
    primaryClass = "hep-ph",
    reportNumber = "CERN-TH-2022-194",
    doi = "10.1103/PhysRevD.107.095041",
    journal = "Phys. Rev. D",
    volume = "107",
    number = "9",
    pages = "095041",
    year = "2023"
}

@article{Branco:1996bq,
    author = "Branco, G. C. and Grimus, W. and Lavoura, L.",
    title = "{Relating the scalar flavor changing neutral couplings to the CKM matrix}",
    eprint = "hep-ph/9601383",
    archivePrefix = "arXiv",
    reportNumber = "FISIST-1-96-CFIF",
    doi = "10.1016/0370-2693(96)00494-7",
    journal = "Phys. Lett. B",
    volume = "380",
    pages = "119--126",
    year = "1996"
}

@article{Trautner:2025yxz,
    author = "Trautner, Andreas",
    title = "{Goofy is the new Normal}",
    eprint = "2505.00099",
    archivePrefix = "arXiv",
    primaryClass = "hep-ph",
    doi = "10.1007/JHEP10(2025)051",
    journal = "JHEP",
    volume = "10",
    pages = "051",
    year = "2025"
}

@article{Weinberg:1976hu,
    author = "Weinberg, Steven",
    title = "{Gauge Theory of CP Violation}",
    reportNumber = "PRINT-76-0555 (HARVARD)",
    doi = "10.1103/PhysRevLett.37.657",
    journal = "Phys. Rev. Lett.",
    volume = "37",
    pages = "657",
    year = "1976"
}

@article{Barroso:2006pa,
    author = "Barroso, A. and Ferreira, P. M. and Santos, R. and Silva, Joao P.",
    title = "{Stability of the normal vacuum in multi-Higgs-doublet models}",
    eprint = "hep-ph/0608282",
    archivePrefix = "arXiv",
    doi = "10.1103/PhysRevD.74.085016",
    journal = "Phys. Rev. D",
    volume = "74",
    pages = "085016",
    year = "2006"
}

@article{Barroso:2012wz,
    author = "Barroso, A. and Ferreira, P. M. and Santos, Rui and Silva, Jo{\~a}o P.",
    title = "{Probing the scalar-pseudoscalar mixing in the 125 GeV Higgs particle with current data}",
    eprint = "1205.4247",
    archivePrefix = "arXiv",
    primaryClass = "hep-ph",
    doi = "10.1103/PhysRevD.86.015022",
    journal = "Phys. Rev. D",
    volume = "86",
    pages = "015022",
    year = "2012"
}

@article{Harlander:2012pb,
    author = "Harlander, Robert V. and Liebler, Stefan and Mantler, Hendrik",
    title = "{SusHi: A program for the calculation of Higgs production in gluon fusion and bottom-quark annihilation in the Standard Model and the MSSM}",
    eprint = "1212.3249",
    archivePrefix = "arXiv",
    primaryClass = "hep-ph",
    reportNumber = "WUB-12-28, LPN12-134",
    doi = "10.1016/j.cpc.2013.02.006",
    journal = "Comput. Phys. Commun.",
    volume = "184",
    pages = "1605--1617",
    year = "2013"
}

@article{Harlander:2016hcx,
    author = "Harlander, Robert V. and Liebler, Stefan and Mantler, Hendrik",
    title = "{SusHi Bento: Beyond NNLO and the heavy-top limit}",
    eprint = "1605.03190",
    archivePrefix = "arXiv",
    primaryClass = "hep-ph",
    reportNumber = "DESY-16-061, KA-TP-14-2016, TTK-16-14",
    doi = "10.1016/j.cpc.2016.10.015",
    journal = "Comput. Phys. Commun.",
    volume = "212",
    pages = "239--257",
    year = "2017"
}

@article{Schienbein:2018fsw,
    author = "Schienbein, Ingo and Staub, Florian and Steudtner, Tom and Svirina, Kseniia",
    title = "{Revisiting RGEs for general gauge theories}",
    eprint = "1809.06797",
    archivePrefix = "arXiv",
    primaryClass = "hep-ph",
    reportNumber = "KA-TP-27-2018",
    doi = "10.1016/j.nuclphysb.2018.12.001",
    journal = "Nucl. Phys. B",
    volume = "939",
    pages = "1--48",
    year = "2019",
    note = "[Erratum: Nucl.Phys.B 966, 115339 (2021)]"
}

@article{Ferreira:2025ate,
    author = "Ferreira, P. M. and Grzadkowski, B. and Ogreid, O. M.",
    title = "{Imaginary scaling invariance of the one-loop effective potential}",
    eprint = "2506.21145",
    archivePrefix = "arXiv",
    primaryClass = "hep-ph",
    doi = "10.1103/9j32-2ggz",
    journal = "Phys. Rev. D",
    volume = "113",
    number = "7",
    pages = "075036",
    year = "2026"
}

@article{deBoer:2025jhc,
    author = "de Boer, Thede and Goertz, Florian and Incrocci, Andrea",
    title = "{The goofy-symmetric Standard Model and the Hierarchy Problem}",
    eprint = "2507.22111",
    archivePrefix = "arXiv",
    primaryClass = "hep-ph",
    month = "7",
    year = "2025"
}

@article{Trautner:2025prm,
    author = "Trautner, Andreas",
    title = "{Goofy transformations and the hierarchy problem}",
    eprint = "2508.02646",
    archivePrefix = "arXiv",
    primaryClass = "hep-ph",
    doi = "10.1016/j.physletb.2026.140190",
    journal = "Phys. Lett. B",
    volume = "873",
    pages = "140190",
    year = "2026"
}

@article{Grzadkowski:2026gkx,
    author = "Grzadkowski, Bohdan and Ogreid, Odd Magne",
    title = "{GOOFy {\textemdash} a systematic approach}",
    eprint = "2602.20849",
    archivePrefix = "arXiv",
    primaryClass = "hep-ph",
    doi = "10.1007/JHEP07(2026)038",
    journal = "JHEP",
    volume = "07",
    pages = "038",
    year = "2026"
}

@article{deBoer:2026ktb,
    author = "de Boer, Thede and Trautner, Andreas",
    title = "{Outer automorphisms are sufficient conditions for RG fixed points}",
    eprint = "2603.12318",
    archivePrefix = "arXiv",
    primaryClass = "hep-th",
    month = "3",
    year = "2026"
}

@article{Kuncinas:2025uty,
    author = "Kun{\v{c}}inas, A. and Osland, P. and Rebelo, M. N.",
    title = "{Systematic analysis of 3HDM symmetries}",
    eprint = "2512.07657",
    archivePrefix = "arXiv",
    primaryClass = "hep-ph",
    doi = "10.1103/hqkk-ffmp",
    journal = "Phys. Rev. D",
    volume = "113",
    number = "11",
    pages = "115054",
    year = "2026"
}

@article{Pilaftsis:2026wyw,
    author = "Pilaftsis, Apostolos",
    title = "{Renormalisation group invariants from scaling and non-overlapping symmetries}",
    eprint = "2605.18341",
    archivePrefix = "arXiv",
    primaryClass = "hep-th",
    doi = "10.1007/JHEP08(2026)039",
    journal = "JHEP",
    volume = "08",
    pages = "039",
    year = "2026"
}

@article{deMedeirosVarzielas:2026jsk,
    author = "de Medeiros Varzielas, I. and Kun{\v{c}}inas, A.",
    title = "{GOOFy-compatible 3HDMs and beyond}",
    eprint = "2608.05304",
    archivePrefix = "arXiv",
    primaryClass = "hep-ph",
    month = "8",
    year = "2026"
}

@article{Ecker:1987qp,
    author = "Ecker, G. and Grimus, W. and Neufeld, H.",
    title = "{A Standard Form for Generalized {CP} Transformations}",
    reportNumber = "UWThPh-1987-23",
    doi = "10.1088/0305-4470/20/12/010",
    journal = "J. Phys. A",
    volume = "20",
    pages = "L807",
    year = "1987"
}

@article{CMS:2021sdq,
    author = "Tumasyan, Armen and others",
    collaboration = "CMS",
    title = "{Analysis of the $CP$ structure of the Yukawa coupling between the Higgs boson and $\tau$ leptons in proton-proton collisions at $ \sqrt{s} $ = 13 TeV}",
    eprint = "2110.04836",
    archivePrefix = "arXiv",
    primaryClass = "hep-ex",
    reportNumber = "CMS-HIG-20-006, CERN-EP-2021-189",
    doi = "10.1007/JHEP06(2022)012",
    journal = "JHEP",
    volume = "06",
    pages = "012",
    year = "2022"
}

@article{ATLAS:2022akr,
    author = "Aad, Georges and others",
    collaboration = "ATLAS",
    title = "{Measurement of the CP properties of Higgs boson interactions with $\tau $-leptons with the ATLAS detector}",
    eprint = "2212.05833",
    archivePrefix = "arXiv",
    primaryClass = "hep-ex",
    reportNumber = "CERN-EP-2022-244",
    doi = "10.1140/epjc/s10052-023-11583-y",
    journal = "Eur. Phys. J. C",
    volume = "83",
    number = "7",
    pages = "563",
    year = "2023"
}

@article{CMS:2026xbb,
    author = "Hayrapetyan, Aram and others",
    collaboration = "CMS",
    title = "{Search for a new heavy scalar resonance decaying to a pair of Z bosons in the four-lepton final state in proton-proton collisions at $ \sqrt{s}=13 $ TeV}",
    eprint = "2605.26462",
    archivePrefix = "arXiv",
    primaryClass = "hep-ex",
    reportNumber = "CMS-HIG-24-002, CERN-EP-2026-116",
    doi = "10.1007/JHEP10(2026)006",
    journal = "JHEP",
    volume = "10",
    pages = "006",
    year = "2026"
}

@article{CMS:2022bcb,
    collaboration = "CMS",
    title = "{Search for high mass resonances decaying into $\mathrm{W^+}\mathrm{W^-}$ in the dileptonic final state with $138\,\text{fb}^{-1}$ of proton-proton collisions at $\sqrt{s}=13\,\text{TeV}$}",
    reportNumber = "CMS-PAS-HIG-20-016",
    year = "2022"
}

@article{CMS:2022rbd,
    collaboration = "CMS",
    title = "{Searches for additional Higgs bosons and vector leptoquarks in $\tau\tau$ final states in proton-proton collisions at $\sqrt{s}=13~\mathrm{TeV}$}",
    reportNumber = "CMS-PAS-HIG-21-001",
    year = "2022"
}

@article{ATLAS:2023vdy,
    author = "Aad, Georges and others",
    collaboration = "ATLAS",
    title = "{Combination of Searches for Resonant Higgs Boson Pair Production Using pp Collisions at s=13{\,}{\,}TeV with the ATLAS Detector}",
    eprint = "2311.15956",
    archivePrefix = "arXiv",
    primaryClass = "hep-ex",
    reportNumber = "CERN-EP-2023-271",
    doi = "10.1103/PhysRevLett.132.231801",
    journal = "Phys. Rev. Lett.",
    volume = "132",
    number = "23",
    pages = "231801",
    year = "2024"
}

@article{CMS:2026zxu,
    author = "Hayrapetyan, Aram and others",
    collaboration = "CMS",
    title = "{Combined measurements and interpretations of Higgs boson production and decay in proton{\textendash}proton collisions at $\sqrt{s} = 13\,\textrm{TeV}$}",
    eprint = "2602.18611",
    archivePrefix = "arXiv",
    primaryClass = "hep-ex",
    reportNumber = "CMS-HIG-21-018, CERN-EP-2026-009",
    doi = "10.1088/1361-6633/ae72c4",
    journal = "Rept. Prog. Phys.",
    volume = "89",
    number = "8",
    pages = "087801",
    year = "2026"
}

@article{ATLAS:2026pdi,
    author = "Aad, Georges and others",
    collaboration = "ATLAS",
    title = "{Combination of Higgs boson measurements at $\sqrt{s} =$ 13 TeV and their interpretations by the ATLAS experiment}",
    eprint = "2608.07332",
    archivePrefix = "arXiv",
    primaryClass = "hep-ex",
    reportNumber = "CERN-EP-2026-228",
    month = "8",
    year = "2026"
}
}

\appendix

\section{$S_3$-GOOFy conserving Yukawa matrices}
\label{ap:yukmat}

In Section~\ref{sec:yuk} eleven Yukawa matrices capable of generating three generations of massive fermions are mentioned. Here their explicit form is provided. As mentioned before, the fact that these are able to reproduce the same CKM matrix possibly signals the existence of basis changes that might relate some, if not all, of these structures. However, that is left for future work.

The Yukawa matrices are presented in a basis in which all their entries are real, obtained through the rephasing discussed in Section~\ref{sec:yuk}.
\begin{itemize}
% ----------------------------------------------------------------------
\item $\alpha=\pi/3$, $\gamma=\pi/3$.
    \begin{align}
        \Gamma_1 &=
        \begin{pmatrix}
        a_{11} & a_{12} & a_{13}\\
        a_{21} & a_{22} & a_{23}\\
        a_{31} & a_{32} & a_{33}
        \end{pmatrix},
        \\[1em]
        \Gamma_2 &= \frac14
        \begin{pmatrix}
        (a_{11}-\sqrt3 a_{12}-\sqrt3 a_{21}+3a_{22})
        & (\sqrt3 a_{11}+a_{12}-3a_{21}-\sqrt3 a_{22})
        & 2(-a_{13}+\sqrt3 a_{23})\\[4pt]
        (\sqrt3 a_{11}-3a_{12}+a_{21}-\sqrt3 a_{22})
        & (3a_{11}+\sqrt3 a_{12}+\sqrt3 a_{21}+a_{22})
        & 2(-\sqrt3 a_{13}-a_{23})\\[4pt]
        2(-a_{31}+\sqrt3 a_{32})
        & 2(-\sqrt3 a_{31}-a_{32})
        & 4 a_{33}
        \end{pmatrix},
        \\[1em]
        \Gamma_3 &= \frac14
        \begin{pmatrix}
        (a_{11}+\sqrt3 a_{12}+\sqrt3 a_{21}+3a_{22})
        & (-\sqrt3 a_{11}+a_{12}-3a_{21}+\sqrt3 a_{22})
        & 2(-a_{13}-\sqrt3 a_{23})\\[4pt]
        (-\sqrt3 a_{11}-3a_{12}+a_{21}+\sqrt3 a_{22})
        & (3a_{11}-\sqrt3 a_{12}-\sqrt3 a_{21}+a_{22})
        & 2(\sqrt3 a_{13}-a_{23})\\[4pt]
        2(-a_{31}-\sqrt3 a_{32})
        & 2(\sqrt3 a_{31}-a_{32})
        & 4 a_{33}
        \end{pmatrix}.
    \end{align}

% ----------------------------------------------------------------------
\item $\alpha=\pi/3$, $\gamma=5\pi/3$.
        \begin{align}
            \Gamma_1 &=
            \begin{pmatrix}
            a_{11} & a_{12} & a_{13}\\
            a_{21} & a_{22} & a_{23}\\
            a_{31} & a_{32} & a_{33}
            \end{pmatrix},
            \\[1em]
            \Gamma_2 &= \frac14
            \begin{pmatrix}
            (a_{11}+\sqrt3 a_{12}-\sqrt3 a_{21}-3a_{22})
            & (-\sqrt3 a_{11}+a_{12}+3a_{21}-\sqrt3 a_{22})
            & 2(-a_{13}+\sqrt3 a_{23})\\[4pt]
            (\sqrt3 a_{11}+3a_{12}+a_{21}+\sqrt3 a_{22})
            & (-3a_{11}+\sqrt3 a_{12}-\sqrt3 a_{21}+a_{22})
            & 2(-\sqrt3 a_{13}-a_{23})\\[4pt]
            2(-a_{31}-\sqrt3 a_{32})
            & 2(\sqrt3 a_{31}-a_{32})
            & 4 a_{33}
            \end{pmatrix},
            \\[1em]
            \Gamma_3 &= \frac14
            \begin{pmatrix}
            (a_{11}-\sqrt3 a_{12}+\sqrt3 a_{21}-3a_{22})
            & (\sqrt3 a_{11}+a_{12}+3a_{21}+\sqrt3 a_{22})
            & 2(-a_{13}-\sqrt3 a_{23})\\[4pt]
            (-\sqrt3 a_{11}+3a_{12}+a_{21}-\sqrt3 a_{22})
            & (-3a_{11}-\sqrt3 a_{12}+\sqrt3 a_{21}+a_{22})
            & 2(\sqrt3 a_{13}-a_{23})\\[4pt]
            2(-a_{31}+\sqrt3 a_{32})
            & 2(-\sqrt3 a_{31}-a_{32})
            & 4 a_{33}
            \end{pmatrix}.
        \end{align}

% ----------------------------------------------------------------------
\item $\alpha=2\pi/3$, $\gamma=\pi/3$.
        \begin{align}
            \Gamma_1 &=
            \begin{pmatrix}
            a_{11} & a_{12} & a_{13}\\
            a_{21} & a_{22} & a_{23}\\
            a_{31} & a_{32} & a_{33}
            \end{pmatrix},
            \\[1em]
            \Gamma_2 &= \frac14
            \begin{pmatrix}
            a_{11}-\sqrt3 a_{12}+\sqrt3 a_{21}-3a_{22}
            & \sqrt3 a_{11}+a_{12}+3a_{21}+\sqrt3 a_{22}
            & 2(-a_{13}-\sqrt3 a_{23})\\[4pt]
            -\sqrt3 a_{11}+3a_{12}+a_{21}-\sqrt3 a_{22}
            & -3a_{11}-\sqrt3 a_{12}+\sqrt3 a_{21}+a_{22}
            & 2(\sqrt3 a_{13}-a_{23})\\[4pt]
            2(-a_{31}+\sqrt3 a_{32})
            & 2(-\sqrt3 a_{31}-a_{32})
            & 4a_{33}
            \end{pmatrix},
            \\[1em]
            \Gamma_3 &= \frac14
            \begin{pmatrix}
            a_{11}+\sqrt3 a_{12}-\sqrt3 a_{21}-3a_{22}
            & -\sqrt3 a_{11}+a_{12}+3a_{21}-\sqrt3 a_{22}
            & 2(-a_{13}+\sqrt3 a_{23})\\[4pt]
            \sqrt3 a_{11}+3a_{12}+a_{21}+\sqrt3 a_{22}
            & -3a_{11}+\sqrt3 a_{12}-\sqrt3 a_{21}+a_{22}
            & 2(-\sqrt3 a_{13}-a_{23})\\[4pt]
            2(-a_{31}-\sqrt3 a_{32})
            & 2(\sqrt3 a_{31}-a_{32})
            & 4a_{33}
            \end{pmatrix}.
        \end{align}

% ----------------------------------------------------------------------
\item $\alpha=2\pi/3$, $\gamma=5\pi/3$.
        \begin{align}
            \Gamma_1 &=
            \begin{pmatrix}
            a_{11} & a_{12} & a_{13}\\
            a_{21} & a_{22} & a_{23}\\
            a_{31} & a_{32} & a_{33}
            \end{pmatrix},
            \\[1em]
            \Gamma_2 &= \frac14
            \begin{pmatrix}
            a_{11}+\sqrt3 a_{12}+\sqrt3 a_{21}+3a_{22}
            & -\sqrt3 a_{11}+a_{12}-3a_{21}+\sqrt3 a_{22}
            & 2(-a_{13}-\sqrt3 a_{23})\\[4pt]
            -\sqrt3 a_{11}-3a_{12}+a_{21}+\sqrt3 a_{22}
            & 3a_{11}-\sqrt3 a_{12}-\sqrt3 a_{21}+a_{22}
            & 2(\sqrt3 a_{13}-a_{23})\\[4pt]
            2(-a_{31}-\sqrt3 a_{32})
            & 2(\sqrt3 a_{31}-a_{32})
            & 4a_{33}
            \end{pmatrix},
            \\[1em]
            \Gamma_3 &= \frac14
            \begin{pmatrix}
            a_{11}-\sqrt3 a_{12}-\sqrt3 a_{21}+3a_{22}
            & \sqrt3 a_{11}+a_{12}-3a_{21}-\sqrt3 a_{22}
            & 2(-a_{13}+\sqrt3 a_{23})\\[4pt]
            \sqrt3 a_{11}-3a_{12}+a_{21}-\sqrt3 a_{22}
            & 3a_{11}+\sqrt3 a_{12}+\sqrt3 a_{21}+a_{22}
            & 2(-\sqrt3 a_{13}-a_{23})\\[4pt]
            2(-a_{31}+\sqrt3 a_{32})
            & 2(-\sqrt3 a_{31}-a_{32})
            & 4a_{33}
            \end{pmatrix}.
        \end{align}

% ----------------------------------------------------------------------
\item $\alpha=4\pi/3$, $\gamma=\pi/3$.
        \begin{align}
            \Gamma_1 &=
            \begin{pmatrix}
            a_{11} & a_{12} & a_{13}\\
            a_{21} & a_{22} & a_{23}\\
            a_{31} & a_{32} & a_{33}
            \end{pmatrix},
            \\[1em]
            \Gamma_2 &= \frac14
            \begin{pmatrix}
            a_{11}-\sqrt3 a_{12}-\sqrt3 a_{21}+3a_{22}
            & \sqrt3 a_{11}+a_{12}-3a_{21}-\sqrt3 a_{22}
            & 2(-a_{13}+\sqrt3 a_{23})\\[4pt]
            \sqrt3 a_{11}-3a_{12}+a_{21}-\sqrt3 a_{22}
            & 3a_{11}+\sqrt3 a_{12}+\sqrt3 a_{21}+a_{22}
            & 2(-\sqrt3 a_{13}-a_{23})\\[4pt]
            2(-a_{31}+\sqrt3 a_{32})
            & 2(-\sqrt3 a_{31}-a_{32})
            & 4a_{33}
            \end{pmatrix},
            \\[1em]
            \Gamma_3 &= \frac14
            \begin{pmatrix}
            a_{11}+\sqrt3 a_{12}+\sqrt3 a_{21}+3a_{22}
            & -\sqrt3 a_{11}+a_{12}-3a_{21}+\sqrt3 a_{22}
            & 2(-a_{13}-\sqrt3 a_{23})\\[4pt]
            -\sqrt3 a_{11}-3a_{12}+a_{21}+\sqrt3 a_{22}
            & 3a_{11}-\sqrt3 a_{12}-\sqrt3 a_{21}+a_{22}
            & 2(\sqrt3 a_{13}-a_{23})\\[4pt]
            2(-a_{31}-\sqrt3 a_{32})
            & 2(\sqrt3 a_{31}-a_{32})
            & 4a_{33}
            \end{pmatrix}.
        \end{align}

% ----------------------------------------------------------------------
\item $\alpha=4\pi/3$, $\gamma=5\pi/3$.
        \begin{align}
            \Gamma_1 &=
            \begin{pmatrix}
            a_{11} & a_{12} & a_{13}\\
            a_{21} & a_{22} & a_{23}\\
            a_{31} & a_{32} & a_{33}
            \end{pmatrix},
            \\[1em]
            \Gamma_2 &= \frac14
            \begin{pmatrix}
            a_{11}+\sqrt3 a_{12}-\sqrt3 a_{21}-3a_{22}
            & -\sqrt3 a_{11}+a_{12}+3a_{21}-\sqrt3 a_{22}
            & 2(-a_{13}+\sqrt3 a_{23})\\[4pt]
            \sqrt3 a_{11}+3a_{12}+a_{21}+\sqrt3 a_{22}
            & -3a_{11}+\sqrt3 a_{12}-\sqrt3 a_{21}+a_{22}
            & 2(-\sqrt3 a_{13}-a_{23})\\[4pt]
            2(-a_{31}-\sqrt3 a_{32})
            & 2(\sqrt3 a_{31}-a_{32})
            & 4 a_{33}
            \end{pmatrix},
            \\[1em]
            \Gamma_3 &= \frac14
            \begin{pmatrix}
            a_{11}-\sqrt3 a_{12}+\sqrt3 a_{21}-3a_{22}
            & \sqrt3 a_{11}+a_{12}+3a_{21}+\sqrt3 a_{22}
            & 2(-a_{13}-\sqrt3 a_{23})\\[4pt]
            -\sqrt3 a_{11}+3a_{12}+a_{21}-\sqrt3 a_{22}
            & -3a_{11}-\sqrt3 a_{12}+\sqrt3 a_{21}+a_{22}
            & 2(\sqrt3 a_{13}-a_{23})\\[4pt]
            2(-a_{31}+\sqrt3 a_{32})
            & 2(-\sqrt3 a_{31}-a_{32})
            & 4a_{33}
            \end{pmatrix}.
        \end{align}

% ----------------------------------------------------------------------
\item $\alpha=5\pi/3$, $\gamma=\pi/3$.
        \begin{align}
            \Gamma_1 &=
            \begin{pmatrix}
            a_{11} & a_{12} & a_{13}\\
            a_{21} & a_{22} & a_{23}\\
            a_{31} & a_{32} & a_{33}
            \end{pmatrix},
            \\[1em]
            \Gamma_2 &= \frac14
            \begin{pmatrix}
            (a_{11}-\sqrt3 a_{12}+\sqrt3 a_{21}-3a_{22})
            & (\sqrt3 a_{11}+a_{12}+3a_{21}+\sqrt3 a_{22})
            & 2(-a_{13}-\sqrt3 a_{23})\\[4pt]
            (-\sqrt3 a_{11}+3a_{12}+a_{21}-\sqrt3 a_{22})
            & (-3a_{11}-\sqrt3 a_{12}+\sqrt3 a_{21}+a_{22})
            & 2(\sqrt3 a_{13}-a_{23})\\[4pt]
            2(-a_{31}+\sqrt3 a_{32})
            & 2(-\sqrt3 a_{31}-a_{32})
            & 4a_{33}
            \end{pmatrix},
            \\[1em]
            \Gamma_3 &= \frac14
            \begin{pmatrix}
            (a_{11}+\sqrt3 a_{12}-\sqrt3 a_{21}-3a_{22})
            & (-\sqrt3 a_{11}+a_{12}+3a_{21}-\sqrt3 a_{22})
            & 2(-a_{13}+\sqrt3 a_{23})\\[4pt]
            (\sqrt3 a_{11}+3a_{12}+a_{21}+\sqrt3 a_{22})
            & (-3a_{11}+\sqrt3 a_{12}-\sqrt3 a_{21}+a_{22})
            & 2(-\sqrt3 a_{13}-a_{23})\\[4pt]
            2(-a_{31}-\sqrt3 a_{32})
            & 2(\sqrt3 a_{31}-a_{32})
            & 4a_{33}
            \end{pmatrix}.
        \end{align}

% ----------------------------------------------------------------------
\item $\alpha=5\pi/3$, $\gamma=5\pi/3$.
        \begin{align}
            \Gamma_1 &=
            \begin{pmatrix}
            a_{11} & a_{12} & a_{13}\\
            a_{21} & a_{22} & a_{23}\\
            a_{31} & a_{32} & a_{33}
            \end{pmatrix},
            \\[1em]
            \Gamma_2 &= \frac14
            \begin{pmatrix}
            (a_{11}+\sqrt3 a_{12}+\sqrt3 a_{21}+3a_{22})
            & (-\sqrt3 a_{11}+a_{12}-3a_{21}+\sqrt3 a_{22})
            & 2(-a_{13}-\sqrt3 a_{23})\\[4pt]
            (-\sqrt3 a_{11}-3a_{12}+a_{21}+\sqrt3 a_{22})
            & (3a_{11}-\sqrt3 a_{12}-\sqrt3 a_{21}+a_{22})
            & 2(\sqrt3 a_{13}-a_{23})\\[4pt]
            2(-a_{31}-\sqrt3 a_{32})
            & 2(\sqrt3 a_{31}-a_{32})
            & 4a_{33}
            \end{pmatrix},
            \\[1em]
            \Gamma_3 &= \frac14
            \begin{pmatrix}
            (a_{11}-\sqrt3 a_{12}-\sqrt3 a_{21}+3a_{22})
            & (\sqrt3 a_{11}+a_{12}-3a_{21}-\sqrt3 a_{22})
            & 2(-a_{13}+\sqrt3 a_{23})\\[4pt]
            (\sqrt3 a_{11}-3a_{12}+a_{21}-\sqrt3 a_{22})
            & (3a_{11}+\sqrt3 a_{12}+\sqrt3 a_{21}+a_{22})
            & 2(-\sqrt3 a_{13}-a_{23})\\[4pt]
            2(-a_{31}+\sqrt3 a_{32})
            & 2(-\sqrt3 a_{31}-a_{32})
            & 4a_{33}
            \end{pmatrix}.
        \end{align}

% ----------------------------------------------------------------------
\item $\alpha=5\pi/3$, $\gamma=2\pi/3$.
        \begin{align}
            \Gamma_1 &=
            \begin{pmatrix}
            a_{11} & a_{12} & a_{13}\\
            a_{21} & a_{22} & a_{23}\\
            a_{31} & a_{32} & a_{33}
            \end{pmatrix},
            \\[1em]
            \Gamma_2 &= \frac14
            \begin{pmatrix}
            a_{11}+\sqrt3 a_{12}+\sqrt3 a_{21}+3a_{22}
            & -\sqrt3 a_{11}+a_{12}-3a_{21}+\sqrt3 a_{22}
            & 2(-a_{13}-\sqrt3 a_{23})\\[4pt]
            -\sqrt3 a_{11}-3a_{12}+a_{21}+\sqrt3 a_{22}
            & 3a_{11}-\sqrt3 a_{12}-\sqrt3 a_{21}+a_{22}
            & 2(\sqrt3 a_{13}-a_{23})\\[4pt]
            2(-a_{31}-\sqrt3 a_{32})
            & 2(\sqrt3 a_{31}-a_{32})
            & 4a_{33}
            \end{pmatrix},
            \\[1em]
            \Gamma_3 &= \frac14
            \begin{pmatrix}
            a_{11}-\sqrt3 a_{12}-\sqrt3 a_{21}+3a_{22}
            & \sqrt3 a_{11}+a_{12}-3a_{21}-\sqrt3 a_{22}
            & 2(-a_{13}+\sqrt3 a_{23})\\[4pt]
            \sqrt3 a_{11}-3a_{12}+a_{21}-\sqrt3 a_{22}
            & 3a_{11}+\sqrt3 a_{12}+\sqrt3 a_{21}+a_{22}
            & 2(-\sqrt3 a_{13}-a_{23})\\[4pt]
            2(-a_{31}+\sqrt3 a_{32})
            & 2(-\sqrt3 a_{31}-a_{32})
            & 4a_{33}
            \end{pmatrix}.
        \end{align}

% ----------------------------------------------------------------------
\item $\alpha=4\pi/3$, $\gamma=2\pi/3$.
        \begin{align}
            \Gamma_1 &= 
            \begin{pmatrix}
            a_{11} & a_{12} & a_{13}\\
            a_{21} & a_{22} & a_{23}\\
            a_{31} & a_{32} & a_{33}
            \end{pmatrix},
            \\[1em]
            \Gamma_2 &= \frac{1}{4}
            \begin{pmatrix}
            a_{11}+\sqrt3 a_{12}-\sqrt3 a_{21}-3a_{22}
            & -\sqrt3 a_{11}+a_{12}+3a_{21}-\sqrt3 a_{22}
            & 2(-a_{13}+\sqrt3 a_{23})\\[4pt]
            \sqrt3 a_{11}+3a_{12}+a_{21}+\sqrt3 a_{22}
            & -3a_{11}+\sqrt3 a_{12}-\sqrt3 a_{21}+a_{22}
            & 2(-\sqrt3 a_{13}-a_{23})\\[4pt]
            2(-a_{31}-\sqrt3 a_{32})
            & 2(\sqrt3 a_{31}-a_{32})
            & 4a_{33}
            \end{pmatrix},
            \\[1em]
            \Gamma_3 &= \frac{1}{4}
            \begin{pmatrix}
            a_{11}-\sqrt3 a_{12}+\sqrt3 a_{21}-3a_{22}
            & \sqrt3 a_{11}+a_{12}+3a_{21}+\sqrt3 a_{22}
            & 2(-a_{13}-\sqrt3 a_{23})\\[4pt]
            -\sqrt3 a_{11}+3a_{12}+a_{21}-\sqrt3 a_{22}
            & -3a_{11}-\sqrt3 a_{12}+\sqrt3 a_{21}+a_{22}
            & 2(\sqrt3 a_{13}-a_{23})\\[4pt]
            2(-a_{31}+\sqrt3 a_{32})
            & 2(-\sqrt3 a_{31}-a_{32})
            & 4a_{33}
            \end{pmatrix}.
        \end{align}

% ----------------------------------------------------------------------
\item $\alpha=2\pi/3$, $\gamma=4\pi/3$.
        \begin{align}
            \Gamma_1 &= 
            \begin{pmatrix}
            a_{11} & a_{12} & a_{13}\\
            a_{21} & a_{22} & a_{23}\\
            a_{31} & a_{32} & a_{33}
            \end{pmatrix},
            \\[1em]
            \Gamma_2 &= \frac{1}{4}
            \begin{pmatrix}
            a_{11}-\sqrt3 a_{12}+\sqrt3 a_{21}-3a_{22}
            & \sqrt3 a_{11}+a_{12}+3a_{21}+\sqrt3 a_{22}
            & 2(-a_{13}-\sqrt3 a_{23})\\[4pt]
            -\sqrt3 a_{11}+3a_{12}+a_{21}-\sqrt3 a_{22}
            & -3a_{11}-\sqrt3 a_{12}+\sqrt3 a_{21}+a_{22}
            & 2(\sqrt3 a_{13}-a_{23})\\[4pt]
            2(-a_{31}+\sqrt3 a_{32})
            & 2(-\sqrt3 a_{31}-a_{32})
            & 4a_{33}
            \end{pmatrix},
            \\[1em]
            \Gamma_3 &= \frac{1}{4}
            \begin{pmatrix}
            a_{11}+\sqrt3 a_{12}-\sqrt3 a_{21}-3a_{22}
            & -\sqrt3 a_{11}+a_{12}+3a_{21}-\sqrt3 a_{22}
            & 2(-a_{13}+\sqrt3 a_{23})\\[4pt]
            \sqrt3 a_{11}+3a_{12}+a_{21}+\sqrt3 a_{22}
            & -3a_{11}+\sqrt3 a_{12}-\sqrt3 a_{21}+a_{22}
            & 2(-\sqrt3 a_{13}-a_{23})\\[4pt]
            2(-a_{31}-\sqrt3 a_{32})
            & 2(\sqrt3 a_{31}-a_{32})
            & 4a_{33}
            \end{pmatrix}.
        \end{align}

\end{itemize}

\section{Rotation to the Physical Basis of the Scalar Sector} \label{sec: rot_mass_basis}

For the potential with the exact GOOFy symmetry,
the rotation to the physical basis of the scalar sector is accomplished in two phases: first we move to the Higgs basis, and then we diagonalise 
the remaining matrix. The full rotation matrix is then given by the following matrix product,
\begin{equation}
    R= U \cdot U_{Higgs}
\end{equation}
\begin{align}
    U_{Higgs} = & 
    \begin{pmatrix}
     \frac{1}{\sqrt{3}} & -\frac{1}{2 \sqrt{3}} & -\frac{1}{2 \sqrt{3}} & 0 & \frac{1}{2} & -\frac{1}{2} \\
     0 & -\frac{1}{2 \sqrt{2}} & \frac{1}{2 \sqrt{2}} & 0 & \frac{\sqrt{\frac{3}{2}}}{2} & \frac{\sqrt{\frac{3}{2}}}{2} \\
     -\sqrt{\frac{2}{3}} & -\frac{1}{2 \sqrt{6}} & -\frac{1}{2 \sqrt{6}} & 0 & \frac{1}{2 \sqrt{2}} & -\frac{1}{2 \sqrt{2}} \\
     0 & -\frac{1}{2} & \frac{1}{2} & \frac{1}{\sqrt{3}} & -\frac{1}{2 \sqrt{3}} & -\frac{1}{2 \sqrt{3}} \\
     0 & -\frac{\sqrt{\frac{3}{2}}}{2} & -\frac{\sqrt{\frac{3}{2}}}{2} & 0 & -\frac{1}{2 \sqrt{2}} & \frac{1}{2 \sqrt{2}} \\
     0 & -\frac{1}{2 \sqrt{2}} & \frac{1}{2 \sqrt{2}} & -\sqrt{\frac{2}{3}} & -\frac{1}{2 \sqrt{6}} & -\frac{1}{2 \sqrt{6}} \\
    \end{pmatrix}
    \\
    U=&
    \begin{pmatrix}
     1 & 0 & 0 & 0 & 0 & 0 \\
     0 & -\frac{A \left(\sqrt{A^2+16 B^2+432 \Lambda_{8}^2}-4 B\right)}{A^2+432 \Lambda_{8}^2} & -\frac{12 \sqrt{3} \Lambda_{8} \left(\sqrt{A^2+16 B^2+432 \Lambda_{8}^2}-4 B\right)}{A^2+432 \Lambda_{8}^2} & 0 & 0 & 1 \\
     0 & -\frac{12 \sqrt{3} \Lambda_{8} \left(\sqrt{A^2+16 B^2+432 \Lambda_{8}^2}-4 B\right)}{A^2+432 \Lambda_{8}^2} & \frac{A \left(\sqrt{A^2+16 B^2+432 \Lambda_{8}^2}-4 B\right)}{A^2+432 \Lambda_{8}^2} & 0 & 1 & 0 \\
     0 & 0 & 0 & 1 & 0 & 0 \\
     0 & \frac{A \left(\sqrt{A^2+16 B^2+432 \Lambda_{8}^2}+4 B\right)}{A^2+432 \Lambda_{8}^2} & \frac{12 \sqrt{3} \Lambda_{8} \left(\sqrt{A^2+16 B^2+432 \Lambda_{8}^2}+4 B\right)}{A^2+432 \Lambda_{8}^2} & 0 & 0 & 1 \\
     0 & \frac{12 \sqrt{3} \Lambda_{8} \left(\sqrt{A^2+16 B^2+432 \Lambda_{8}^2}+4 B\right)}{A^2+432 \Lambda_{8}^2} & -\frac{A \left(\sqrt{A^2+16 B^2+432 \Lambda_{8}^2}+4 B\right)}{A^2+432 \Lambda_{8}^2} & 0 & 1 & 0 \\
    \end{pmatrix}
\end{align}
where the quantities $A,B~\text{and}~C$ were defined in Eq. \eqref{eq: ABC}.

Writing each doublet as
\beq
\Phi_a\,=\,\begin{pmatrix}
    \phi_a^+ \\ \frac{1}{\sqrt{2}} (v_a  + \rho_a + i \eta_a)
\end{pmatrix}
\eeq
the rotation between neutral real and imaginary doublet components and the mass eigenstates is therefore,
\begin{equation}
    \begin{pmatrix}
        \rho_1\\
        \rho_2\\
        \rho_3\\
        \eta_1\\
        \eta_2\\
        \eta_3
    \end{pmatrix} 
    = R 
    \begin{pmatrix}
        h_{SM}\\
        H_1\\
        H_2\\
        G\\
        H_3\\
        H_4
    \end{pmatrix} \,. 
    \label{eq: physical_states_scalars}
\end{equation}
Notice that in this equation $H_{1,2}$ are degenerate in mass and identified with the mass state $h_1$ in~\eqref{eq:massesex}. Likewise,
$H_{3,4}$ are degenerate in mass and identified with the mass state $h_2$ in the same equation. Inverting~\eqref{eq: physical_states_scalars}
gives us
\begin{equation}
h_{SM} \,=\,    \frac{1}{2} \left(\eta_{2}-\eta_{3}+\frac{2 \rho_{1}}{\sqrt{3}}-\frac{\rho_{2}}{\sqrt{3}}-\frac{\rho_{3}}{\sqrt{3}} \right) \,.
\end{equation}

\end{document}